\documentstyle[11pt,apjpt4,tighten,amstex]{article}
\begin{document}
\def\gtsima{$\, \buildrel > \over \sim \,$}
\def\ltsima{$\, \buildrel < \over \sim \,$}
\def\simgt{\lower.5ex\hbox{\gtsima}}
\def\simlt{\lower.5ex\hbox{\ltsima}}

\def\sm{$\sim\,$}
\def\smgt{$\simgt\,$}
\def\smlt{$\simlt\,$}
\def\smeq{$\simeq\,$}
\def\onesigma{$1\,\sigma$}
\def\nhat{\ifmmode {\hat{\bf n}}\else${\hat {\bf n}}$\fi}

\def\degs{\ifmmode^\circ\,\else$^\circ\,$\fi}
\def\kps{\ifmmode{\rm km}\,{\rm s}^{-1}\else km$\,$s$^{-1}$\fi}
\def\kms{\ifmmode{\rm km}\,{\rm s}^{-1}\else km$\,$s$^{-1}$\fi}
\def\ksmpc{\ifmmode{\rm km}\,{\rm s}^{-1}\,{\rm Mpc}^{-1}\else km$\,$s$^{-1}\,$Mpc$^{-1}$\fi}
\def\kmsmpc{\kms\ {{\rm Mpc}}^{-1}}
\def\etal{{\sl et al.}}
\def\ie{{\it i.e.}}
\def\eg{{\it e.g.}}
\def\apriori{{\em a priori}}
\def\aposteriori{{\em a posteriori}}
\def\halpha{H$\alpha$}
\def\h1{$h^{-1}$}
\def\dnsigma{$D_n$-$\sigma$}
\font\tensm=cmcsc10
\def\hii{H\kern 2.0pt{\tensm ii}}	
\def\hi{\ifmmode{\rm H\kern 2.0pt{\tensm I}}\else H\kern 2.0pt{\tensm I}\fi}	
\def\potent{{\small POTENT}}
\def\potiras{{\small POTIRAS}}
\def\simpot{{\small SIMPOT}}

\def\dinv{d_{{\rm inv}}}
\def\deltainv{\Delta_{{\rm inv}}}
\def\muinv{\mu(\dinv)}
\def\delinv{\Delta_{{\rm inv}}}
\def\onehalf{\frac{1}{2}}
\def\sigz{\sigma_z}
\def\sigeta{\sigma_{\eta}}
\def\sigr{\sigma_r}
\def\sigv{\sigma_v}
\def\sigin{\sigma_{{\rm in}}}
\def\sigout{\sigma_{{\rm out}}}
\def\sigM{\sigma_M}
\def\sigrb{\sigma_{\rm RB}}
\def\etazero{\eta^0}
\def\etapr{\eta^{0\prime}}
\def\nsigeta{\frac{1}{\sqrt{2\pi}\,\sigeta}}
\def\nsigr{\frac{1}{\sqrt{2\pi}\,\sigr}}
\def\nsigv{\frac{1}{\sqrt{2\pi}\,\sigv}}
\def\nsigM{\frac{1}{\sqrt{2\pi}\,\sigM}}
\def\nsig{\frac{1}{\sqrt{2\pi}\,\sigma}}
\def\nsigD{\frac{1}{\sqrt{2\pi}\,\Delta}}
\def\nsigDinv{\frac{1}{\sqrt{2\pi}\,\delinv}}
\def\nsigzero{{1 \over \sqrt{2\pi}\,\sigma_0}}
\def\nsigm0{{1 \over \sqrt{2\pi}\,\sigma_{0m}}}
\def\nsigxi{\frac{1}{\sqrt{2\pi}\,\sigma_\xi}}
\def\nsigz{\frac{1}{\sqrt{2\pi}\,\sigz}}
\def\2overpi{2 \over \pi}
\def\expmM{\exp\!\left(-\frac{\left[m-(M(\eta)+\mu(r))\right]^2}{2\sigtf^2} \right)}
\def\expmMm0{\exp\!\left(-\frac{(m-(M(\eta_0)+\mu(r)))^2}{2\sigma_{0m}^2} \right)}
\def\expeta{e^{-\frac{(\eta-\eta_0)^2}{2\sigeta^2}}}
\def\expet{\exp\!\left(-\frac{\left[\eta - \eta^0(m-\mu(r))\right]^2}{2\sigeta^2}\right)}
\def\expetm{\exp\!\left(-\frac{[\eta - \eta^0(m-\mu)]^2}{2\sigeta^2}\right)}
\def\expmm{\exp\!\left(-\frac{(m-m(\eta))^2}{2\sigma^2}\right)}
\def\expmu{\exp\!\left(-\frac{(\mu(r) - \mu(m,\eta))^2}{2 \sigma^2} \right)}
\def\expmuD{\exp\!\left(-\frac{(\mu(r) - \mu(d))^2}{2 \sigma^2} \right)}
\def\explnrD{\exp\!\left(-\frac{\left[\ln r/d \right]^2}{2 \Delta^2} \right)}
\def\explnrDinv{\exp\!\left(-\frac{\left[\ln r/\dinv \right]^2}{2 \delinv^2} \right)}
\def\explnvD{e^{-\frac{\left[\ln\frac{v_c}{\vd}\right]^2}{2 \Delta^2}}}
\def\explnu{e^{-\frac{\left[\ln\frac{v_r-u}{v_c}\right]^2}{2 \Delta^2}}}
\def\explnx{e^{-\frac{\left(\ln x\right)^2}{2 \Delta^2}}}
\def\explnxinv{e^{-\frac{\left(\ln x\right)^2}{2 \delinv^2}}}
\def\expvr{\exp\!\left(-\frac{[v_r - (v_c+ v_p(v_c))]^2}{2\sigr^2}\right)}
\def\expcz{\exp\!\left(-\frac{[cz - (r+ u(r))]^2}{2\sigv^2}\right)}
\def\expxi{\exp\!\left(-\frac{(\xi - \xi(m,\eta))^2}{2 \sigma_{\xi}^2} \right)}
\def\expz{\exp\!\left(-\frac{(m_z-m_z(m,\eta))^2}{2\sigz^2} \right)}
\def\xlim{\xi_\ell}
\def\xint{\int_{\xlim}^\infty\,}
\def\xxint{\int^{\xlim}_{-\infty}\,}
\def\erf{{\rm erf}}
\def\infint{\int_{-\infty}^\infty}
\def\inf0int{\int_0^\infty}
\def\mlim{m_\ell}
\def\mlint{\int_{-\infty}^{\mlim}\,}
\def\Aint{\int_{-\infty}^\cala\,}
\def\onplserf{\left[1+\erf\left(\frac{\xi(m,\eta)-\xlim}{\sqrt{2}\,\sigma_\xi}\right)\right]}
\def\oneplserf{\left[1+\erf\left(\frac{\xi(m,\eta^0(m-\mu(r)))-\xlim}{\sqrt{2}\,\sigma_\xi}\right)\right]}
\def\sigxi{\sigma_\xi}
\def\atan{{\rm tan}^{-1}\,}
\def\mij{m_{ij}}
\def\etaij{\eta_{ij}}	
\def\dij{d_{ij}}
\def\vij{v_{ij}}
\def\muij{\mu_{ij}}

\def\dugc{\ifmmode D_{{\rm UGC}}\else$D_{{\rm UGC}}$\fi}
\def\logdugc{\ifmmode \log\dugc\else$\log\dugc$\fi}
\def\deso{\ifmmode D_{{\rm ESO}}\else$D_{{\rm ESO}}$\fi}
\def\logdeso{\ifmmode \log\deso\else$\log\deso$\fi}
\def\iras{{\sl IRAS\/}}
\def\potent{{\small POTENT\/}}
\def\itf{{\small ITF\/}}
\def\velmod{{\small VELMOD\/}}
\def\bfv{{\bf v}}
\def\verr{\bfv_{{\rm err}}}
\def\bfw{{\bf w}}
\def\bfwlg{\bfw_{{\rm LG}}}
\def\bfV{{\bf V}}
\def\bfr{{\bf r}}
\def\bfs{{\bf s}}
\def\bfq{{\bf q}}
\def\bfu{{\bf u}}
\def\bfx{{\bf x}}
\def\bfk{{\bf k}}
\def\bfz{{\bf z}}
\def\bfn{{\bf n}}
\def\sigtf{\sigma_{{\rm TF}}}
\def\vev#1{{\left\langle#1\right\rangle}}

\title{Maximum-Likelihood Comparisons of Tully-Fisher \\
and Redshift Data: Constraints on $\Omega$ and Biasing}
\author{Jeffrey A.\ Willick$^a$, Michael A.\ Strauss$^{b,e}$,
Avishai Dekel$^{c,f}$, and Tsafrir Kolatt$^d$\\
\bigskip
{\small
$^a$ Dept.\ of Physics, Stanford University,
Stanford, CA 94305-4060
{\tt (jeffw@@perseus.stanford.edu)} \\
$^b$ Dept.\ Astrophysical Sciences, Princeton University,
Princeton, NJ 08544
{\tt (strauss@@astro.princeton.edu)} \\
$^c$ Racah Institute of Physics, The Hebrew University of Jerusalem,
Jerusalem 91904, Israel
{\tt (dekel@@astro.huji.ac.il)} \\
$^d$  Harvard-Smithsonian Center for Astrophysics, 60 Garden Street, Cambridge,
MA 02138, and\\
UCO/Lick Observatory, University of California, Santa Cruz, CA 95064
{\tt (tsafrir@@ucolick.org)} \\
$^e$ Alfred P.\ Sloan Foundation Fellow
$^f$ Center for Particle Astrophysics, University of California,
Berkeley, CA 94720 \\
}}

\begin{abstract}
We compare Tully-Fisher (TF) data for 838 galaxies
within $cz = 3000\ \kms$ from the Mark III catalog
to the peculiar velocity and
density fields
predicted from the 1.2 Jy \iras\ redshift survey.
Our goal is to test the relation between the galaxy density and 
velocity fields predicted by gravitational instability theory and
linear biasing, and
thereby to estimate 
$\beta_I\equiv\Omega^{0.6}/b_I,$ where $b_I$ is the linear
bias parameter for \iras\ galaxies on a 300 \kms\ scale. 
Adopting the \iras\ velocity and density fields
as a prior model,
we maximize the likelihood of the raw TF observables,
taking into account the full range of selection effects 
and properly treating  
triple-valued zones in the redshift-distance relation.           
Extensive tests with realistic simulated galaxy catalogs 
demonstrate that
the method produces unbiased estimates of $\beta_I$ and its error.
When we apply the method
to the real data, we model the presence of a small
but significant 
velocity quadrupole residual (\sm 3.3\% of Hubble flow), which we argue is due
to density fluctuations incompletely sampled by \iras.
The method then yields
a maximum likelihood estimate $\beta_I=0.49\pm 0.07$
(\onesigma\ error). 
We discuss the constraints on $\Omega$ and biasing that follow
from this estimate of $\beta_I$ if we assume a
COBE-normalized CDM power spectrum.
Our model also yields the one dimensional noise in the velocity
field, including \iras\ prediction errors, 
which we find to
be $125\pm 20\ \kms.$ 
 
We define a $\chi^2$-like statistic,
$\chi^2_\xi,$
that measures the coherence of residuals between the
TF data and the \iras\ model. In contrast with
maximum likelihood, this statistic can
identify poor fits, but is
relatively insensitive to the best $\beta_I.$
As measured by $\chi^2_\xi,$
the \iras\ model does
not fit the data well without accounting
for the residual quadrupole;
when the quadrupole is added the fit {\em is} acceptable
for $0.3 \leq \beta_I \leq 0.9$.
We discuss this in view of the
Davis, Nusser, \& Willick 
analysis that questions the consistency of the TF and \iras\ data.
\end{abstract}

\section{Introduction}
\label{sec:intro}
One of the most important tasks facing observational cosmology is
determination of the density parameter $\Omega.$
Along with the Hubble constant $H_0$ and the 
cosmological constant $\Lambda,$ the density parameter
fixes the global structure of spacetime.
One approach to the problem uses the classical cosmological tests
of the geometry of the universe, such as the apparent
magnitudes as a function of redshift of standard candles
(\eg, Type Ia Supernovae, Perlmutter \etal\ 1996). 
While promising, this approach is sensitive to
the possible evolution of the standard candles with redshift. 
Moreover, it is difficult to disentangle the
effects of $\Lambda$ and $\Omega$ in such tests (Dekel, Burstein,
\& White 1997).
Alternatively, one may carry out {\em dynamical\/} measurements
of $\Omega$ in the local ($z\simlt 0.05$) universe,
in which both evolution and
the geometrical effects of 
the cosmological constant may
be safely neglected.

Low-redshift tests of $\Omega$ are based on dynamical measurements of
the mass of gravitating matter on some characteristic size scale.
For example, measurements of rotation curves
(Rubin 1983) or the motions of
satellite galaxies
(Zaritsky \etal\ 1993) yield the masses of ordinary
spirals within \sm 10--200 kpc of their centers.
The velocity dispersions (Carlberg \etal\ 1996), X-ray temperatures (White
\etal\ 1993), and gravitational
lensing effects (Tyson \& Fischer 1995; Squires \etal\ 1996) of
rich clusters of galaxies provide
mass estimates on \sm 1 Mpc scales.
In general, these and other dynamical analyses
of matter
in the highly clustered regime 
have pointed to a mass density corresponding
to $\Omega \simeq 0.2 \pm 0.1$ (e.g.,
Bahcall, Lubin, \& Dorman 1995). 
This value exceeds
that implied by known sources of luminosity ($\Omega_{{\rm lum}}
\simlt 0.01;$ Peebles 1993)
or inferred from primordial nucleosynthesis ($\Omega_{{\rm baryon}}
\simlt 0.05;$ Turner \etal\ 1996),
and thus points to the existence of nonbaryonic dark matter.
However, it is well below the Einstein-de Sitter      
value of $\Omega=1$ that is favored by simplicity 
and coincidence arguments (e.g., Dicke 1970).
The natural expectation from the inflation scenario is that 
the universe is flat, $\Omega + \Omega_\Lambda =1$,                 
where $\Omega_\Lambda\equiv\Lambda/3H_0^2$ is the effective energy  
density contributed by a cosmological constant  (Guth
1981; Linde 1982; Albrecht \& Steinhardt 1982).                   
However, if $\Omega\simeq 0.2,$ this inflationary prediction requires
$\Omega_\Lambda\simeq 0.8,$ which
conflicts
with upper limits
obtained from studies of gravitational lensing (Carroll, Press, \& Turner 1992;
Maoz \& Rix 1993; Kochanek 1996).

It is possible, however, that $\Omega$ could be close
to or exactly equal to unity despite evidence
to the contrary from
dynamical tests on \sm 1 Mpc scales. 
This could occur if the dark matter
is poorly traced by dense concentrations of luminous
matter such as galaxies and galaxy clusters. 
If so, dynamical tests
on scales $\simgt 10$ Mpc 
are necessary to
obtain an unbiased estimate of $\Omega.$
Such tests involve measurements of
the coherent, large-scale peculiar velocities of
galaxies. 
According to gravitational instability
theory (cf.\ Eq.~\ref{eq:vpdelta}), 
these motions are related in an $\Omega$-dependent
way to the large-scale distribution
of mass. If the latter, in turn, can be inferred from the observed
distribution of galaxies on large scales, one might hope
to derive an estimate of $\Omega$ that
is free from the pitfalls of
small-scale dynamical analyses.

This program requires
a comparative analysis of two types of data sets.
The first consists of radial
velocities and redshift-independent
distance estimates for large samples of galaxies.
The largest such compilation to date is
the Mark III catalog
(Willick \etal\ 1997), which contains
distance estimates for \sm 3000 spiral galaxies from
the Tully-Fisher (1977; TF) relation, and for 544
elliptical galaxies from
the \dnsigma\ relation (Djorgovski \& Davis 1987; Dressler \etal\ 1987).
The second type of data set is a full-sky redshift
survey with well-understood selection criteria.
Several large redshift
surveys exist (cf.\
Strauss \& Willick 1995,
hereafter SW, and Strauss 1996a, for reviews); 
the one which most nearly meets the requirements
of full-sky coverage and well-understood selection is
the \iras\ 1.2 Jy survey (Fisher \etal\ 1995).
The basic idea behind the comparison is as follows. In the
linear regime (mass density fluctuations $\delta \equiv
\delta\rho/\rho_0 \ll 1$),
the global relationship between the peculiar
velocity field $\bfv(\bfr)$ and
the mass-density fluctuation field $\delta(\bfr)$ is given by
gravitational instability theory:
\begin{equation}
\bfv(\bfr) = \frac{f(\Omega)}{4 \,\pi} \int d^3\bfr'\,\frac{\delta(\bfr')
(\bfr' - \bfr)}{\left|\bfr' - \bfr\right|^3}\,,
\label{eq:vpdelta}
\end{equation}
where $f(\Omega)\approx\Omega^{0.6}$ (Peebles 1980).\footnote{We measure
distances $\bfr$ in velocity units (\kms). In such a system of units,
the Hubble Constant is
equal to unity by definition, and does not affect the amplitude of predicted
peculiar velocities.}
If mass density fluctuations are equal to
galaxy number density fluctuations,
at least on the scales ($\simgt$ few Mpc) over which it is possible to
define continuous density fields, then
the redshift survey data yield a map of
$\delta(\bfr)$ (after correction for peculiar velocities;
Appendix~\ref{sec:iras}). By Eq.~(\ref{eq:vpdelta}),
one then derives a predicted peculiar velocity field
$\bfv(\bfr)$ as a function of $\Omega.$ The
TF or \dnsigma\ data provide the observed peculiar velocities.
The best estimate of $\Omega$ is the one
for which the predicted and observed peculiar velocities
best agree.

Two obstacles make this comparison a difficult one.
The first,
already alluded to, is fundamental: one observes galaxy number
density ($\delta_g$) rather than mass density ($\delta$) fluctuations.
A model is required for relating the first to the second.
The simplest approximation is
{\em linear\/} biasing,
\begin{equation}
\delta_g(\bfr) = b\, \delta(\bfr),
\label{eq:linear-bias}
\end{equation}
in which the bias parameter $b$ is assumed
to be spatially constant.
Substituting Eq.~(\ref{eq:linear-bias}) in
Eq.~(\ref{eq:vpdelta}) yields
\begin{equation}
\bfv(\bfr) = \frac{\beta}{4 \,\pi} \int d^3\bfr'\,\frac{\delta_g(\bfr')
(\bfr' - \bfr)}{\left|\bfr' - \bfr\right|^3}\,,
\label{eq:vpdeltag}
\end{equation}
where $\beta\equiv f(\Omega)/b.$
Thus, under the dual assumptions of
linear dynamics and linear biasing, 
comparisons of peculiar velocity and redshift
survey data, by themselves, can yield the parameter $\beta$ but not
$\Omega.$
One might hope to break the $\Omega$-$b$ degeneracy by generalizing
Eq.~(\ref{eq:vpdelta}) to the nonlinear dynamical regime (cf.\
Dekel 1994, \S~2, or Sahni \& Coles 1996, for a review).
However, such generalizations are difficult to implement
in practice;
furthermore, nonlinear extensions to Eq.~(\ref{eq:linear-bias}) will
enter to the same order as
nonlinear dynamics (we discuss this issue further in \S~\ref{sec:nonlinear}).
Thus, without a more realistic {\em a priori\/} model of the relative
distribution of galaxies versus mass, it is prudent to limit the
goals of the peculiar velocity-redshift survey comparison to
testing gravitational instability theory 
and determining 
$\beta.$ One may then adduce external information 
on the value of $b$ to place constraints on $\Omega$ itself.

The second obstacle is the sheer technical difficulty
of the problem. 
The redshift-independent
distances obtained from methods such as TF 
are large (\sm 20\%; Willick
\etal\ 1996), and are subject to potential
systematic errors due to statistical bias effects
(Dekel 1994; 
SW, \S~6).
Furthermore,
we measure the galaxy density field $\delta_g$ in {\em redshift\/} space,
whereas it is the real-space density that yields peculiar
velocities via Eq.~(\ref{eq:vpdeltag}). The relationship between
the two depends on the peculiar velocity field itself. Self-consistent
methods, in which $\bfv_p$ is both the desired end product and
a necessary intermediate ingredient in the calculation, must therefore be
developed for predicting peculiar velocities from redshift surveys
(Appendix~\ref{sec:iras}).
For these reasons, reliable comparisons of peculiar velocity
and redshift survey data require
extremely careful statistical analyses.

This problem has inspired a number
of independent approaches
in recent years. The \potent\ method (Dekel, Bertschinger, \&
Faber 1990; Dekel 1994; Dekel \etal\ 1997) was the first effort at a
rigorous treatment of peculiar velocity data. 
Dekel \etal\ (1993) compared
the \potent\ reconstruction of the Mark II peculiar velocity data (Burstein 1989)
to the 
\iras\ 1.936 Jy redshift survey (Strauss \etal\ 1992b),
finding $\beta_I=1.28^{+0.75}_{-0.59}$ at 95\% confidence.\footnote{Because
the bias parameter can differ for different galaxy samples, the value of
$\beta$ can differ as well.
We will use $\beta_I$ for
the \iras\ redshift survey 
and $\beta_{{\rm opt}}$
for an optical survey. Because optical galaxies are about 30\%
more clustered than \iras\ galaxies (SW), the conversion
is $\beta_I\simeq 1.3\beta_{{\rm opt}}.$ When speaking
generically about the velocity-density relation, we will place no subscript
on $\beta.$}
An improved treatment 
using the Mark III peculiar velocities (Willick \etal\ 1997) and the
\iras\ 1.2 Jy survey (Fisher \etal\ 1995) 
yields $\beta_I=0.86\pm 0.15$ (Sigad \etal\ 1997, hereafter \potiras).
Hudson \etal\ (1995) compared the optical redshift survey data of
Hudson (1993) to the \potent\ reconstruction
based on a preliminary version of the Mark III catalog,
finding $\beta_{{\rm opt}}=0.74\pm 0.13$ (\onesigma\ errors).
These results from \potent\ were obtained using 1200\ \kms\ Gaussian
smoothing. 
A distinct approach, which differs from
\potiras\
in the statistical biases to which it is vulnerable
(SW), and which typically uses much smaller
smoothing, is to predict galaxy 
peculiar velocities and thus 
distances
from the density field, and then use 
these predictions to minimize the scatter in
the TF or \dnsigma\ relations
(Strauss 1989;
Hudson 1994; Roth 1994; Schlegel 1995;
Shaya, Peebles, \& Tully 1995; Davis, Nusser, \& Willick 1996,
hereafter DNW).
This second kind of analysis has produced
estimates 
of $\beta_I$ in the range \sm 0.4--0.7,
lower than the values obtained from \potiras.
We further clarify the distinction between the two methods
in \S~\ref{sec:method_intro}, and discuss possible reasons for the
discrepancies 
in \S~\ref{sec:beta-discussion}.

In this paper, we present a new maximum-likelihood method for comparing
TF data to the predicted peculiar velocity
and density fields 
in order to estimate $\beta.$
Its chief strength
is an improved treatment of 
nearby galaxies $(cz \le 3000\ \kms)$, 
and we limit the analysis to this range. 
The TF data we use 
comprise a subset of the Mark III catalog of Willick \etal\ (1997).
The predicted peculiar velocities are obtained using
new reconstruction methods (Appendix~\ref{sec:iras}) from the
\iras\ 1.2 Jy redshift survey.\footnote{The
original \iras\ 1.936 Jy survey was presented in a series of six
papers (Strauss \etal\ 1990; Yahil \etal\ 1991; Davis, Strauss \&
Yahil 1991; Strauss \etal\ 1992abc), numbered 1, 2, 3, 4, 5, and 7,
respectively.  The missing paper 6 was to be the comparison of the
observed and predicted velocities, to be based on Chapter 3
of Strauss (1989).
However, it has taken us until now to come up with
statistically rigorous ways of doing this comparison. The
long-lost \iras\ Paper 6 has thus been
incorporated into Dekel \etal\ (1993), DNW,
Sigad \etal\ (1997), and especially this paper.}
The outline of this paper is as follows.
In \S~\ref{sec:method}, we first review the strengths
and weaknesses of existing approaches, and then describe 
our new method in detail.
In \S~\ref{sec:mock}, we present tests
of the method using mock catalogs. In \S~\ref{sec:real},
we apply the method to the Mark III catalog and obtain an
estimate of $\beta_I.$ In \S~\ref{sec:resid}, we
analyze residuals from our maximum likelihood solution in
order to assess whether \iras\ predictions give a statistically
acceptable fit to the Mark III data. In \S~\ref{sec:discussion},
we further discuss and summarize our principal results.  
This paper is the product of nearly three years work
and contains considerable detail. We recommend that
readers interested primarily in results and interpretation
skim \S~2, and then read \S~3.1, 4.4, 4.5, 5.1, 5.2, and 6.4.

\section{Description of the Maximum Likelihood Method}
\label{sec:method}
\subsection{Alternative Approaches to the Peculiar Velocity-Density
Comparison}
\label{sec:method_intro}

Before presenting our method in detail, we briefly
review the principal alternatives.
Two approaches are fairly
paradigmatic, and serve to illustrate
the main issues and motivate our approach.
These are the \potent\ method of Dekel and coworkers
(e.g., Dekel 1994; Dekel \etal\ 1997) mentioned in \S~\ref{sec:intro},
and the \itf\ method of Nusser \& Davis (Nusser \& Davis 1995; DNW).

The \potent\ algorithm is designed to reconstruct,
from sparse and noisy radial peculiar velocity estimates,
a smooth three-dimensional peculiar velocity field
and the associated 
mass density field.
The method is based on the property 
that the smoothed velocity field
of gravitating systems 
is the       
gradient of a potential.
The divergence of Eq.~(\ref{eq:vpdeltag}) is
\begin{equation}
\nabla \cdot \bfv = -\beta \delta_g\,.
\label{eq:delvdelg}
\end{equation}
Thus, $\beta$ is the slope of the correlation between $\nabla \cdot \bfv,$ obtained
from \potent, and $\delta_g$ obtained from redshift survey data.
This is the basis of the \potiras\ approach\footnote{Dekel \etal\ (1993) and Sigad
\etal\ (1997) actually use a non-linear extension to Eq.~(\ref{eq:delvdelg}).} 
to determining $\beta_I,$ discussed above
(\S~\ref{sec:intro}).

\potent\ has several advantages 
as a reconstruction method. 
It yields model-independent, three-dimensional velocity
and density fields 
well-suited for comparison with theory and for visualization. 
It works in the space of TF-inferred distances, 
i.e., it is a {\em Method I\/}
approach to velocity analysis (cf.\ SW, \S~6.4.1).
Unlike {\em Method II\/} approaches (see below), it does
not assume that there is a unique distance corresponding to a given redshift. 
In regions where galaxies at
different distances are superposed in redshift space,
\potent\ is capable of recovering the true velocity field.
The \potiras\ comparison 
between the mass  and  galaxy density fields 
is entirely {\em local} (Eq.~\ref{eq:delvdelg}), whereas predicted
peculiar velocities are highly nonlocal
(Eq.~\ref{eq:vpdeltag}). Locality ensures that
biases due to unsampled regions are minimized.

The liabilities of \potent\ are
closely related to its strengths. In order to
construct a model-independent velocity field it must have
redshift-independent distances as input. Such distances
require properly calibrated TF relations.
In particular, the TF distances
for samples that probe different regions of the sky
must be brought to a uniform system, which is a difficult
procedure (cf.\ Willick \etal\ 1995, 1996, 1997).
Errors made 
in calibrating and homogenizing
the TF relations will propagate into the \potent\ velocity field.
Because \potent\ works in inferred distance space, it is
subject to inhomogeneous Malmquist bias
(Dekel, Bertschinger, \& Faber 1990).
Minimizing this bias
requires significant smoothing of the input data.
\potent\ currently employs a Gaussian
smoothing scale of 1000--1200\ \kms\ (Dekel 1994;
Dekel \etal\ 1997), making it
relatively insensitive to
dynamical effects on small scales.
As a result, the current \potent\ 
applications are not particularly effective at
extracting
detailed 
information from the velocity field
in the local ($cz\simlt 3000\ \kms$) universe.

DNW take a different approach. 
They work with the ``inverse'' form of  the TF relation       
(Dekel 1994, \S~4.4; SW, \S~6.4.4), and thus refer to their method as \itf.
They express peculiar velocity as a function
of redshift-space, rather than real-space, position; in the terminology of
SW, \itf\ is thus a {\em Method II\/} analysis,
largely impervious to inhomogeneous Malmquist bias.
DNW expand the redshift-space peculiar velocity field 
in a set of independent basis functions, or {\em modes,}
whose coefficients are solved for simultaneously with the 
parameters of a global inverse TF relation
via $\chi^2$ minimization of TF residuals. The TF data
are never converted into inferred distances
and thus do not require 
pre-calibrated TF relations.
The \iras-predicted velocity field
is expanded in the same set of basis functions,
allowing a mode-by-mode comparison of
predicted and observed peculiar velocities.
This ensures that one is comparing quantities
that have undergone the same spatial smoothing,
a desirable characteristic of the fit.

As with \potent, the strengths of \itf\ 
are connected with certain disadvantages.
Because it is a Method II approach,
multivalued or flat zones in the
redshift-distance relation (see below) 
necessarily bias the
\itf\ analysis.
It neglects the
role of small-scale velocity noise, which
is non-negligible for galaxies within 1000 \kms. 
These features make \itf,
like \potent, a relatively ineffective tool for probing
the very local region. Last and most importantly,
the \itf\ method as implemented
by DNW
requires that the
raw magnitude and velocity width data
from several distinct data sets be carefully matched
before being input to the algorithm. 
Any systematic errors
incurred in matching the raw data from different parts of the
sky will induce large-scale, systematic errors in the derived
velocity field. 
Thus, although
\itf\ 
does not need input TF
distances, it is vulnerable to \apriori\ calibration errors just
as \potent\ is.

\subsection{VELMOD}
\label{sec:velmod}

The approach we take in this paper, ``\velmod,''
is a maximum likelihood method designed to surmount several of the difficulties
that face
\potiras\ 
and \itf.
\velmod\ generalizes and improves upon the Method II approach to velocity
analysis. Method II
takes as its basic input the {\em TF observables\/}
(apparent magnitude and
velocity width) and redshift of a galaxy, and asks,
what is the probability of observing the former, given the
value of the latter? It then maximizes this probability over the
entire data set with respect to parameters describing the
TF relation and the velocity field. The underlying assumption 
of Method II is that a galaxy's redshift, in combination with
the correct model of the velocity field, yields its true distance,
which then allows the probability
of the TF observables to be computed. This analytic approach 
was originally developed by Schechter (1980), and was 
later used by Aaronson \etal\ (1982b), Faber \& Burstein (1988),
Strauss (1989), Han \& Mould (1990),
Hudson (1994), Roth (1994), and Schlegel (1995),
among others.

The main problem with Method II is its assumption that
a unique redshift-distance mapping is possible.
This assumption breaks down for two reasons. First,
redshift is a ``noisy'' realization of distance
plus predicted peculiar velocity---both because
of true velocity noise generated on very small ($\simlt\,1$ Mpc) scales,
and because of the inaccuracy of the
velocity model (even for the correct $\beta$) due to nonlinear effects and
shot noise in the density field. Second, even in the
absence of noise, the redshift-distance relation
can, in principle, be multivalued: more than one distance
along the line of sight can correspond to a given redshift.
\velmod\ accounts for all of these effects statistically by
replacing the unique 
distance of Method II
with the joint probability distribution of
redshift and distance. 
This distribution is constructed to allow for both
noise and multivaluedness. The distance dependence
is then integrated out (\S~\ref{sec:velmod-detail}), yielding the
correct probability distribution of the TF observables given redshift.

There are two additional advantages to the \velmod\ approach.
First, it requires neither \apriori\ calibration
of the TF relations (as does \potent) nor matching
of the input data from disparate samples (as does \itf).
An individual TF calibration for each independent
sample occurs naturally as part of the analysis.
Second, it
does not require smoothing of the input TF
data, and thus allows as high-resolution an analysis as
the data intrinsically permit. This second feature,
along with its allowance for velocity noise and triple-valued
zones, makes
\velmod\ well-suited for probing
the local ($cz\simlt 3000\ \kms$) velocity field.
An analysis of local data is desirable because
random and systematic errors in both the \iras\ and TF data
are less important nearby than far away. 

\subsubsection{Mathematical Details}
\label{sec:velmod-detail}
We now describe the method in detail.
We assume that the relevant distance indicator is the
TF relation;
with minor changes the formalism could be adapted to
comparable distance indicators such as \dnsigma.
We use the terminology of Willick (1994) and
Willick \etal\ (1995): briefly, we
denote by $m$ and $\eta \equiv \log v_{\rm rot} - 2.5$ a galaxy's corrected
apparent magnitude and velocity width parameter, respectively;
by $cz$ its Local Group frame radial velocity (``redshift'') in \kms;
and by $r$
its true distance in \kms. 
We define the distance modulus as $\mu\equiv 5\log r,$
and absolute magnitudes as $M=m-\mu.$  We write
the forward and inverse TF relations as linear
expressions, $M(\eta)=A-b\eta$ and $\eta^0(M)=-e(M-D),$
and denote their rms scatters $\sigtf$ and $\sigeta,$
respectively.

We seek
an exact expression
for the probability that a galaxy at redshift $cz$
possesses TF observables $(m,\eta)$ given
a model of the peculiar velocity and density fields.\footnote{The
dependence of all quantities on the line of sight direction
will remain implicit.}
We first consider the joint probability
distribution of the TF observables, redshift, and
an {\em unobservable} quantity, the true distance $r.$
Later, we will integrate over $r$ to obtain the
probability distribution of the observables.
We may write
\begin{equation}
P(m,\eta,cz,r)=P(m,\eta|r)\times P(cz|r)\times P(r)\,.
\label{eq:pmetazr}
\end{equation}
The splitting into conditional probabilities
reflects the fact that the TF observables and
the redshift couple with one another only
via their individual dependences on
the true distance $r.$

The first of the
three terms on the right hand side of Eq.~(\ref{eq:pmetazr})
depends on the luminosity function, the sample selection
function, and the TF relation. We can express it
in one of two ways, depending on whether we are using
the forward or inverse form of the TF relation:
\begin{enumerate}
\item Forward relation:
\begin{equation}
P(m,\eta|r) \propto \phi(\eta)S(m,\eta,r) {1 \over \sigtf}\expmM
\label{eq:pmetagr_f}
\end{equation}
\item Inverse relation:
\begin{equation}
P(m,\eta|r) \propto \Phi(m-\mu(r))S(m,\eta,r){1 \over \sigma_\eta}\expet\,,
\label{eq:pmetagr_i}
\end{equation}
\end{enumerate}
where $\phi(\eta)$ and $\Phi(M)$ are the (closely related)
velocity width distribution function and luminosity function,
$S(m,\eta,r)$ is the sample selection function, and we have assumed
Gaussian scatter of the TF relation (an assumption
validated by Willick \etal\ 1997).
Detailed derivations of these expressions are given
by Willick (1994).\footnote{Willick
(1994) assumed that the selection function depended only on
the TF observables. Here, we acknowledge the possibility
of an explicit distance dependence; the origin of
such a dependence was discussed by SW, \S~6.5.3.}
In Eqs.~(\ref{eq:pmetagr_f})
and~(\ref{eq:pmetagr_i}) we have written only proportionalities,
as the normalization is straightforward and will occur
at a later point in any case.

The third term on the right hand side of Eq.~(\ref{eq:pmetazr})
is simply the \apriori\ probability of observing an object
at distance $r,$
\begin{equation}
P(r)\propto r^2 n(r)\,,
\label{eq:pofr}
\end{equation}
where $n(r) \propto 1 + \delta_g(\bfr)$ is
the number density of the species of galaxies that
makes up the sample.
The second term on the right hand side of
Eq.~(\ref{eq:pmetazr}), $P(cz|r),$
is the one which couples
the TF observables to the velocity field model. We
assume that, for the correct \iras\
velocity field reconstruction (i.e., for the correct value of $\beta_I$ and
other velocity field parameters to be described below),
the redshift is normally distributed about
the value 
predicted from the velocity model:
\begin{equation}
P(cz|r) = \nsigv\expcz\,,
\label{eq:pzr}
\end{equation}
where $u(r)\equiv \hat \bfr \cdot \left[\bfv(\bfr) - \bfv({\bf 0})\right]$ 
is the radial
component of the predicted peculiar velocity field in the Local Group frame
(cf.\ Eq.~\ref{eq:cz-r}).
We treat the velocity noise $\sigv$ as a free
parameter in our analysis; we discuss its origin in detail in
\S~\ref{sec:mocksigv}. 
Although $\sigv$ must be position or density dependent at
some level, we
treat it as spatially constant in this paper, except in the Virgo
cluster (\S~\ref{sec:virgo}). 

Substituting Eqs.~(\ref{eq:pmetagr_f}) or~(\ref{eq:pmetagr_i}),
(\ref{eq:pofr}), and~(\ref{eq:pzr}) into Eq.~(\ref{eq:pmetazr})
yields the joint probability distribution $P(m,\eta,cz,r).$
To obtain the joint probability distribution of the observable
quantities, one integrates over the (unobserved) line-of-sight distance, i.e.,
\begin{equation}
P(m,\eta,cz) = \int_0^\infty P(m,\eta,cz,r)\,dr\,.
\label{eq:pmetaz}
\end{equation}
In practice, it is not optimal to base a
likelihood analysis on the joint distribution $P(m,\eta,cz)$
because of its sensitivity to terms, such as the luminosity
function, the sample selection function, and the density field, that are not
critical
for our purposes. Instead, the desired probability distributions
are the {\em conditional} ones:
\begin{enumerate}
\item Forward TF relation:
\[ P(m|\eta,cz)\,=\,\frac{P(m,\eta,cz)}{\infint P(m,\eta,cz)\,dm} \]
\begin{equation}
= \frac{\int_0^\infty\!dr\,r^2 n(r)\,P(cz|r)\,S(m,\eta,r) \expmM}
{\int_0^\infty\!dr\,r^2 n(r)\,P(cz|r)\,\infint\!dm\,S(m,\eta,r) \expmM}\,;
\label{eq:pmgetaz}
\end{equation}
\item Inverse TF relation:
\[ P(\eta|m,cz)\,=\,\frac{P(m,\eta,cz)}{\infint P(m,\eta,cz)\,d\eta} \]
\begin{equation}
= \frac{\int_0^\infty\!dr\,r^2 n(r)\,\Phi(m-\mu(r))\,P(cz|r)\,S(m,\eta,r)
\expet}
{\int_0^\infty\!dr\,r^2
n(r)\,\Phi(m-\mu(r))\,P(cz|r)\,\infint\!d\eta\,S(m,\eta,r) \expet}\,,
\label{eq:petagmz}
\end{equation}
\end{enumerate}
where $P(cz|r)$ is given by Eq.~(\ref{eq:pzr}).  Although neither of
these expressions is independent of the density field $n(r)$ or the
selection function $S$, 
their appearance in both the numerator and denominator much
reduces their sensitivity to them. A similar statement holds
for the luminosity function $\Phi$ in Eq.~(\ref{eq:petagmz}).  
The velocity width distribution function
$\phi$ has, however,
dropped out entirely from the forward relation probability.
We discuss these points further in \S~\ref{sec:velmod-discussion}.

\def\like{{\cal L}}
\def\likeforw{\like_{{\rm forw}}}
Equations~(\ref{eq:pmgetaz}) and~(\ref{eq:petagmz}) are the conditional
probabilities whose products over all galaxies in the sample we wish to
maximize.
In practice,
we do so by {\em minimizing} the quantities
\begin{equation}
\likeforw = -2\sum_i \ln P(m_i|\eta_i,cz_i)
\label{eq:deflamforw}
\end{equation}
or
\begin{equation}
\like_{{\rm inv}} = -2\sum_i \ln P(\eta_i|m_i,cz_i)\,,
\label{eq:deflaminv}
\end{equation}
where the index $i$ runs over all objects in the TF sample. 
We
have assumed that the probabilities for each galaxy are
independent; we validate this assumption
\aposteriori\ (cf.\ \S~\ref{sec:residcorr}).

\subsubsection{Further discussion of the \velmod\ likelihood}
\label{sec:velmod-discussion}
The physical meaning of the \velmod\ likelihood expressions is
clarified by considering them in a suitable limit. 
If we take 
$\sigma_v$ to be ``small,'' in a sense to be made precise below,
the integrals in Eqs.~(\ref{eq:pmgetaz}) and~(\ref{eq:petagmz})
may be approximated using standard techniques.
If in addition we
neglect sample selection ($S = 1$) and density variations
($n(r) = \rm constant$), and assume that the redshift-distance
relation is single-valued, we find for the forward relation:
\begin{equation}
P(m|\eta,cz)\simeq \frac{1}{\sqrt{2\pi}\sigma_e}\,\exp\left\{-\frac{1}{2\sigma_e^2}
\left(m-\left[M(\eta)+5\log w + \frac{10}{\ln 10}\Delta_v^2\right]
\right)^2\right\}\,,
\label{eq:pmapprox}
\end{equation}
where $w$ is the solution to the equation $cz=w+u(w),$ i.e., it is the
distance inferred from the redshift and peculiar velocity model;
$\Delta_v\equiv\sigma_v/[w(1+u^\prime)],$ where $u^\prime=\left(\partial u/\partial r\right)_{r=w},$ 
is the effective
logarithmic velocity dispersion; 
and 
\begin{equation}
\sigma_e\equiv\left[\sigtf^2
+\left(\frac{5}{\ln 10}\right)^2\Delta_v^2\right]^{1/2}
\label{eq:sigtf_eff} 
\end{equation}
is the effective
TF scatter, including the contribution due to $\sigma_v.$
An analogous result holds for the inverse relation.
The criterion
$\Delta_v^2 \ll 1,$ which quantifies the statement
that $\sigv$ is ``small,'' must be satisfied 
to derive Eq.~(\ref{eq:pmapprox}).

Eq.~(\ref{eq:pmapprox}) shows that
the probability distribution $P(m|\eta,cz)$ preserves the
Gaussian character of the real-space TF probability distribution
$P(m|\eta,r)$ in this limit.
However, the expected value of $m$ is shifted
from the ``na\"\i ve'' value $M(\eta)+5\log w$ by an
amount $\sim 4.3\Delta_v^2.$ This shift is in fact nothing
more than the homogeneous Malmquist bias due to small-scale
velocity noise; it differs in detail from the usual Malmquist
expression (i.e., that which affects a Method I analysis) 
because it arises from the Gaussian
(rather than log-normal) probability distribution, Eq.~(\ref{eq:pzr}).
Furthermore, the effective scatter $\sigma_e$ is larger than $\sigtf$,
because the velocity dispersion introduces
additional distance error and thus magnitude scatter.
The effects associated with velocity noise diminish
with distance ($\Delta_v\propto r^{-1}$), however;
the velocity Malmquist effect vanishes in the limit of large
distances, 
in contrast with the distance-independent Malmquist effect for Method I,
and the effective scatter approaches
the TF scatter. At large enough distance 
the \velmod\ likelihood
approaches a simple Gaussian TF distribution with expected
apparent magnitude $M(\eta)+5\log w,$ and \velmod\ reduces
to standard Method II.

Indeed, Eq.~(\ref{eq:pmapprox}) enables us to define the regime
in which \velmod\ represents a significant modification of Method II.
The distance 
$r_{{\rm II}}$ at which the
velocity noise effects become unimportant is determined by 
$r_{{\rm II}}\gg \sigma_v/\Delta_{{\rm TF}}(1+u^\prime),$
where $\Delta_{{\rm TF}}=\ln 10\,\sigtf/5$ is the fractional
distance error due to the TF scatter ($\Delta_{{\rm TF}}\simeq 0.2$
for the samples used here). For $\sigma_v=125\ \kms,$ the value we
find for the real data (\S~\ref{sec:results}), this shows
that in the {\em unperturbed\/} Hubble flow, where $u^\prime=0,$
velocity noise effects become unimportant beyond \sm 1500 \kms.
However, at about this distance, in many
directions, the Local Supercluster
significantly retards the Hubble flow, $u^\prime\simeq -0.5,$ so
that the effective $\sigma_v$ is about twice its nominal value.
Thus, \velmod\ in fact differs substantially from Method II to
roughly twice the Virgo distance. This fact guided our decision
to apply \velmod\ only out to 3000 \kms\ (cf.\ \S~\ref{sec:real}). 

Eq.~(\ref{eq:pmapprox}) also demonstrates that maximizing
likelihood (minimizing $\likeforw$) is not
equivalent to $\chi^2$ minimization, even under the adopted assumptions
of constant density and negligible selection effects,
because of the factor $\sigma_e^{-1}$ in
front of the exponential factor. This factor couples
the velocity model (i.e., the values of $w$ and $u^\prime(w)$)
to the velocity noise. 
In particular, {\em maximizing the \velmod\ likelihood
is not equivalent to minimizing TF scatter} (cf.\
\S~\ref{sec:results}), except in the limit that $\sigma_v$ is set to
zero. 
 
The assumptions required for deriving Eq.~(\ref{eq:pmapprox}) remind
us that there are two other factors which distinguish \velmod\
from standard Method II. First, for realistic samples one cannot
assume that $S=1.$ The presence of the selection function in
Eqs.~(\ref{eq:pmetagr_f})
and~(\ref{eq:pmetagr_i}) is essential for evaluating true
likelihoods, and we have fully incorporated these effects into
our analysis.\footnote{Selection effects are not specific to
\velmod\ per se, however. They can and should be modeled in any
Method II-like analysis. In particular, they do not vanish
in the $\Delta_v\longrightarrow 0$ limit.}
Second, the galaxy density $n(r)$ is not effectively constant
along most lines of sight. Thus, \velmod, like Method I but unlike
Method II, 
requires that $n(r)$ be modeled.
We do so here by using the \iras\ density field itself, which
is a good approximation to the number density of the spiral
galaxies in the TF samples. 
The density field has a non-negligible effect on the \velmod\
likelihood whenever it changes rapidly on the scale
of the effective velocity dispersion $\sigv/(1+u^\prime).$

The most significant differences between \velmod\ and Method II 
thus occur in regions where $u^\prime \longrightarrow -1$
(flat or triple-valued zones), or when the density varies
particularly sharply. In practice, both these effects
occur in the vicinity of large density enhancements
such as the Virgo cluster.  
We illustrate this in Figure~1,
which shows the redshift-distance relation,
and the corresponding value of $P(r|cz) \propto P(cz|r)P(r)$ in the vicinity of
triple-valued zones.  When looking at these panels, keep in mind that the
\velmod\ likelihood is given by multiplying $P(r|cz)$
and the TF probability factor $P(m|\eta,r)$ and integrating
over the entire line of sight.
Panels (a) and (b) depict the situation near the core
of a strong cluster, and panels (c) and (d) farther from the center. 
In each case, the
cloud of points represents the velocity noise, here taken to be
$\sigma_v = 150 ~\kms.$ In panel (a), the
redshift of 1200
\kms\ crosses the redshift-distance diagram at three distinct
distances.  The quantity $P(r | cz)$ 
shows three distinct peaks.  The highest redshift one is
the strongest because of the $r^2$ weighting in
Eq.~(\ref{eq:pofr}). In panel (b), the redshift of 1700 \kms\ is such
that the object just misses being triple-valued; however, the finite
scatter in the redshift-distance diagram means that there is still
appreciable probability that the galaxy be associated with the
near-crossing at $cz \sim 900~ \kms$.  In panel (c), the
redshift-distance diagram goes nearly flat for almost 600 \kms; a
redshift that comes close to that flat zone has a probability
distribution that is quite extended.  Finally, panel (d) shows a galaxy
whose redshift crosses the redshift-distance diagram in a region in
which it is quite linear, and the probability distribution has a
single narrow peak without extensive tails.  
\begin{figure}[t!]
\centerline{\epsfxsize=4.0 in \epsfbox{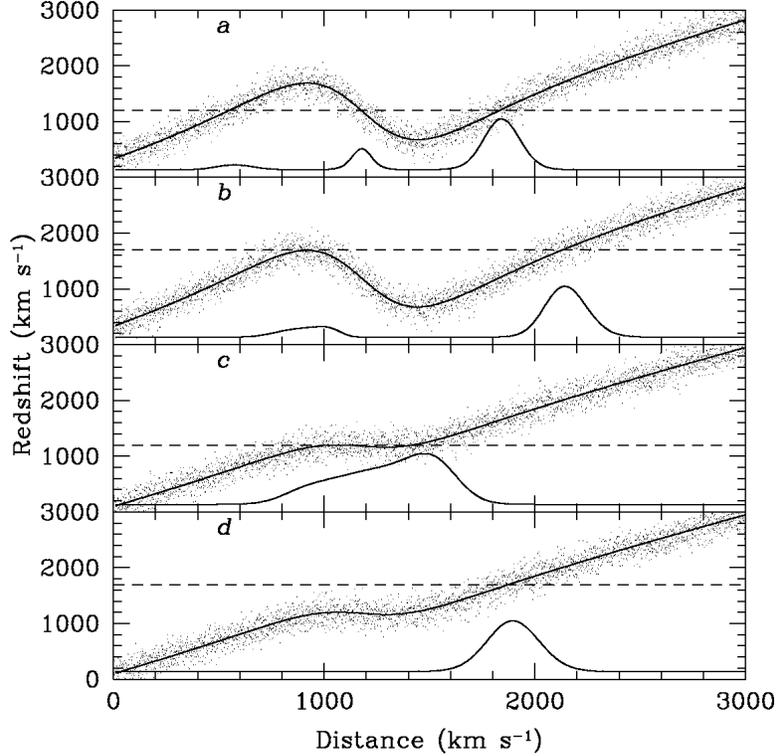}}
\caption{{\small The effects of triple-valued or
flat zones. The S-shaped curves
show the relation between redshift and distance along two lines of
sight to a cluster. (a) A galaxy with a redshift of 1200 \kms\ can lie at
three distinct distances. When the small-scale noise inherent in any
velocity field model, as indicated by the scattered points, is taken into
account, the quantity $P(r|cz),$ shown as the three-peaked curve at the
bottom, gets smoothed out.
(b) A galaxy at 1700 \kms\ along the same line of sight
intersects the redshift-distance curve at only one point, but comes
close enough to it elsewhere to give a second peak to the $P(r|cz)$
curve. (c) Further from the cluster, the redshift-distance curve
becomes flat, giving a broad peak to $P(r|cz)$. (d) At a redshift
sufficiently far from a triple-valued zone, $P(r|cz)$ has only one
narrow peak.
}}
\label{fig:tvz}
\end{figure}

Two final details 
deserve brief mention. First,
the integrals
over $m$ and $\eta$ that appear in the denominators of
Eqs.~(\ref{eq:pmgetaz}) and~(\ref{eq:petagmz}) 
may be done analytically for the case of ``one-catalog selection''
studied by Willick (1994, \S~4.1), which indeed applies for the
samples used in this paper (Willick \etal\ 1996).
The numerical integrations required to evaluate
Eqs.~(\ref{eq:pmgetaz}) and~(\ref{eq:petagmz})
are thus one-dimensional only. 
Second, as noted above, the velocity width distribution function $\phi(\eta)$
drops out of Eq.~(\ref{eq:pmgetaz}),
but the luminosity function $\Phi(M)$ does not drop out of
Eq.~(\ref{eq:petagmz}). Thus, inverse \velmod\ requires that we
model the luminosity function of TF galaxies. This is an annoyance
at best, and could introduce biases, if we model it incorrectly, at worst.
We have thus chosen to implement only forward \velmod\ in this paper.
On the other hand, inverse \velmod\ enjoys the virtue that inverse Method II
approaches do generally: to the degree that the selection function $S$ is
independent of
$\eta$ and $r,$ it drops out of Eq.~(\ref{eq:petagmz}).
In a future
paper, we will apply the small-$\sigma_v$ approximation to \velmod\ for
more extensive samples to larger distances. For that analysis the
inverse approach will be used as well.

\subsubsection{Implementation of \velmod}

The probability distribution
$P(m|\eta,cz)$ (Eq.~\ref{eq:pmgetaz}) 
is dependent on a number of free
parameters, most importantly $\beta_I$.  
However,
because $\beta_I$ enters at an earlier stage---in
the reconstruction
of the underlying density and velocity fields from \iras\ (Appendix A)---it
is on a different footing from other parameters. Thus, rather than
treating $\beta_I$ as a continuous free parameter, \velmod\ is
run sequentially for the ten discrete values $\beta_I=0.1,\ldots,1.0$ for the
real data, and for the nine discrete values $\beta_I=0.6,\ldots,1.4$
for the mock catalog data (\S~\ref{sec:mock}).\footnote{The choice
of these values of $\beta_I$ was based on the need to bracket
the ``true'' value: 1.0 in the mock catalogs,
and, as it turns out, \sm 0.5 for the real data.} For each $\beta_I,$
probability is maximized ($\like_{{\rm forw}}$ is minimized) with
respect to the remaining free parameters. These parameters
are:
\begin{enumerate}
\item The TF parameters $A$,
$b,$ and $\sigtf$ for each sample in the analysis.
Here we limit the analysis to the Mathewson \etal\
(1992; MAT) and Aaronson \etal\ (1982a; A82) samples, as we discuss 
in \S~\ref{sec:selection}. Thus, there are a total
of 6 TF parameters that are varied. 
Note that the TF scatters are not simply calculated
\aposteriori. The statistic $\like_{{\rm forw}}$ depends
on their values and they are varied to minimize it.
\item The small-scale velocity dispersion $\sigma_v.$
The quantities $\sigma_v$ and $\sigtf$ can trade off
to a certain extent (cf.\ Eq.~\ref{eq:pmapprox}).  However, their
relative importance depends on distance. Sufficiently nearby ($\simlt
1000\ \kms$),
$\sigma_v$
is as large or a larger source of error than the TF scatter
itself. Thus it is determined in 
this local region.  Beyond \sm 2000 \kms, the
TF scatter dominates the error, and it is determined at these distances.
Because the samples populate a range of distances, the two
can be determined separately, with relatively little covariance.
\item We also allow for a {\em Local Group random velocity vector\/} $\bfwlg.$
The \iras\ peculiar velocity predictions are
given in the
Local Group frame (Eq.~\ref{eq:cz-r}). That is, the computed Local Group peculiar velocity
vector has
been 
subtracted from all other peculiar velocities. However,
just as we expect all external galaxies to have a noisy as well
as a systematic component to their peculiar velocity, so we must
expect the Local Group to have one as well, especially considering the
uncertainties in the conversion from heliocentric to Local Group
frame. We allow for this by 
writing $u(r)=u_\iras(r)-\bfwlg\cdot \hat \bfr,$ where $u_\iras(r)$
is given as described in Appendix~\ref{sec:iras}, and the three Cartesian components
of $\bfwlg$
are varied in each \velmod\ run at a given $\beta_I.$
We note briefly that this procedure is self-consistent only
as long as $|\bfwlg|$ is at most comparable to $\sigma_v.$ 
In practice,
we will find that for $\beta_I$ near its best value, the amplitude of
$\bfwlg$ is trivially small.
\item Finally, we allow for the existence of a quadrupole velocity
component that is not included in the \iras\ velocity field. 
The justification for such a velocity component will
be discussed in \S~\ref{sec:quadrupole} and Appendix B. 
The quadrupole is specified by
five independent parameters, although we will not take them as 
free in the final analysis (we discuss this further
in \S~\ref{sec:quadrupole}).
\end{enumerate}

Thus there are $3\times 2 + 1 + 3  = 10$ free parameters that are varied for
any given value of $\beta_I,$ when the
quadrupole is held fixed. 
Thus, for any value of $\beta_I,$ we give the data the fairest chance
it possibly has to fit the \iras\ model. In particular, the TF
relations for the two separate samples used are not ``precalibrated''
in any way. This ensures that TF calibration 
in no way prejudices the value of
$\beta_I$ we derive.

\section{Tests With Simulated Galaxy Catalogs}
\label{sec:mock}
In this section, we test the \velmod\ method on simulated data sets.
Kolatt \etal\ (1996) have produced simulated catalogs that mimic the
properties of both the \iras\ redshift survey and the Mark III
samples.  We briefly review the salient points here.

The mass density distribution of the simulated universe
is based on 
the distribution of \iras\ galaxies in the real universe. This was achieved
by, first, taking the present redshift distribution of \iras\ galaxies 
and solving for a 500 \kms\ smoothed real-space distribution   
via an iterative procedure that applies nonlinear corrections 
and a power-preserving filter (Sigad \etal\ 1997).           
The smoothed, filtered \iras\ density field was then
``taken back in time'' using the
Zel'dovich-Bernoulli 
algorithm of Nusser \& Dekel (1992)
to obtain the linear initial density field.
The method of constrained realization (Hoffman \& Ribak 1991;
Ganon \& Hoffman 1993) 
was used to restore
small-scale power
down to galactic scales.
The resulting initial conditions were then evolved forward as an $\Omega=1$
$N$-body simulation using the PM code of Gelb \& Bertschinger (1994). The
present-day density field resulting from this procedure is displayed in
Figure 6 of Kolatt \etal\ (1996). 

We generated a suite of 20 mock Mark III and mock \iras\ catalogs
from this simulated universe.\footnote{The 20 catalogs (of both types) are
different
statistical realizations of the same simulation. As a result, our simulations
fully probe the effects of statistical variance (due to distance indicator
scatter, spatial inhomogeneities, etc.) but do not include those of cosmic
variance. However, as we shall argue in \S~\ref{sec:discussion},
we expect that cosmic variance will have minimal effect on
our $\beta$-determination.}
Each mock Mark III TF
sample was constructed to mimic the distribution on the sky and
in redshift space of the corresponding real sample, and the TF
relations and scatters of the mock samples were chosen
to be similar to the observed ones. 
The mock TF samples were
subject to selection criteria similar to
those imposed on the real samples.
The mock \iras\ redshift catalogs were generated so as to resemble
the actual \iras\ 1.2 Jy redshift survey. 
They have the
true \iras\ selection and luminosity functions applied, and lack data in
the \iras\ excluded zones (cf.\ Strauss \etal\ 1990).
These data were then put through exactly the same code to derive
peculiar velocity and density fields
as is used for the real data (Appendix~\ref{sec:iras}).
To simplify interpretation of
the mock catalog tests, the mock \iras\ galaxies were generated
with probability proportional to the mass density itself.
Thus, the mock \iras\
galaxies are unbiased relative to the mass; i.e., for
the mock catalogs $b_I=1,$ and therefore the {\em true\/} value of
$\beta_I$ for the simulated data is
unity. 

\subsection{Accuracy of the $\beta$ Determination}
\label{sec:beta-accuracy}

The \iras\ velocity field reconstructions may
be produced using a variety of smoothing scales,
and we have used 300 and 500 \kms\ Gaussian smoothing.
We found, however, that at 500 \kms\ smoothing \velmod\ returned
a mean $\beta_I$ biased high by \sm 20\%;
the predicted peculiar velocities 
were too small, and
a too-large $\beta_I$ was needed to compensate. Our discussion
from this point on will refer to 300
\kms\ smoothing,
which, as we now describe, we found to yield
correct peculiar velocities and an unbiased estimate of $\beta_I.$

\begin{figure}[t!]
\centerline{\epsfxsize=4.0 in \epsfbox{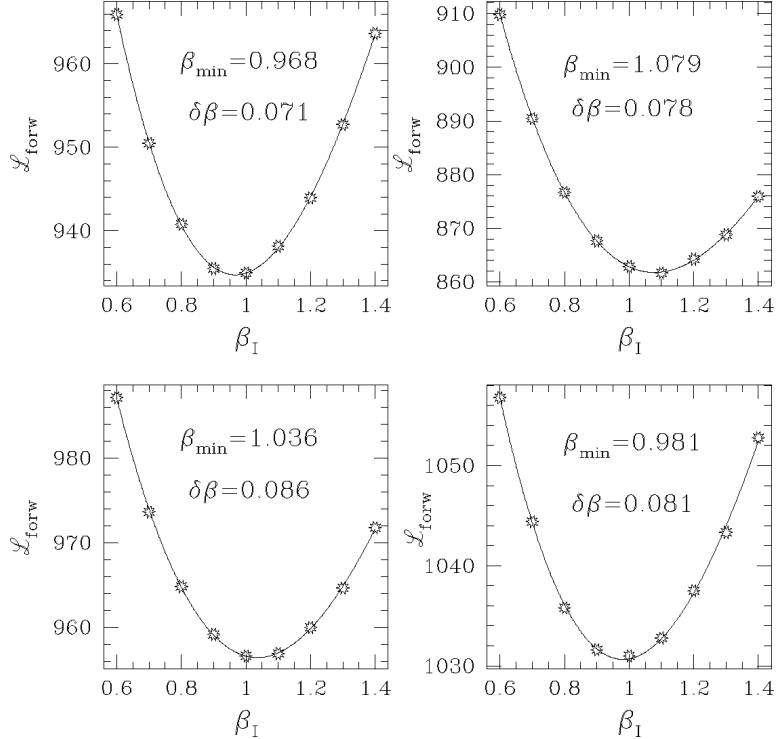}}
\caption{{\small Plots of the likelihood 
statistic, $\like_{{\rm forw}},$
versus $\beta_I$ for \velmod\ runs using four of the mock catalogs.
(The true value of $\beta_I$ for the mock catalogs is unity, as
discussed in the text.)
Also indicated on the plots are $\beta_{{\rm min}},$ 
the maximum likelihood
values
for $\beta_I,$ and the average of its two one-sided errors
$\delta\beta_\pm.$ The
solid lines
drawn through the points are the cubic fits used to
determine $\beta_{{\rm min}}$ and $\delta\beta_\pm.$
}}
\label{fig:velmod_mock4}
\end{figure}

\velmod\ was run on the 20 mock catalogs, and likelihood ($\like_{{\rm forw}}$)
versus $\beta_I$ curves were generated
for each. As with the real data (\S~\ref{sec:real}), we used
only the A82 and MAT TF samples; we limited the analysis
to $cz\leq 3500\ \kms$.{}\footnote{The real data analysis extended
only to 3000 \kms, but because there are fewer nearby TF galaxies in
the mock catalogs, we extended the mock analysis to a slightly larger distance.}
The curves were fitted with a cubic equation of the form
\begin{equation}
\like =\like_0 + q\,(\beta_I-\beta_{{\rm min}})^2
+ p\,(\beta_I-\beta_{{\rm min}})^3
\label{eq:cube}
\end{equation}
to determine $\beta_{{\rm min}},$ the value of $\beta_I$ for
which $\like_{{\rm forw}}$ is minimized.
This is the maximum likelihood value of $\beta_I.$
Four representative $\like_{{\rm forw}}$
versus $\beta_I$ plots are shown in Figure~2,
along with the cubic fits.  
We estimate the $1\,\sigma$ errors $\delta\beta_\pm$ in
our maximum likelihood estimate
by noting the values $\beta \pm \delta \beta_\pm$ at which $\like =
\like_0 + 1.$  Given 
the presence of the cubic term in Eq.~(\ref{eq:cube}), this is not 
necessarily rigorous, but 
we can test our errors by defining 
the $\chi^2$-like statistic
\begin{equation}
\chi^2 = \sum\,(\beta_{{\rm min}} - 1)^2/ \delta\beta_{\pm}^2\,,
\label{eq:chi2err}
\end{equation}
where $\delta\beta_+$ was used if $\beta_{{\rm min}}\leq 1,$ and
$\delta\beta_-$ was
used if $\beta_{{\rm min}}>1.$ For 
the 20 mock catalogs, it was found that $\chi^2 = 21.2.$
Thus, our tests were
consistent with the statement that the error estimates obtained
from the change in the likelihood statistic near its minimum
are true 1\,$\sigma$ error estimates.
Although we formally derive two-sided error bars, 
the upper and lower errors differ little, and
when we discuss the real data (\S~\ref{sec:real})
we will give only the average of the two.
The weighted mean value of $\beta_{{\rm min}}$ over the mock catalogs 
was $0.984$, 
with an error in the mean of $\sim 0.08/\sqrt{20}=0.017.$ Thus, the mean $\beta_{{\rm min}}$
is within
$\sim 1\,\sigma$ from the true answer. We conclude
that there is no
statistically significant bias in the \velmod\ estimate of $\beta_I.$
The results of this and other tests we carried out using the
mock catalogs are summarized in Table~\ref{table:mock}.

\subsection{Accuracy of the Determination of $\sigma_v$ and $\bfwlg$}
\label{sec:mocksigv}
The mock catalogs also enable us to determine the reliability
of the small-scale velocity dispersion $\sigv$
derived from \velmod.  This quantity may be viewed as the quadrature
sum of true velocity noise ($\sigv^{n}$) and \iras\ velocity prediction errors 
($\sigv^{I}$) resulting from
shot-noise and imperfectly modeled nonlinearities.  (For the real data,
there is an additional contribution from redshift measurement errors,
which are zero in the mock catalog.) 
We can measure both $\sigv^n$ and $\sigv^I$ directly from the mock
catalogs.  To measure
velocity noise, we determined $\sigma_u,$ the rms value of pair
velocity differences $cz(\bfr_i)-cz(\bfr_j)$ of mock catalog TF galaxies
within 3500 \kms\ outside of the mock Virgo core, for $|\bfr_i
- \bfr_j| \le r_{{\rm max}}.$ 
We found $\sigma_u$ to be insensitive to the precise value
of $r_{{\rm max}}$ provided it was $\simlt 150\ \kms,$ implying   
that we are not including the gradient of the true velocity field on these
scales.  Taking $r_{{\rm max}}=150\ \kms,$ we found $\sigma_u=71\ \kms,$ 
corresponding to $\sigv^{n}=\sigma_u/\sqrt{2} = 50\ \kms.$ This value
is so small because the PM code does not properly model 
particle-particle interactions on small scales. 

We measured the \iras\ prediction
errors $\sigv^{I}$ as follows. For each mock TF particle (again,
within 3500 \kms\ and outside the mock Virgo core), 
we computed
an \iras\ predicted redshift $cz^{I}_i=r_i+u(r_i)
+fr_i - \bfwlg\cdot \nhat_i,$
where $r_i$ was the true distance of the object, $u(r_i)$ was the
\iras-predicted radial peculiar velocity in the Local Group frame
(for $\beta_I=1$), $f$ was a zero-point error in the \iras\ model 
(cf.\ \S~\ref{sec:tfmock}),
and $\bfwlg$ was the mock Local
Group peculiar velocity, which (just as in the real data) is not known precisely
and was also treated as a free parameter. We then minimized the mean squared
difference between $cz^{I}_i$ and the actual redshifts $cz_i$ over the entire
TF sample with respect to $f$ and $\bfwlg.$ 
The rms value of $(cz_i - cz^{I}_i)$
at the minimum was then our estimate of the 
quadrature sum of \iras\ prediction error and
true velocity noise, which we found to be $98\pm 2\ \kms$ after
averaging over the 20 mock catalogs.  Subtracting
off the small value of $\sigv^{n}$ found above, we obtain $\sigv^{I}\simeq 
84\ \kms.$ This surprisingly small value is indicative of the
high accuracy of the \iras\ predictions for nearby galaxies not
in high density environments.  

The value $\sigma_v = 98 \ \kms$ 
is somewhat smaller than the real universe value of 
$\sigv=125\ \kms$  (\S~\ref{sec:results}). 
Because we wanted the mock catalogs
to reflect the errors in the real data, we added
artificial velocity noise of $110 \ \kms$ to the redshift of each mock TF
galaxy before applying the \velmod\
algorithm, increasing $\sigv$ to $147\ \kms.${}\footnote{In retrospect, 
we added more noise than was necessary, but at the time we had
a higher estimate of the real universe $\sigv.$}
The mean value of $\sigv$ from the \velmod\ runs on the 20 mock
catalogs was $\vev{\sigv}=148.7\ \pm 4.6\ \kms,$ in excellent
agreement with the expected value. 
We conclude that \velmod\ produces an unbiased estimate of the
$\sigv,$ just as it does of $\beta_I.$  The rms error in the
determination of $\sigv$ from a single realization is $\sim 20 \
\kms$. 

The calculation in which we minimized 
$(cz_i - cz^{I}_i)^2$ also yielded estimates of the Cartesian components of
Local Group random velocity vector $\bfwlg.$ Their mean values
over 20 mock
catalogs is given in Table~\ref{table:mock}, together with the corresponding
mean values
returned from \velmod\ over the 20 mock catalog runs.  The two
are in impressive agreement. 
These values reflect an offset between the CMB to LG
transformation assigned to the simulation and the
average value of $\bfwlg$ assigned by the mock \iras\
reconstruction for $\beta_I=1$. 
We conclude that
\velmod\ properly measures the Cartesian components of $\bfwlg$
to within \sm 50 \kms\ accuracy per mock catalog.

\subsection{The TF Parameters Obtained from \velmod}
\label{sec:tfmock}

The mock catalogs also enable us to test the accuracy of the
TF parameters determined from the likelihood maximization procedure.
The comparison of input and output values is given in
Table~\ref{table:mock}.  The results for the TF slope and
scatter 
are consistent with the statement that \velmod\
returns unbiased values of these TF parameters for
each of the two samples. The fact that the TF scatters
and $\sigma_v$
are unbiased means that \velmod\ correctly measures
the overall variance in peculiar velocity predictions.

The TF zero points returned by \velmod\ are systematically in
error by 2--3 standard deviations.  This can be traced to a bias in
the \iras-predicted peculiar velocities; the mean value of the
quantity $f$ in
\S~\ref{sec:mocksigv} over 20 realizations was $0.018 \pm 0.007$. 
This bias makes the \iras-predicted
distances $d_\iras\simeq cz-u_\iras$ too large
by a factor of $1.018,$ or \sm 0.04 mag. To bring
the TF and \iras\ distances into agreement, the
TF zero points must decrease by this amount, which in fact
they do (cf.\ Table~\ref{table:mock}).
Thus, \velmod\
determines the TF zero points in such a way as
to compensate for a small systematic error in the
\iras\ predictions. 
We expect such an error to
be present in the real data as well, but it will be completely
absorbed into the TF zero points, and our derived value of $\beta_I$ will
be unaffected. 
  
\subsection{Properties of the \velmod\ Likelihood}
\begin{figure}[t!]
\centerline{\epsfxsize=4.0 in \epsfbox{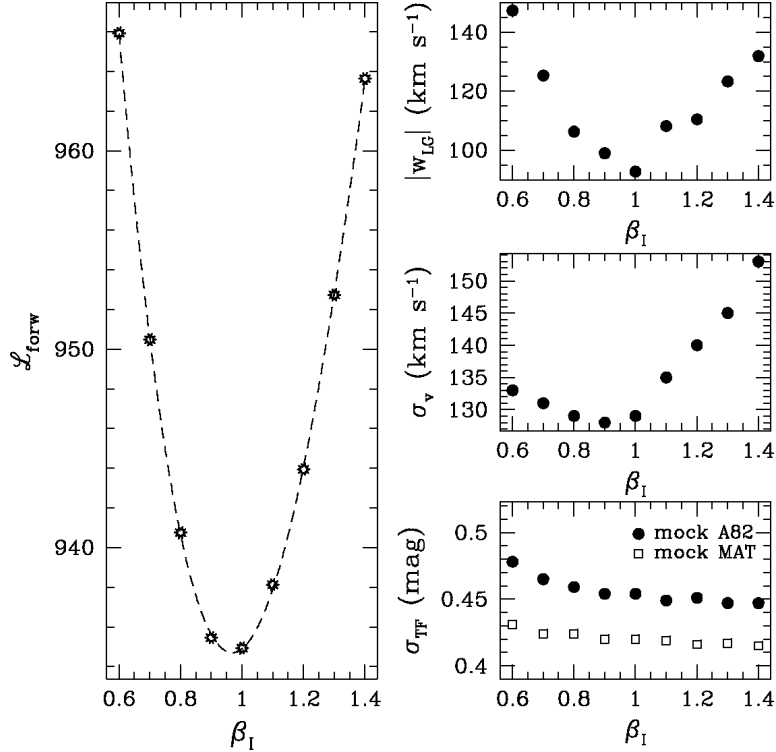}}
\caption{{\small Some of the parameters obtained from running \velmod\
on a single mock catalog. The left hand panel shows the
likelihood statistic along with the cubic fit used to
determine its minimum. The right hand panels show the
amplitude of the Local Group random velocity vector,
the velocity noise $\sigma_v,$ and the TF scatters for
the mock A82 and mock MAT samples. Note that the Local
Group velocity vector has its minimum amplitude for
$\beta_I\simeq 1.$ Note also that the TF scatters do
not track the likelihood curve, primarily because the
velocity noise $\sigma_v$ also measures the inaccuracy
of the fit. This demonstrates that minimizing TF scatter is
not equivalent to maximizing likelihood, as it is for
standard Method II.}}
\label{fig:mockparam}
\end{figure}

The mock catalogs may also be used to illustrate some important
features of the \velmod\ analysis. An example of these is shown
in Figure~3. The left hand panel shows $\likeforw$
versus $\beta_I$ for one of the 20 catalogs. The right hand panels
show how three other quantities vary with $\beta_I$ in the same
\velmod\ run: the amplitude of the LG random velocity $\bfwlg$ (top panel),
the velocity noise $\sigma_v,$ and the TF scatter $\sigtf$ for
each of the two mock TF samples (A82 and MAT) considered (bottom
panel). Note first that the amplitude of the LG velocity vector
is minimized near the true value of $\beta_I.$ This was generally
seen in the mock catalogs; it reflects the fact that the fits at
the wrong values of $\beta_I$ try to compensate for wrong peculiar velocity
predictions with Local Group motion. 
If $\bfwlg$
were held
fixed at its maximum likelihood value, or set equal to zero, the
$\likeforw$ versus $\beta_I$ curves would have sharper minima
and the $\beta$-uncertainty would be reduced (cf.\
\S~\ref{sec:results}). Unfortunately, we 
cannot do this for the real universe because we do not know $\bfwlg$
\apriori. Nevertheless, there we will find similar behavior;
$\bfwlg$ has a minimum near the best-fit value of $\beta_I$ for the real
universe. 

The figure also shows that $\sigma_v$ is a weak function of
$\beta_I$, but goes to a minimum at $\beta_I \approx 1$.  Its value at the
minimum for this realization is 127 \kms, within $1\,\sigma$ of the
correct value (Table~\ref{table:mock}). 
The
$\sigtf$ are also weak functions of $\beta_I$, but are in good
agreement with the input values 
near the maximum likelihood values of $\beta_I.$ Most importantly,
the figure demonstrates that maximizing likelihood {\em does not
necessarily correspond to minimizing TF scatter}, as we argued in
\S~\ref{sec:velmod-discussion}.  The TF scatters
in fact decrease monotonically with increasing $\beta_I.$ 
As they do, $\sigv$ increases to compensate.  
However,
one cannot simply minimize $\sigtf,$ or even a simple combination
of $\sigtf$ and $\sigv,$ to obtain an unbiased $\beta_I.$ One must
instead maximize likelihood as defined in \S~\ref{sec:velmod-detail}.

\section{Application to the Mark III Catalog Data}
\label{sec:real}
\subsection{Sample Selection}
\label{sec:selection}
\def\czlg{cz_{{\rm LG}}}
To apply \velmod\ to the real TF data, we needed first to identify
a suitable subsample of the Mark III catalog.
As discussed in \S~\ref{sec:velmod-discussion}, 
we elected to restrict the TF sample
to $\czlg \leq 3000\ \kms$, where here and throughout, we correct
heliocentric redshifts to the Local Group frame following Yahil,
Tammann, \& Sandage (1977).  We thus use
the Aaronson \etal\ (1982a; A82) and Mathewson \etal\ (1992;
MAT) TF samples, which are rich in local galaxies, restricted to this
redshift interval. 
The two cluster samples in the Mark
III catalog, HMCL and W91CL, 
contain only clusters at greater redshifts
and thus are not used here.
Finally, only a small fraction of the W91PP and CF samples is
found at $\czlg \leq 3000\ \kms$ (\sm 2\% of W91PP, \sm 15\% of
CF). This small number of additional galaxies
was not worth the additional 6 free parameters that would
be required for the likelihood maximization procedure
(\S~\ref{sec:method}). 

We made several further cuts on the data, as follows:
\begin{enumerate}
\item An RC3 $B$-magnitude limit of $m_B=14.0$ mag was adopted for
A82. As discussed by Willick \etal\ (1996), A82 galaxies
within 3000 \kms, and subject to this magnitude limit,
are well described by the ``one-catalog''
selection function of Willick (1994) that enters into the likelihood
equations (\S~\ref{sec:method}).
\item An ESO $B$-band diameter limit of $d_{{\rm ESO}}=1.6$
arcminute was adopted for MAT. As discussed by Willick \etal\ (1996),
this allows the MAT subsample to be described
by the one-catalog selection function of Willick (1994).
\item Only galaxies with axial ratios $\log (a/b)\geq 0.1$
were included. This cut, corresponding to an inclination limit
of $i\geq 38\degs$ (Willick \etal\ 1997), 
reduces TF scatter due to velocity width errors.
\item Galaxies with $\eta<-0.45$ (rotation velocities
less than about 55 \kms) were excluded. In practice, this
criterion applied only to the MAT sample, which contains numerous very
low-linewidth galaxies. The need for excluding such
objects was discussed by Willick \etal\ (1996).
\item Two objects within the Local Group, defined as having
raw forward TF
distances $<100\ \kms,$ were excluded.
No lower bound was placed on the {\em redshifts\/}
of sample objects, however.
\end{enumerate}

This left a sample of 856 A82 and MAT galaxies.
As discussed by Willick \etal\ (1996, 1997), real samples exhibit
a mainly Gaussian distribution of TF residuals, but with an admixture of a few percent
of non-Gaussian outliers.
We excluded eighteen additional galaxies
(4 in A82, 14 in MAT), or \sm 2\% of our sample,
because of their extremely large residuals from the TF relation.
Finally, then, 838 galaxies, 300 in A82 and
538 in MAT, were
used in the \velmod\ analysis. Of these, 53 are objects found
in both samples (though with different raw
data), and thus are used twice in the analysis.

\subsection{Velocity-Width Dependence of the TF Scatter}

It has been noted by a number of authors (Federspiel, Sandage, \&
Tammann 1994; Willick \etal\ 1997; Giovanelli \etal\ 1997) that
$\sigtf$ exhibits a velocity-width (or, equivalently, a luminosity)
dependence: luminous, rapidly rotating galaxies have smaller TF scatter than 
faint, slowly rotating ones. Willick \etal\ (1997) showed that this
effect could be parameterized by $\sigtf(\eta)=\sigma_0 - g\eta,$
with different values of $\sigma_0$ (the scatter for a typical, $\eta
= 0$, galaxy) and $g$ for each sample. 
For the
MAT sample, they found $g=0.33,$ while for A82 they
found $g=0.14.$ In the \velmod\ analysis, we treated $\sigma_0$
as a free parameter for both samples, but fixed the values of $g$ to the
Willick \etal\ (1997) values. For the remainder of this paper,
when we refer to $\sigtf$ we are actually referring
to the $\sigma_0$ for the respective samples. We note that 
a significant likelihood increase was achieved by adopting this
variable TF scatter, but that the derived value of $\beta_I$ was
essentially unchanged.  The mock catalogs were generated
and analyzed with $g = 0$. 

\subsection{Treatment of Virgo}
\label{sec:virgo}

To simplify the analysis,
we have taken the small-scale velocity noise,
$\sigma_v,$ to be independent of position. Clearly,
this assumption must fail in the immediate vicinity of a rich cluster.
The Virgo cluster is the only rich cluster within 3000 \kms.
Thus, we must 
artificially ``cool down'' the galaxies near Virgo.
We do so as follows:
if a galaxy lies within 10\degs\
of the Virgo core (taken to be
$l=283.78\degs,$ $b=74.49\degs$) on the sky, within 1500 \kms\ of
its mean Local Group redshift (taken to be $1035.1\ \kms,$ following
Huchra 1985),
and has a raw TF distance from Willick \etal\ (1997) between 800 and
2100 \kms, 
its Local Group redshift is set to the mean Virgo value.
Twenty objects used in the \velmod\ analysis meet these criteria. We
similarly collapsed mock Mark III objects associated with the Virgo
cluster in the mock catalogs.

\subsection{Implementation of a Quadrupole Flow}
\label{sec:quadrupole}

In the discussion of \velmod\ in \S~\ref{sec:method}, it was
assumed that the \iras-predicted
velocity field, for the correct value of $\beta_I,$
is as good a model as can be obtained. 
However, there can be additional contributions to the local flow field
from structures beyond the volume surveyed ($R \leq 12,800\
\kms$), as well as from shot noise- and Wiener-filter-induced
differences between the 
true and derived density fields beyond
3000\ \kms\ but within the \iras\ volume
(cf.\ Appendix B). 

Fortunately, the nature of this contribution is such that we can
straightforwardly model its general form, and thus treat it as a quasi-free parameter
(see below) in the \velmod\ fit.
Let us write the error in
the \iras-predicted velocity field due to 
incompletely sampled fluctuations 
as $\verr(\bfr).$
Because the total peculiar velocity field, $\bfv+\verr$ must satisfy
Eq.~(\ref{eq:delvdelg}), and because $\bfv$ does so by construction
(Eq.~\ref{eq:vpdeltag}), it follows that $\verr$ must have zero
divergence.
Moreover, if we suppose that $\verr$ corresponds to the growing
mode of the linear peculiar velocity field, it must have zero curl as
well. These properties will be satisfied if $\verr$ is given by the
gradient of a velocity potential $\phi$ that satisfies Laplace's equation.
Such a potential may be expanded in a multipole series, each term of
which vanishes at the origin (where, by construction, $\verr$ must itself
vanish).

The leading term in the resulting expansion of $\verr$ is a
{\em monopole,} $\verr^{(0)}(\bfr)=A\bfr,$ or Hubble flow-like term.
However, such a
term is
degenerate with the zero point of the TF
relation (\S~\ref{sec:tfmock}), and is thus undetectable.
The next term
in the expansion is a {\em dipole,} $\verr^{(1)}={\bf B},$ or bulk flow
independent of position. Like the monopole term, however, the
dipole term is undetectable, because we work in the frame of
the Local Group. Whatever bulk flow is generated by distant density
fluctuations is shared by the Local Group as well.
\def\vq{\bfv_Q}
The leading term in the expansion of $\verr(\bfr)$ to which our method
is sensitive is therefore a {\em quadrupole\/} term. Such a term 
represents the tidal field of mass density fluctuations not traced by
the \iras\ galaxies. 
We may write the quadrupole velocity component as
\begin{equation}
\vq(\bfr) = {\cal V}_Q\,\bfr,
\label{eq:quad}
\end{equation}
where ${\cal V}_Q$ is a $3\times 3$ matrix. 
In order for both the divergence and curl of $\vq(\bfr)$ to vanish,
${\cal V}_Q$ must be a traceless, symmetric matrix. Consequently,
it has only five independent elements, two diagonal and three
off-diagonal. 

We could allow for the presence of such a quadrupole in
\velmod\ by treating these five elements as free parameters.
However, this is a dangerous procedure, because the
modeled quadrupole would then have the freedom to fit the
quadrupole {\em already present\/} in the \iras\ velocity
field, which is generated by observed density fluctuations.  
We wish to allow for the external
quadrupole, but we do not want it to fit the $\beta$-dependent
quadrupolar component of the \iras-predicted velocity field.
In other words,
we want the external quadrupole to be that required
for the true value of $\beta_I,$ which we do not know \apriori,
rather than the ``best fit'' value at any given $\beta_I. $
This problem would indeed be very serious if inclusion of
the quadrupole made a large difference
in the derived value of $\beta_I.$ Fortunately, however, it
does not. As we show below, we obtain
a maximum likelihood value $\beta_I=0.56$ when
the quadrupole is not modeled. When we treat all five components
of the quadrupole as free parameters for each $\beta_I,$
we obtain $\beta_I=0.47.$\footnote{This value differs 
from the value of
0.49 quoted in the Abstract because we will {\em not\/} allow the
quadrupole to be free parameters at each value of $\beta_I$.}
Because the best-fit quadrupole
is relatively insensitive
to $\beta_I,$
we can estimate the external quadrupole
by averaging the fitted values of
the five independent components obtained for 
$\beta_I=0.1,0.2,\ldots,1.0.$
In this way, we ``project out'' the $\beta_I$-independent
part of the quadrupole. In our final \velmod\ run, we use
this average external quadrupole at each value of $\beta_I.$
Throughout, we ignore the very small effect that this quadrupole might
have on the derived \iras\ density field.
\begin{figure}[t!]
\centerline{\epsfxsize=4.0 in \epsfbox{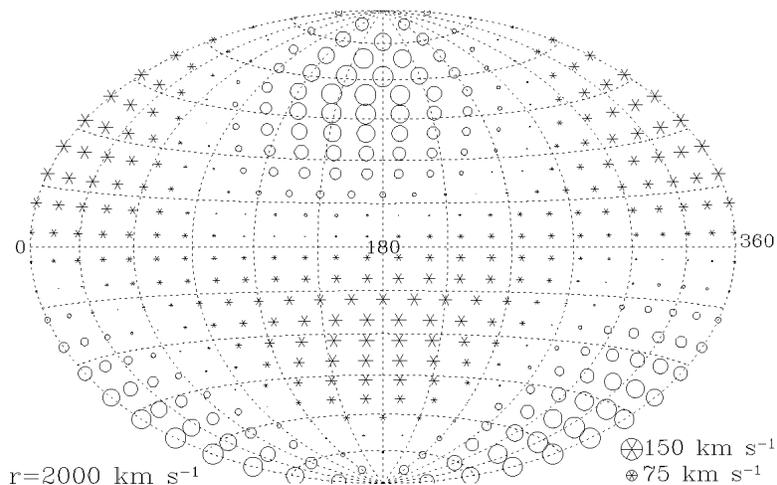}}
\caption{{\small The external quadrupolar velocity
field used in the \velmod\ analysis, plotted in Galactic coordinates.
Open symbols indicate negative radial
velocities, stars positive radial velocities.
The amplitude of the quadrupole is shown for
a distance $r=2000\ \kms.$ As indicated by Eq.~(\ref{eq:quad}),
the quadrupole flow increases linearly with
distance at a given position on the sky. The maximum
amplitude of the quadrupole at this distance
is 147 \kms, which occurs at $l\simeq 165\degs,$
$b\simeq 55\degs,$ as well as on the opposite
side of the sky.}}
\label{fig:velquad}
\end{figure}

In Figure~4, this quadrupole field
is plotted on the sky in Galactic coordinates for
a distance of 2000\ \kms. The inflow
due to the quadrupole, which occurs near the
Galactic poles, is of greater amplitude
than the outflow, which occurs at low Galactic
latitude. The quadrupole reaches its maximum
amplitude at $l\simeq 165\degs,$ $b\simeq 55\degs,$
in the direction of the Ursa Major cluster,
as well as on the opposite side of the sky.
In \S~\ref{sec:resid}, when we plot \velmod\
residuals on the sky with and without the
quadrupole, the need for the quadrupole
field shown in Figure~4 will
become clear. Indeed, we will show in \S~\ref{sec:resid}
that the \velmod\ fit is statistically acceptable
only when the quadrupole is included. 
Table~\ref{table:real} tabulates the numerical
values of the independent elements of ${\cal V}_Q$ that
generate this flow.  The rms value of this quadrupole over the sky is
3.3\%, pleasingly close to the value we expect 
from theoretical considerations 
(Appendix~\ref{sec:quad-theory}). 

When both the quadrupole and the Local Group random velocity
vector are modeled, the radial peculiar velocity $u(r)$ that enters into
the likelihood analysis (see Eq.~\ref{eq:pzr}) is given by
\begin{equation}
u(r) =  \left(\bfv_\iras(\bfr) + {\cal V}_Q \bfr - \bfwlg\right)\cdot
\hat\bfr\,.
\label{eq:urquad}
\end{equation}
We emphasize again that while the three components of
the Local Group random velocity $\bfwlg$ are treated as
free parameters in \velmod, the five independent parameters
of ${\cal V}_Q$ are not, with the exception
of a single run we used to obtain and then average their fitted
values at each $\beta_I.$ In the final run, from which
we derive the estimate of $\beta_I$ quoted in the Abstract, 
the quadrupole velocity field shown in Figure~4
was used at each value of $\beta_I.$

\begin{figure}[t!]
\centerline{\epsfxsize=4.0 in \epsfbox{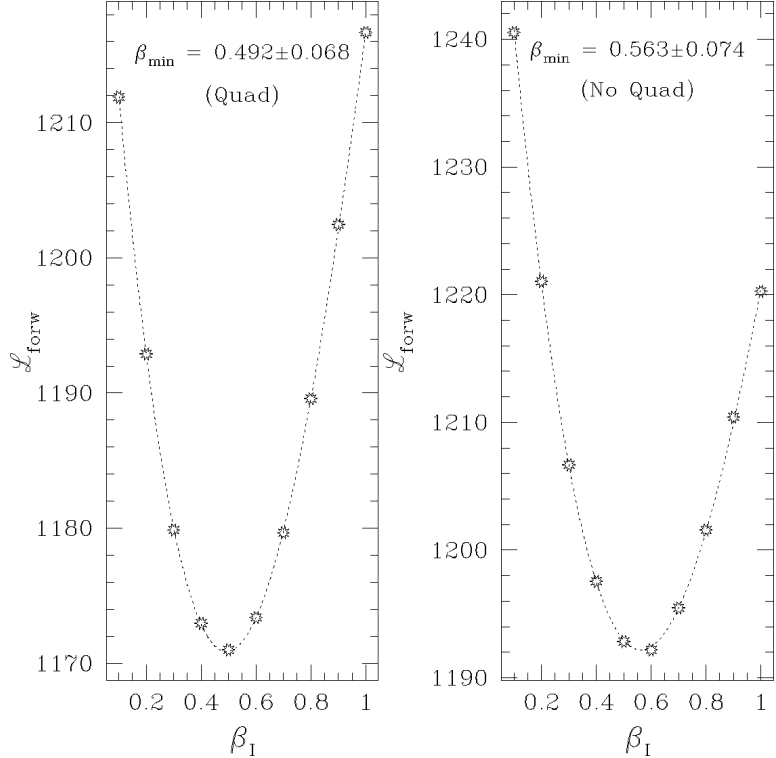}}
\caption{{\small The \velmod\ likelihood statistic, $\likeforw$
(Eq.~\ref{eq:deflamforw}), plotted as a function of
$\beta_I$ for the real data. In the left hand plot,
an external quadrupole is  modeled, as described
in the text. In the right hand plot, no external quadrupole
is included in the velocity field. Cubic fits to
the likelihood points are shown as dotted lines. The
minima of the fitted curves, $\beta_{{\rm min}},$ are the
maximum likelihood estimates of $\beta_I$ in each case.
Note the very different values of the vertical axes
of the two plots; this indicates the large increase
in formal likelihood when the quadrupole is included.}}
\label{fig:like31}
\end{figure}
 
\begin{figure}[t!]
\centerline{\epsfxsize=4.0 in \epsfbox{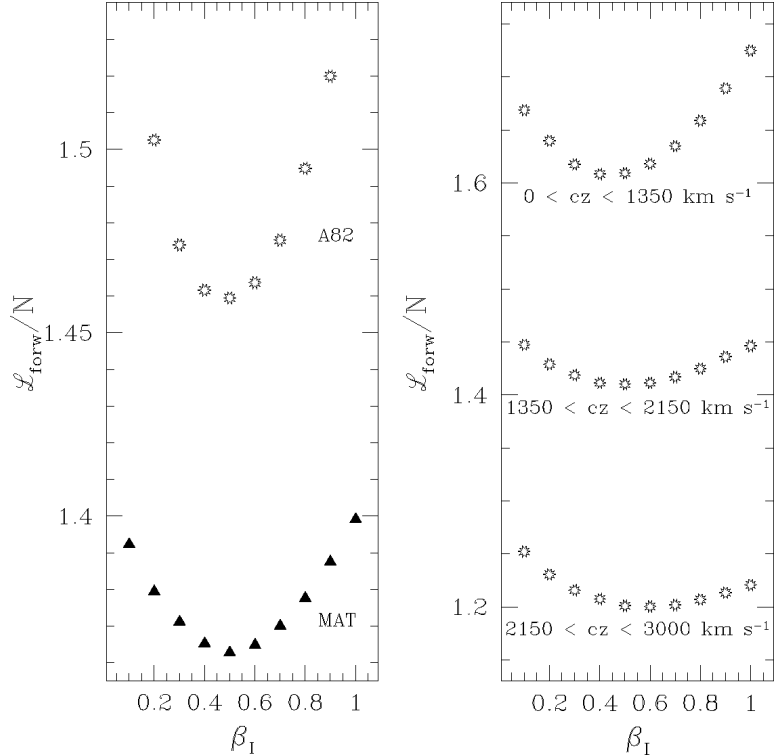}}
\caption{{\small Breakdown of the \velmod\ likelihood statistic
among subsamples. The left hand panel plots
likelihood per point versus $\beta_I$ for the A82 and MAT samples
individually.
%Note that each sample exhibits a minimum in $\likeforw$
%at about the same value of $\beta_I.$
The right hand panel plots
likelihood per point versus $\beta_I$ for three different redshift
intervals containing roughly the same number of objects. In each
case, the minimum occurs within \sm 0.1 in $\beta_I$ of the global minimum
in $\likeforw$ at $\beta_I=0.492.$}}
\label{fig:like_obj}
\end{figure}
 
\begin{figure}[t!]
\centerline{\epsfxsize=4.0 in \epsfbox{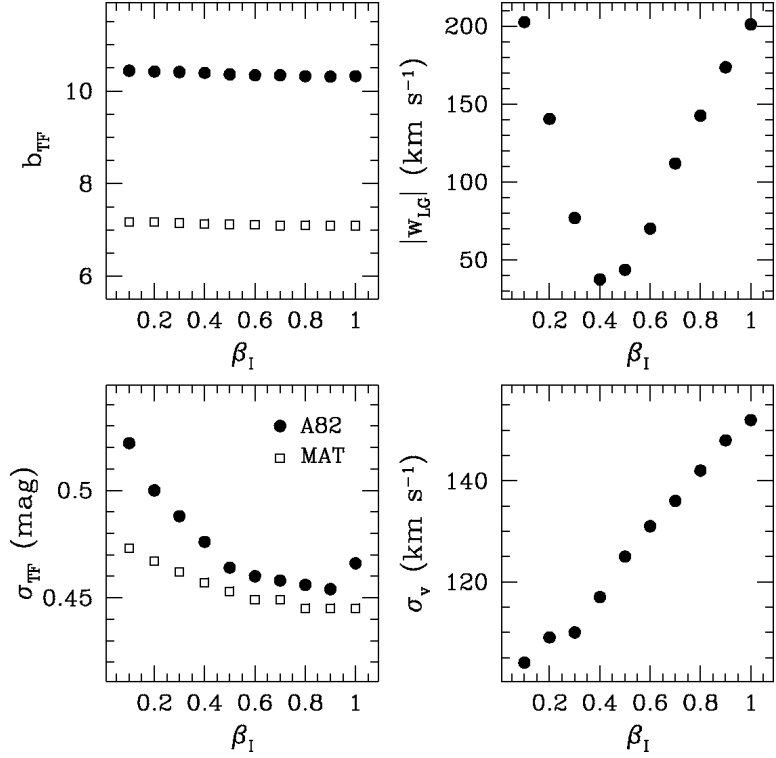}}
\caption{{\small Left hand panels: the TF slopes
(top) and scatters (bottom), for the A82 and
MAT samples, derived from \velmod\
as a function of $\beta_I.$
Right hand panels: the
amplitude of the Local Group random velocity vector (top)
and the velocity noise $\sigma_v$ (bottom) derived
from \velmod\ as a function of $\beta_I.$ All plots
correspond to the run in which the quadrupole was modeled.}}
\label{fig:realparam}
\end{figure}

\begin{figure}[t!]
\centerline{\epsfxsize=4.0 in \epsfbox{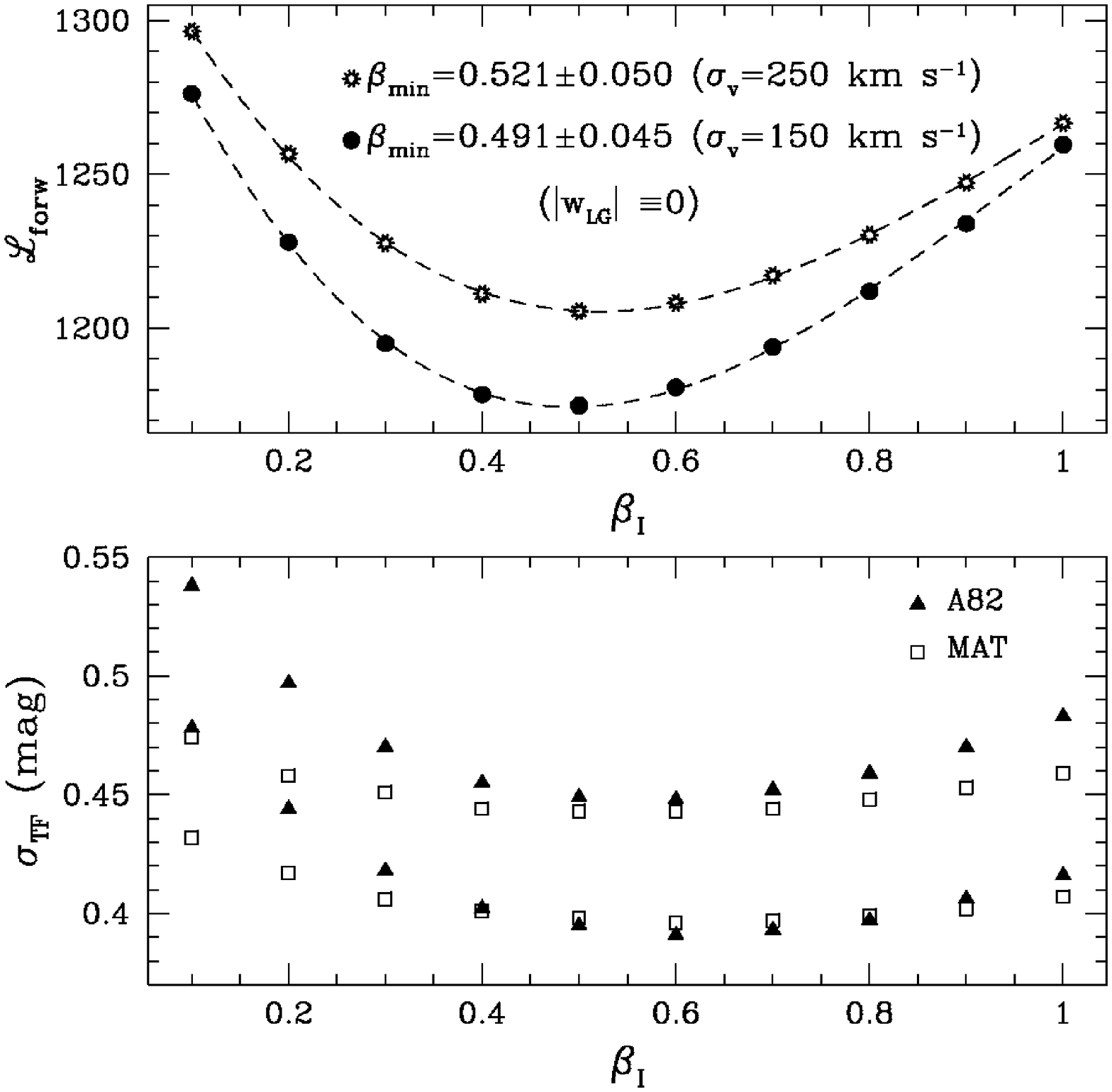}}
\caption{{\small Top panel: The \velmod\ likelihood
statistic $\likeforw$ as a function of $\beta_I$
for a two runs in which the Local Group velocity vector was forced to
vanish, and the velocity noise parameter $\sigma_v$ was held fixed
at 150\ and 250\ \kms.
Although the formal likelihoods of the fit
are worse than that of the full fit 
(compare with Figure~\ref{fig:like31}),
particularly for $\sigv=250\ \kms,$
the maximum likelihood estimates of $\beta_I$ are nearly unchanged.
Bottom panel: variation in the TF scatters $\sigtf,$
for the A82 and MAT samples, as a function of $\beta_I$ 
for these \velmod\ runs.
The larger values correspond to the $\sigv=150\ \kms$ run.
Note that the minimum derived TF scatters do not necessarily
correspond to maximum likelihood; see
text for further details.}}
\label{fig:wlg0}
\end{figure}
 
\subsection{Results}
\label{sec:results}

The outcome of applying \velmod\ to the A82+MAT
subsample described above is presented in Figure~5.
The \velmod\ likelihood curves are shown both with
and without the external quadrupole included.
The formal likelihood is vastly improved
when the quadrupole is included in the fit: since
the likelihood statistic $\likeforw$ is defined
as $-2\ln\left[P({\rm data}|\beta)\right],$ the
\sm 20 point reduction in the minimum of $\likeforw$, minus the five
extra degrees of freedom when the quadrupole is modeled corresponds to
a probability
increase of a factor $\sim e^{7.5}\simeq 2000.$ 
The improvement in formal likelihood through the
addition of the quadrupole is so pronounced that we take
the maximum likelihood value of $\beta_I$ from
that fit, $0.492\pm 0.068,$ as our best estimate.
%\footnote{In
%\S~\ref{sec:mock} we described the calculation of two-sided
%error bars $\delta\beta_\pm.$ Using these, our result
%may be described as $\beta_I=0.492^{+0.069}_{-0.066}.$ 
%The upper and lower errors are so close as to make
%the distinction unimportant, and from here on
%we give only averages of the two.}
However, 
the maximum
likelihood estimate of $\beta_I$ when the quadrupole
is neglected, $0.563\pm 0.074,$ differs from our
best value at only the 1-$\sigma$ level. While the
quadrupole is important, it does not qualitatively
affect our conclusions about the likely value of $\beta_I.$

We can make several additional tests of the robustness of our
results. Figure~6 
shows how the likelihoods per object break down for fits to different
cuts on the sample; see also
Table~\ref{table:real}. The left hand 
panel plots $\likeforw/N$ versus $\beta_I$ for the A82
and MAT samples separately, where $N=300$ for A82 and
$N=538$ for MAT.
Cubic fits
to the individual sample likelihoods yield
$\beta_I=0.489\pm 0.084$ and $0.498\pm 0.107$ for
A82 and MAT respectively. This agreement is remarkable,
given that there are only 53 galaxies in common between the two
samples. 
Note that the $\beta$-uncertainty
is larger for the MAT sample, even though it contains
nearly twice as many objects as the A82 sample. This
is because the MAT objects typically lie at
larger distances
than do A82 objects, a property of the likelihood fit we now illustrate.

The right hand panel of Figure~6 plots
$\likeforw/N$ versus $\beta_I$ for three subsamples in different
redshift ranges. As Table~\ref{table:real} shows, the agreement in the
derived values of $\beta_I$
is quite good. Changing the specific redshift intervals
used for this test does not significantly change the results. 
Note that the $\beta$-resolution decreases as one goes
to higher redshift, despite the fact that there are nearly equal numbers
of objects in each of the three redshift bins.
This is because the likelihood
is sensitive mainly to the fractional distance
error in the \iras\ prediction. 
Hence, nearby galaxies are more diagnostic of
incorrect peculiar velocity predictions, and
thus of $\beta_I.$

The fact that $\likeforw/N$ decreases with redshift 
should not be interpreted as
meaning that more distant objects are better fit by the velocity
model. This decrease instead reflects a property
of the \velmod\ likelihood implicit in Eq.~(\ref{eq:pmapprox}),
which shows that
the expectation
value of $\likeforw/N$ is $\sim 1 + \ln(2\pi)
+ \ln\left[\sigtf^2+\left(2.17
\sigma_v/[w(1+u'(w))]\right)^2\right],$ 
which increases with decreasing $cz$ in general.
This effect will be particularly pronounced in
flat zones ($u' \sim -1$) in the redshift-distance relation, which
are found in the Local Supercluster, which is why there
is a marked difference between $\likeforw/N$ for the A82
and MAT samples (the former preferentially populates the
Local Supercluster region).

In Figure~7 we
plot for the real data the same quantities plotted
for a mock catalog in Figure~3,
as well as the TF slopes.
The slopes are extremely insensitive
to $\beta_I.$ This indicates that the \iras\
assigns low and high linewidth galaxies nearly
same relative distances at all $\beta_I.$
Significantly, the
amplitude of the fitted Local Group velocity vector
is minimized near the maximum likelihood value of $\beta_I$, just as
we saw with the mock catalog. 
This indicates once again that the fit attempts to compensate
for a poor velocity field at very low and high $\beta_I$
by moving the Local Group.
The mock catalogs showed us that the errors on the Cartesian components of
$\bfwlg$ are of order 50 \kms (Table~\ref{table:mock}).
Thus, the small value of $|\bfwlg|$ 
obtained from \velmod\ indicates
that the Yahil \etal\ (1977) transformation to the Local Group
barycenter is correct to within \sm 50 \kms, and that
the Local Group has random velocity $\simlt 50\ \kms$ relative to the mean
peculiar velocity field in its neighborhood.

The lower right panel of Figure~7 shows
that $\sigma_v$ increases monotonically with $\beta_I.$
Its maximum likelihood value is 125 \kms. This is a
remarkably small number, when one considers that it includes
the effect not only of random velocity noise but also of
\iras\ prediction error. In particular, if our estimate of the \iras-prediction
errors derived from our mock catalog experiments (\S~\ref{sec:mocksigv}),
\sm 84\ \kms, are roughly correct, our value for $\sigv$ implies
that the true 1-dimensional velocity noise is $\simlt 100\ \kms.$
This result is
consistent with past observations that the velocity field
is ``cold'' (cf., Sandage 1986; Brown \& Peebles 1987; Burstein 1990; Groth,
Juszkiewicz, \& Ostriker 1989; Strauss, Cen, \& Ostriker 1993;
Strauss, Ostriker, \& Cen 1997).
Finally, the lower left panel demonstrates again what was
seen earlier with the mock catalogs (Figure~3),
namely, that maximizing probability does not correspond to
minimizing TF scatter. In large measure, this is because there
is a tradeoff between the variance due to the velocity noise
$\sigma_v$ and that due to the TF scatter. As $\beta_I$ approaches
1, $\sigma_v$ gets steadily larger; $\sigtf$ gets correspondingly smaller,
despite the fact that the high $\beta_I$ models are worse fits to the TF data.
The TF scatters level out or rise only at $\beta_I\simeq 1.$

A final test of robustness involves
eliminating the freedom in the \velmod\ fit provided by the parameters
$\bfwlg$ and $\sigma_v.$ One could argue
that these parameters are like the quadrupole: they ``are what they are,''
and we should not allow them to absorb the fit inaccuracies at the
wrong value of $\beta_I.$ To assess this, we carried out two \velmod\
runs in which $\bfwlg$ was assumed to vanish identically.
In the first run, we fixed the value of $\sigma_v$ at 150 \kms,
and in the second at 250 \kms.
The quadrupole was held fixed at
its best fit value; the free parameters in this fit were
limited to $\beta_I$ and the three TF parameters 
for each of the two samples. The results of this exercise are shown in
Figure~8 and in Table~\ref{table:real}. 
The derived values of $\beta_I$ differ
inconsequentially from our best estimate obtained from the full fit.
This shows that allowing ourselves the freedom to fit both $\bfwlg$ and
$\sigma_v$ does not materially affect the derived value of $\beta_I.$
The formal uncertainties in $\beta_I$ are much reduced relative to
the full fit because formerly free parameters have been held fixed.
For $\sigma_v=150\ \kms$ the formal likelihood is
worse than for the full fit, but only at the $\sim 2\,\sigma$
level. This reflects the fact that $\sigma_v=150\ \kms$
and $\bfwlg=0$ themselves differ by only \sm $1\,\sigma$
from their maximum likelihood values, according to
our error estimates from \S~\ref{sec:mocksigv}. 
However,
the formal likelihood for the $\sigma_v=250\ \kms$
run is considerably worse (by a factor of
$\sim 10^{-7}$) than
for the full fit. This shows that we rule out such a
large $\sigma_v$ at high significance.

The bottom panel of Figure~8 shows the
fitted values of $\sigtf$ as a function of $\beta_I$
for each of the two values of $\sigma_v$ and
for each of the two TF samples.
The TF scatters now track the likelihood much better than they
did in the full fit (bottom panel of Figure~7);
with $\sigv$ fixed, maximizing likelihood is more nearly
equivalent to minimizing TF scatter. However, they are still not the
same thing: likelihood maximization occurs for $\beta_I\simeq 0.5,$
whereas
TF scatter is minimized at $\beta_I\simeq 0.6.$
This is 
due to the nonlocal nature of the probability
distribution described by Eq.~(\ref{eq:pmgetaz}) (cf.\
Figure~1). The likelihood 
of a given data point depends on the peculiar velocity and density
fields all along the line of sight interval allowed by the
TF and velocity dispersion probability factors, not merely on
how close the TF-inferred and \iras-predicted
distances are to one another.

The bottom panel of Figure~8 also shows that
the TF scatter one derives from \velmod\ depends on
the value of $\sigma_v.$  The full fit
told us that \iras\ errors plus true velocity noise amount
to \sm 125 \kms.  The values of $\sigtf$ obtained in the
full fit (Table~\ref{table:real})
absorbed the remaining variance. Changing $\sigma_v$
to 150 \kms\ reduces the TF scatters by about 0.01 mag.
With $\sigma_v$ fixed
at 250 \kms, however, we find 0.39 and 0.40 mag for the
A82 and MAT TF scatters. While these latter values are
certainly
underestimates, the large changes
demonstrate that it is very difficult to estimate $\sigtf$ to
high accuracy because of its covariance, however slight,
velocity noise. 
This is one reason that it is inadvisable
to use the value of $\sigtf$ obtained from fitting TF
data to peculiar velocity models as a measure of the
goodness of fit. We return to this issue in \S~\ref{sec:resid} below.

\subsection{\velmod\ Results using 500 \kms\ Smoothing}
\label{sec:smoo500}

In our mock catalog tests, we found that using 500 \kms\ smoothing
in the \iras\ reconstruction resulted in \sm 20\% overestimates of $\beta_I$
(\S~\ref{sec:beta-accuracy}).
However, because the mock catalog may not faithfully reproduce
the dynamics of the real universe, it is useful to see how much 
$\beta_I$ changes for the real data when the 500\ \kms-smoothed
\iras\ reconstructions are used. We carried out two such \velmod\ runs,
one with and one without the quadrupole. 
(We determined the quadrupole the same way as for the 300\ \kms\ smoothed 
reconstruction, and found that the two differ little.) 
The resulting maximum likelihood estimates of $\beta_I$
are listed in
Table~\ref{table:real}. The larger smoothing results in
an increase in $\beta_I,$ as expected. However, 
the 500 \kms\ result, $\beta_I=0.544\pm 0.071,$
is within $1\,\sigma$ of our favored result
obtained at 300 \kms\ smoothing. If we reduce this
value by 20\% in accord with the bias seen in the
mock catalogs, we obtain $\beta_I=0.45\pm 0.07,$ also
within $1\,\sigma$ of our preferred result. Our choice
of a 300 \kms\ smoothing scale is thus unlikely to have
led us seriously astray, even if the mock catalogs are imperfect guides.

\subsection{Consistency of the Mark III and \velmod\ TF Relations}
\label{sec:consistency}

\begin{figure}[t!]
\centerline{\epsfxsize=4.0 in \epsfbox{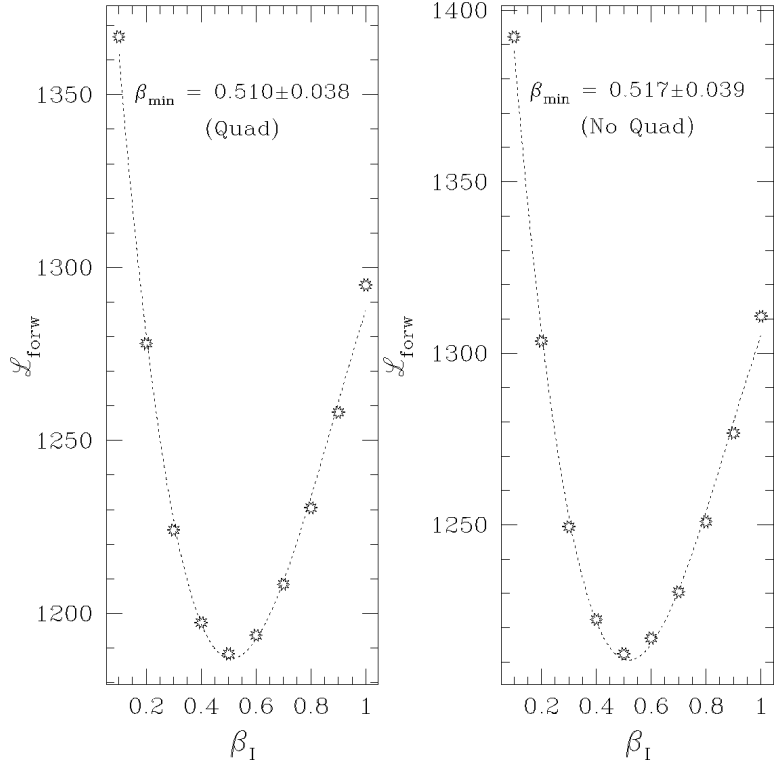}}
\caption{{\small Same as Figure~\ref{fig:like31}, except
that the A82 and MAT TF parameters have been fixed
at their Mark III catalog values.
The extremely small formal error bars result from
fixing the TF parameters and are unrealistic. The
maximum likelihood values of $\beta_I$ differ negligibly
from those obtained when the TF parameters are free.}}
\label{fig:velm_markiii}
\end{figure}

In constructing
the Mark III catalog, Willick \etal\ (1996, 1997)
required that the TF distances for
objects common to two or more samples agree in the mean.
As noted above, \velmod\ yields an independent TF calibration
for each sample included in the analysis.  As a further
consistency check, we can ask
whether the \velmod\ TF calibrations for the A82 and
MAT samples are also 
mutually consistent.

We compared A82 and MAT TF distances using the \velmod\ TF
relations
for 75 objects common to the two samples. We limited the
comparison to objects whose A82 versus MAT TF distance moduli
differ by 0.8 mag or less. 
(Not all of these objects 
were part of the
\velmod\ analysis, as some did not meet the
criteria outlined in \S~\ref{sec:selection}). We found that
the \velmod\ calibrations yield an
average distance modulus difference (in the
sense MAT$-$A82) $\left\langle \Delta\mu
\right\rangle = -0.056\pm 0.046$ mag; the Mark III TF
calibrations yield $\left\langle \Delta\mu\right\rangle =
0.018\pm 0.046$ mag.
The corresponding
{\em median\/} distance modulus differences are
$-0.015$ mag (\velmod) and $0.035$ mag (Mark III).  Thus, as measured
by the criterion of generating mutually consistent TF distances
among samples, \velmod\ gives the correct result. 
In Table~\ref{table:real} we list the \velmod\ TF parameters
and their Mark III counterparts. We see that
the A82 zero points, slopes, and scatters 
derived from the two methods are in almost perfect agreement. 
The MAT zero points and scatters also agree to well within
the errors.
The MAT slopes show a somewhat larger discrepancy. However, the
two slopes are nearly within their mutual error bars; moreover, 
the MAT sample use here is only about half as large as
that used by Willick \etal\ (1996) in deriving the MAT TF slope.
In any case, this slope difference,
even if real, is of no consequence for determination of $\beta_I,$
as we now show.

As a final test of \velmod-Mark III consistency, we ran \velmod\ without
allowing the TF parameters to vary, instead holding them fixed at
their Mark III values. We did so both with and without the quadrupole,
while holding $\sigma_v$ fixed at $150\ \kms$ and setting $\bfwlg\equiv 0$
(note from Figure~8 that these latter velocity parameters
yield the same $\beta_I$ as when they are allowed to vary freely). 
The results of this exercise are shown in Figure~9
and tabulated in Table~\ref{table:real}. As can be seen, 
while there is a large formal
likelihood decrease relative to the best solution, 
{\em using the Mark III TF relations has a negligible effect on
the value of $\beta_I$ obtained from \velmod.} In particular, use
of the Mark III TF relations does not bring our \velmod\ result
appreciably closer to the \potiras\ result, $\beta_I=0.86,$ of Sigad \etal\ (1997).
We discuss this issue further in \S~\ref{sec:velmod-potiras}.
Note that, in contrast with full \velmod, neglect
of the quadrupole now has no effect on the derived $\beta_I,$
although its inclusion still results in a significant likelihood
increase. The indicated formal error bars on  
$\beta_I$ should not be taken literally here, because
fixing the TF zero points prevents them from compensating 
for \iras\ zero point errors (cf.\ \S~\ref{sec:tfmock}).

\section{Analysis of the Residuals: Do the Predictions Match the Observations?}
\label{sec:resid}
The \velmod\ analysis
can tell us which velocity field models---which values of $\beta_I,$
$\sigma_v,$ $\bfwlg,$ and quadrupole parameters---are ``better'' than
others. However, as with maximum likelihood approaches generally,
it cannot by itself tell us which, if any, of these models is an acceptable
fit to the data. This is because we do not have precise, \apriori\ knowledge
of the two sources of variance,
the velocity noise $\sigma_v$ and the TF scatter $\sigtf.$
We have instead treated these quantities as free parameters and
determined their values by maximizing likelihood.
As a result, a standard $\chi^2$ 
statistic will be 
\sm 1 per degree of freedom even if the fit is poor.

We can, of course, ask whether the values of $\sigtf$ and
$\sigv$ obtained from \velmod\ agree with independent estimates.
It is reassuring that they do. We find $\sigtf\simeq 0.46$
mag for both the A82 and MAT samples, within the
range estimated by Willick \etal\ (1996) by methods
independent of peculiar velocity models. This agreement
is of limited significance, however.
TF scatter is very sensitive
to non-Gaussian outliers
(\S~\ref{sec:selection}), and thus to precisely
which objects have been excluded.
Furthermore,
the MAT subsample used here is only about half as large
as the MAT subsample used by Willick \etal\ (1996) to
estimate its scatter.
The \velmod\ result for the velocity noise,
$\sigma_v\simeq 125\ \kms,$ is remarkably small, and appears
consistent with recent studies based on independent methods
(e.g., Miller, Davis, \& White 1996; Strauss,
Ostriker, \& Cen 1997).  
Indeed, because $\sim 90\ \kms$ may be attributed to
\iras\ velocity prediction errors (\S~\ref{sec:mocksigv}), our value of $\sigma_v$
suggests a true 1-D velocity noise of $\simlt 90\ \kms.$ 
Still, the small
$\sigma_v$ is not necessarily diagnostic;
for demonstrably poor models (e.g., $\beta_I\leq 0.2$)
we find an even smaller value of $\sigma_v.$ 
An alternative approach is
thus required for identifying a poor fit.

Consider fitting
a straight line
$y=ax+b$ by least squares to data
$(x_i,y_i)$ whose errors are unknown. One obtains $a,$ $b,$ and
also the rms scatter about the fit. Because the scatter
is derived from the fit, the $\chi^2$ statistic
is \sm 1 per degree of freedom by construction. However,
if the straight line is a {\em bad\/} fit---if, say, the relation
between $y$ and $x $ is actually quadratic---then
the residuals from the fit will exhibit {\em coherence.}
Coherent residuals in
excess of what is expected from the observed scatter 
would signify that a model is a poor fit.
In this section, we will make such an assessment for the \velmod\
residuals. We will first define a suitable residual and plot
it on the sky. We will demonstrate coherence and incoherence
of the residuals, for ``poor'' and ``good'' models respectively,
by plotting residual autocorrelation functions. Motivated
by these considerations, we will
define and compute a statistic that measures goodness of fit.

\begin{figure}[t!]
\centerline{\epsfxsize=4.5 in \epsfbox{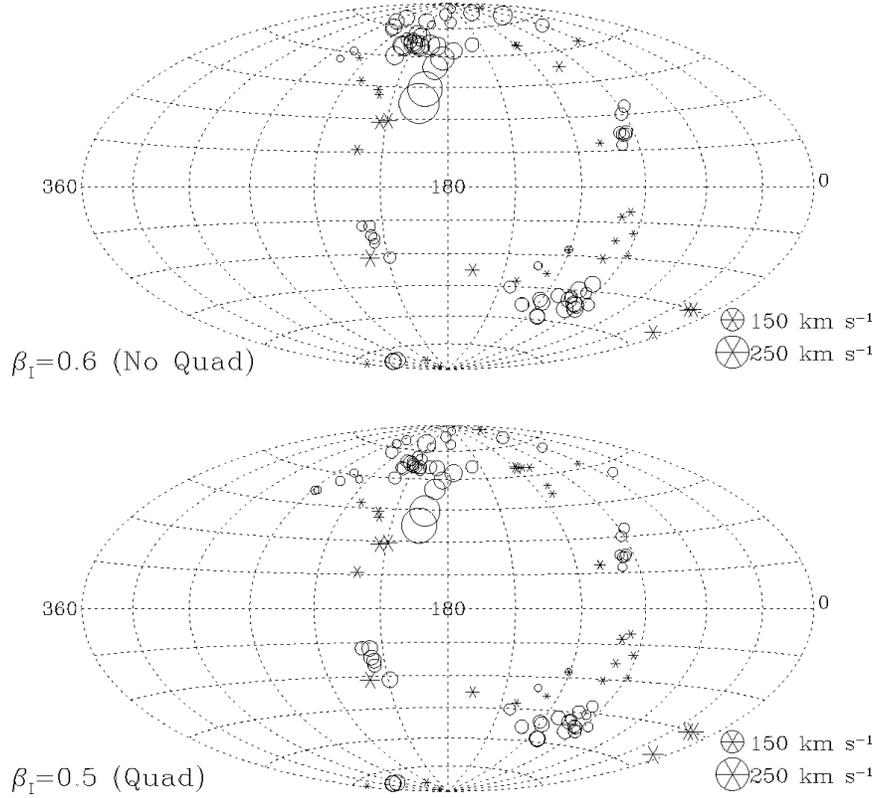}}
\caption{{\small \velmod\ velocity residuals plotted on the
sky in Galactic coordinates,
for objects with $0 < \czlg \leq 1000\ \kms.$ The top panel
is for the $\beta_I=0.6$ run without the quadrupole. The
bottom panel is for the $\beta_I=0.5$ run with the quadrupole
modeled. Open circles indicate objects that are inflowing
relative to the \iras\ prediction; stars indicate outflowing objects.}}
\label{fig:resid01}
\end{figure}
 
\begin{figure}[t!]
\centerline{\epsfxsize=4.5 in \epsfbox{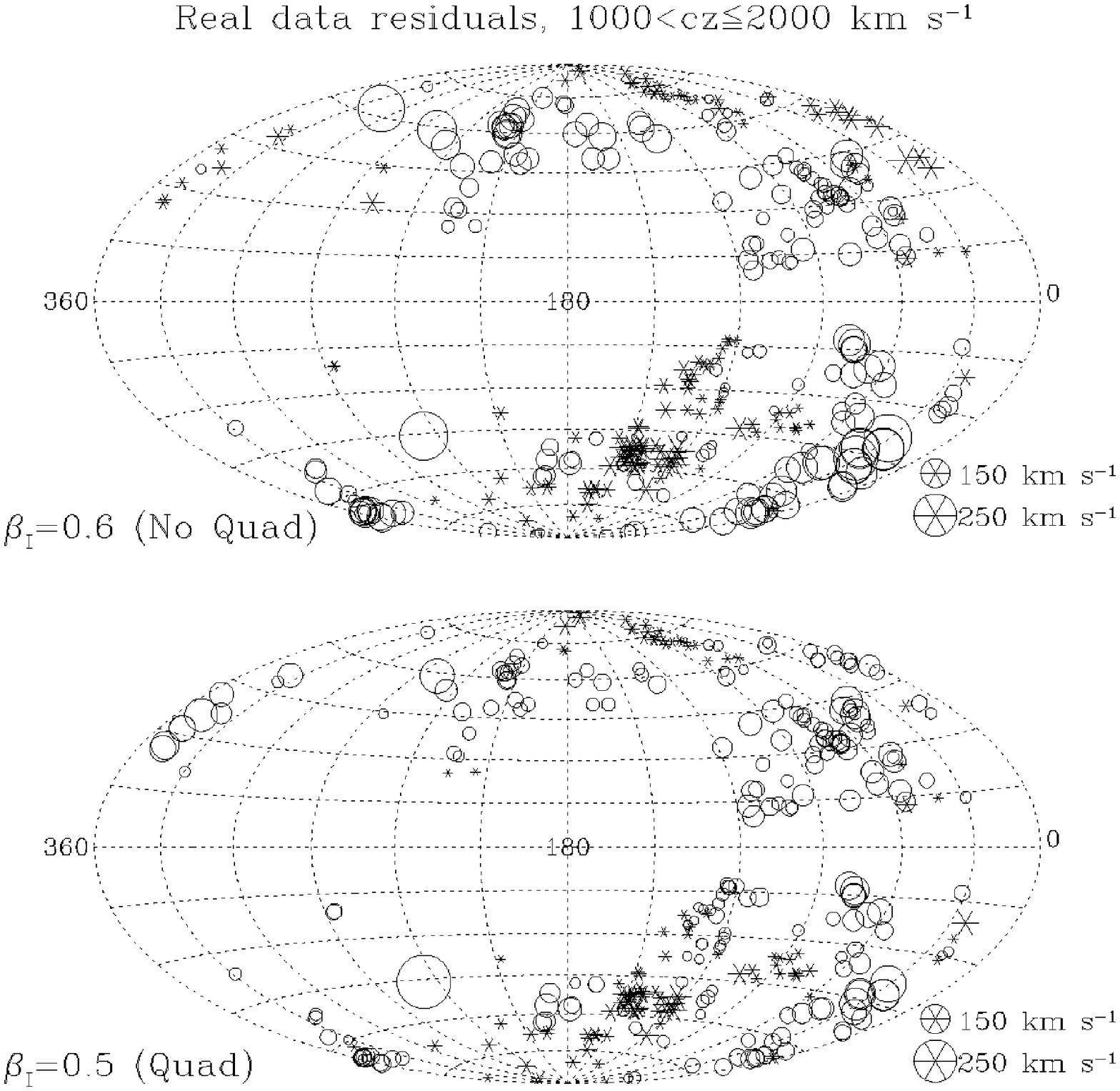}}
\caption{{\small Same as Figure~\ref{fig:resid01}, but
for objects with $1000 < \czlg \leq 2000\ \kms.$
}}
\label{fig:resid12}
\end{figure}
 
\begin{figure}[t!]
\centerline{\epsfxsize=4.5 in \epsfbox{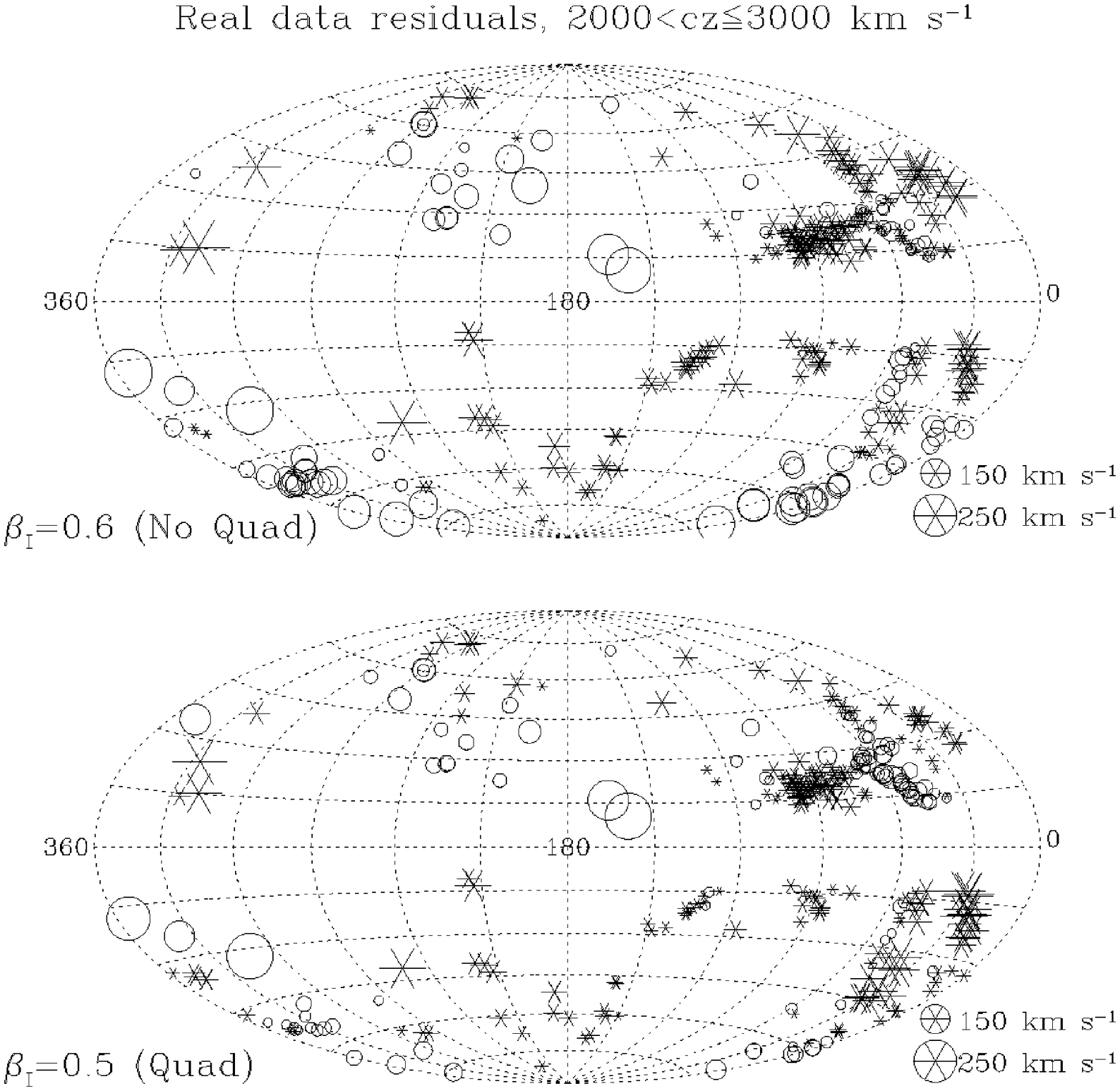}}
\caption{{\small Same as Figure~\ref{fig:resid01}, but
for objects with $2000 < \czlg \leq 3000\ \kms.$
}}
\label{fig:resid23}
\end{figure}
\subsection{Sky Maps of \velmod\ Residuals}

\velmod\ does not assign galaxies
a unique distance
(\S~\ref{sec:velmod-detail}).
Thus, there is no unique measure of the amount by
which their observed and predicted
peculiar velocities differ. 
However, 
there is a well-defined
{\em expected apparent magnitude\/} for each object,
\begin{equation}
E(m|\eta,cz) = \infint \! m\, P(m|\eta,cz)\,dm\,,
\label{eq:magresid}
\end{equation}
where $P(m|\eta,cz)$ is given by Eq.~(\ref{eq:pmgetaz}).
Similarly, the rms dispersion about this
expected value is
\begin{equation}
\Delta m = \sqrt{E(m^2|\eta,cz) - \left[E(m|\eta,cz)\right]^2}\,.
\label{eq:rmsmag}
\end{equation}
Note that $\Delta m$ is not equal to the TF scatter; it also includes
the combined effects of velocity noise, peculiar velocity
gradients, and density changes along the line of sight. 
At large distances, $\Delta m$ tends toward $\sigtf$
(although dispersion bias can make it smaller; cf.\ Willick 1994).
With the above definitions, one can define
a {\em normalized magnitude 
residual\/} for each galaxy
\begin{equation}
\delta_m = \frac{m - E(m|\eta,cz)}{\Delta m}\,.
\label{eq:normresid}
\end{equation}
The normalized residual has the virtue of having
unit variance for all objects.
In contrast, the variance of the unnormalized magnitude residual $m -
E(m|\eta, cz)$ depends on
distance
(velocity noise is more important for nearby objects),
while the variance of a peculiar velocity residual formed from
the magnitude residual (Eq.~\ref{eq:smooresid} below) grows
with distance. The normalized magnitude residual $\delta_m$ is
a measure of the correctness of the \iras\ velocity model.
If, in a given region, $\delta_m>0$ in the mean, galaxies
in that part of space must be more distant than \iras\
predicts them to be---i.e., they have negative radial
peculiar velocity relative to the \iras\ prediction.
Regions in which $\delta_m<0$ in the mean have positive
radial peculiar velocities relative to \iras.

\def\du{\delta u}
We will use the normalized magnitude residual below in our
quantitative analysis of residuals, but let us first visualize this
residual field on the sky, by converting $\delta_m$ into the
corresponding radial peculiar velocity residual $\du.$ Were
we to do this for each galaxy individually,
the \sm 20\% distance
errors due to TF scatter 
would completely hide systematic departures
from the \iras\ model. Instead, we will compute smoothed
velocity residuals. This procedure is most well-behaved
if we first smooth $\delta_m$ and then convert the
result into $\du.$ We first place
each galaxy at the distance $d$ assigned it
\footnote{We take
$d$ to be the ``crossing point distance'' $w$ defined
in \S~\ref{sec:velmod-discussion}.
In the case of triple-valued zones, we take
the central distance.}
by the \iras\
velocity model; this is
a redshift space distance so our calculation
is unaffected by Malmquist bias. Then for each
galaxy $i$, we compute a smoothed residual
$\delta_{m,i}^s$ as the weighted sum of the
residuals $\delta_m$ of itself and its neighbors $j$,
where the weights are $w_{ij}=\exp\left(-d_{ij}^2/2S_i^2\right),$
and $d_{ij}$ is the \iras-predicted distance between galaxies
$i$ and $j.$ We take the smoothing length $S_i$
to be $S_i=d_i/5.$
The smoothed residual $\delta_{m,i}^s$ 
is converted into a smoothed velocity
residual according to
\begin{equation}
\du_i^s = d_i \left[1-f_i 10^{0.2(\delta_{m,i}^s\times \Delta
m_i)}\right]\,,
\label{eq:smooresid}
\end{equation}
where $\Delta m_i$ is given by Eq.~(\ref{eq:rmsmag}).
The quantity $f_i$ is
given by $\exp\left(-\Delta_i^2/2\right),$
where $\Delta_i=0.46\Delta m_i/\sqrt{\sum_j w_{ij}};$
it guarantees that $\du_i^s,$
which is log-normally distributed, has
expectation value zero if 
$\Delta m_i$ (which is normally distributed) does
(cf.\ Willick 1991, \S~6.3, for details).

In Figures~10, 11, and~12
we plot \velmod\ velocity residuals on the sky for the
redshift ranges 0--1000 \kms, 1000--2000 \kms, and 2000--3000 \kms\
respectively. In each figure, the top panel shows residuals from the
$\beta_I=0.6$
(no quadrupole) fit, and the bottom panel shows
residuals from the $\beta_I=0.5$ (quadrupole modeled) fit,
the \velmod\ runs closest to the maximum likelihood value
of $\beta_I$ for each case. The plots reveal
why the addition of the quadrupole results in a large
increase of likelihood. In each redshift range, the no-quadrupole fits
show coherent negative velocity
residuals in both the Ursa Major region ($l\simeq 150\degs,$
$b\simeq 65\degs$), and at $b\simeq -60\degs,$
$l\simlt 30\degs$ and $l\simgt 330\degs.$ In both of
these regions, the addition of the quadrupole greatly
reduces the amplitude of the residuals. In other parts
of the sky, smaller but still significant coherent
residuals are reduced with the addition of the quadrupole.
This shows
that the pattern of departure from the pure \iras\ velocity field
is well-modeled by a quadrupolar flow of modest
amplitude, and therefore
has the simple physical interpretation we discussed in
\S~\ref{sec:quadrupole}.

In the
bottom panels, it is difficult to find any
well-sampled region within $2000\ \kms$
where $|\du| \simgt 100\ \kms.$ This is all the more
remarkable because the TF errors themselves are of
order 300\ \kms\ per galaxy at a distance of 1500 \kms.
Figure~12 does show several high-amplitude
residuals. However, at
2500 \kms, the TF residual for a single object is 500 \kms,
so when the effective number of galaxies per smoothing length is only
a few, velocity residuals
of several hundred \kms\ are expected from TF scatter
only. In well-sampled regions, one sees that in general
$|\du|\simlt 150\ \kms,$
the only exception being a patch of large ($\simgt 250\
\kms$) positive
residuals at $l\simeq 330\degs,$ $b\simeq -20^\circ.$ In the
$b>0\degs$ part of the Great Attractor region at $l\simeq 300\degs,$
the residuals are $<100\ \kms$ even in this highest redshift shell.
This is significant, given the oft-heard claims that the \iras\ model
cannot fit the observed flow into the Great Attractor. 

\begin{figure}[t!]
\centerline{\epsfxsize=4.5 in \epsfbox{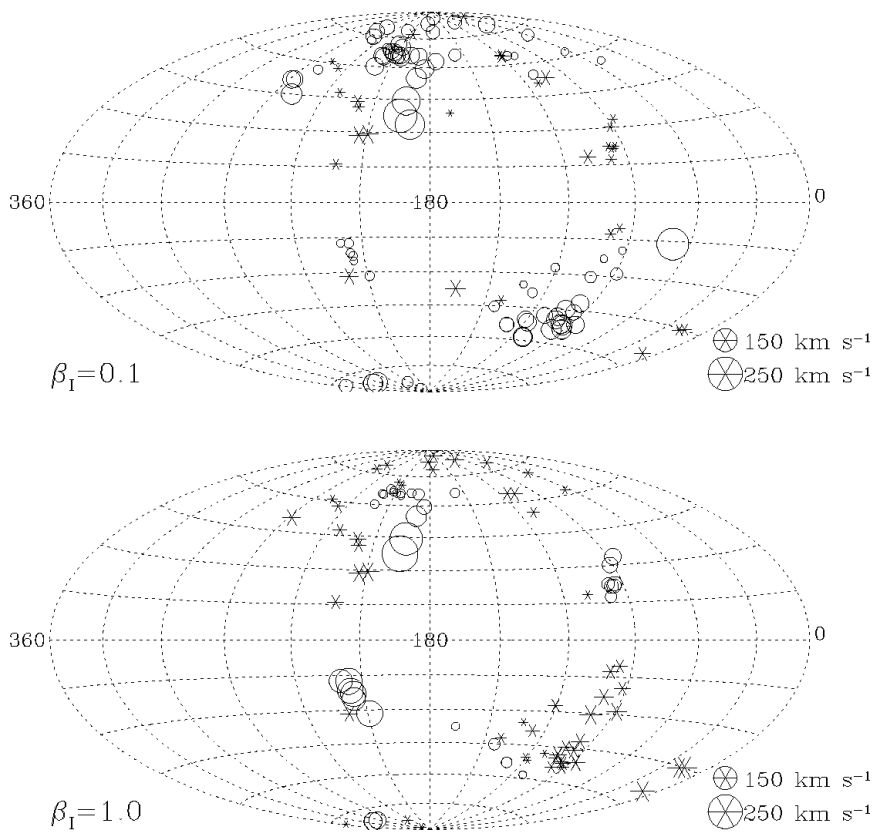}}
\caption{{\small Same as Figure~\ref{fig:resid01}, except
now results for $\beta_I=0.1$ and $\beta_I=1.0$ are shown.
In each case, the quadrupole is the same as it was for
the best fit model ($\beta_I=0.5$).
}}
\label{fig:resid101_01}
\end{figure}
 
\begin{figure}[t!]
\centerline{\epsfxsize=4.5 in \epsfbox{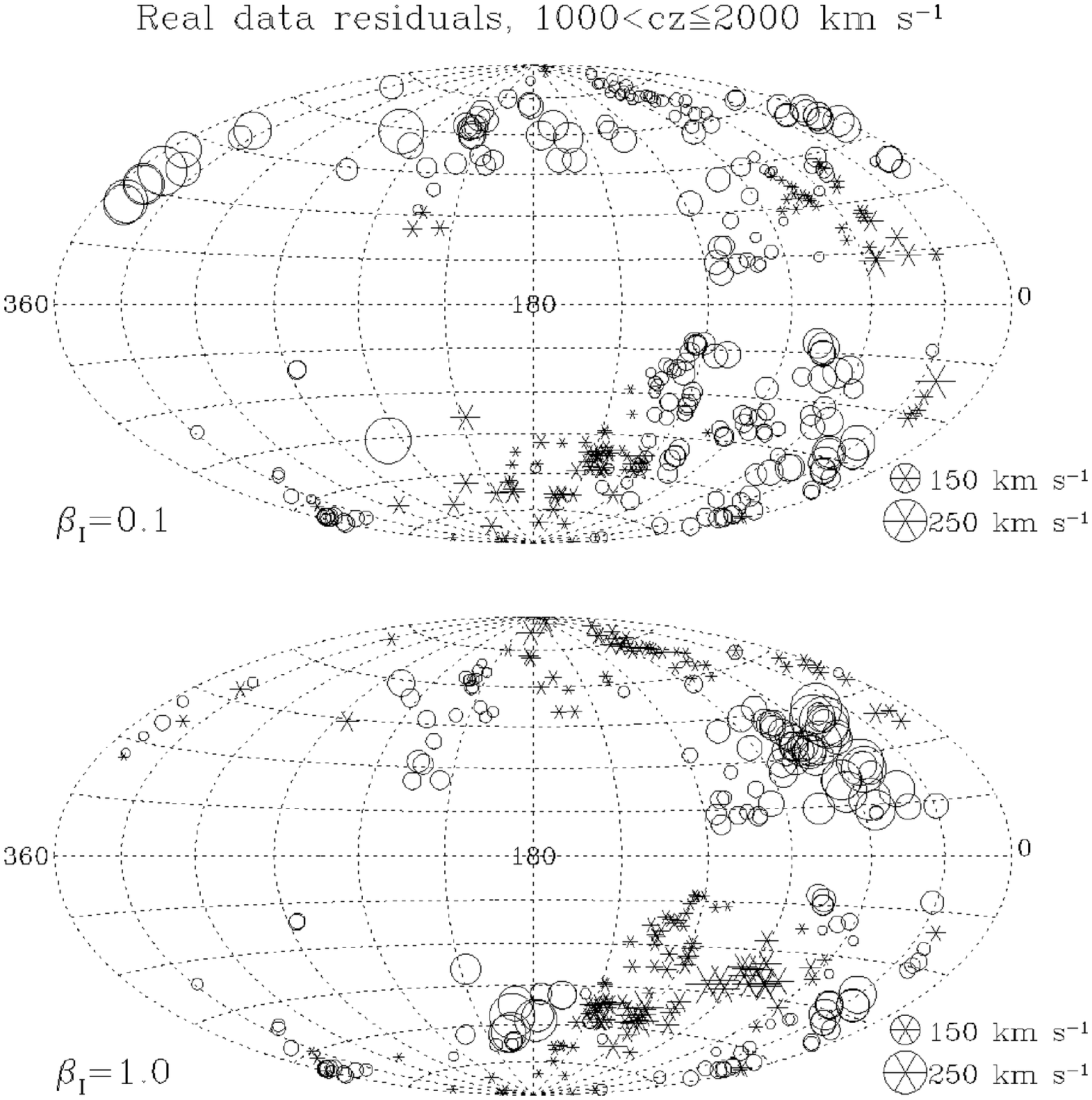}}
\caption{{\small Same as the previous figure, but for
objects with $1000 < cz_{{\rm LG}} \leq 2000\ \kms.$
}}
\label{fig:resid101_12}
\end{figure}
 
\begin{figure}[t!]
\centerline{\epsfxsize=4.5 in \epsfbox{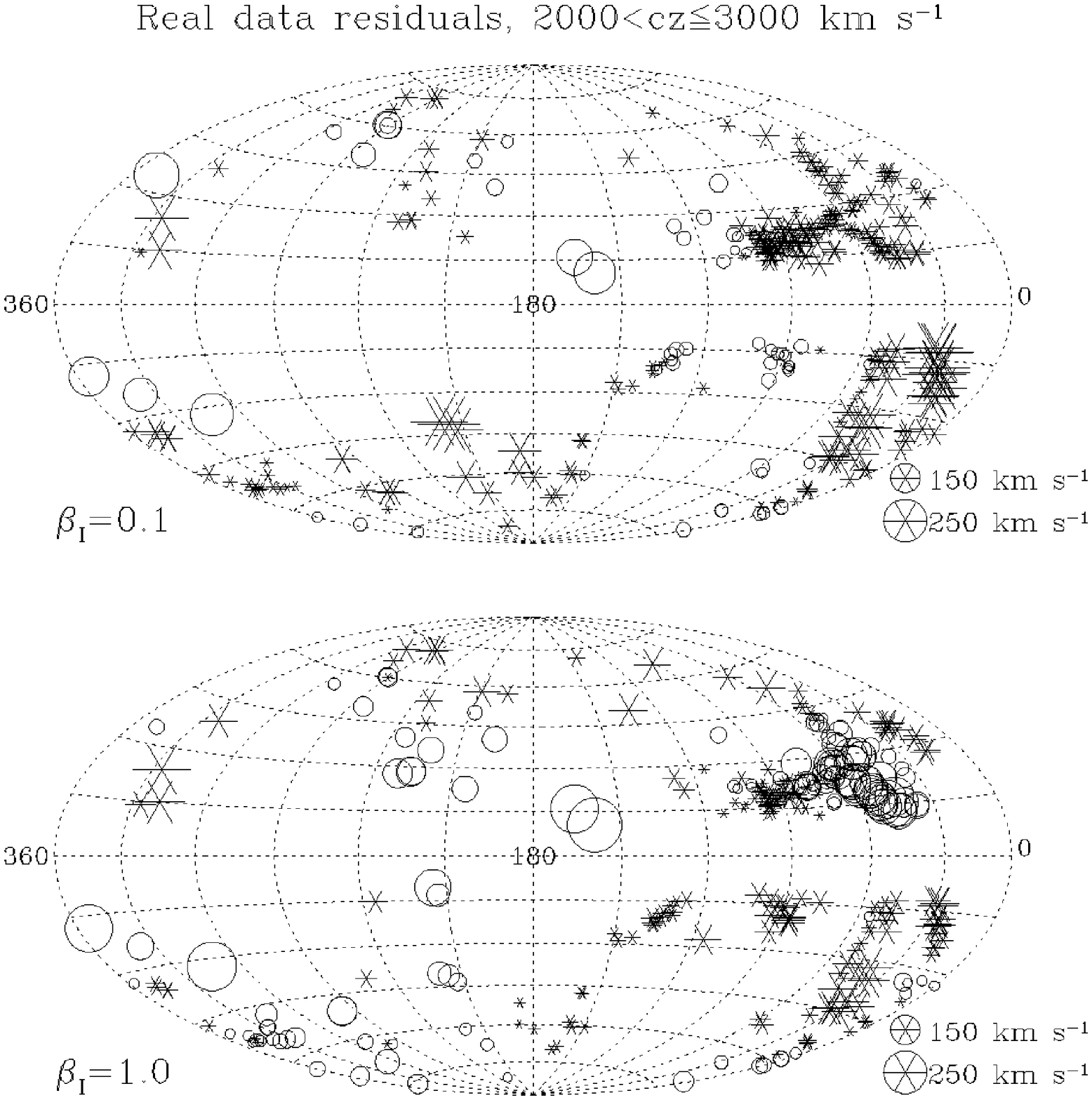}}
\caption{{\small Same as the previous figure, but for
objects with $2000 < cz_{{\rm LG}} \leq 3000\ \kms.$
}}
\label{fig:resid101_23}
\end{figure}
In Figures~13, 14,
and~15 we again plot \velmod\ residuals
on the sky for the three redshift ranges, 
now for the two values of $\beta_I$
most strongly disfavored by the likelihood statistic
in the range studied, 
$\beta_I=0.1$ (top panels) and $\beta_I=1.0$ (bottom panels).
In each plot, the quadrupole of Figure~4
has been included.
These plots, which should be compared with the
bottom panels of Figures~10,
11, and~12,
demonstrate why very low and high $\beta_I$
do not fit the TF data well. In each redshift range,
these models exhibit large, coherent residuals. For $\beta_I=0.1,$
we see large negative peculiar
velocities relative to \iras\ in the Ursa Major
region at $cz\leq 2000\ \kms.$ 
Indeed, the residual plot for $\beta_I = 0.1$ (with quadrupole
included) shows many of the same features as the no-quadrupole model
with $\beta_I = 0.6$, 
because
the \iras\ field itself contributes some of the needed
quadrupole. However, the \iras\ contribution 
scales with $\beta_I,$ and
is thus inadequate at low $\beta_I.$  
At $\beta_I=1.0$ many of the systematic residuals associated
with the quadrupole are gone, especially in Ursa Major.
However, other regions show highly significant residuals: at $l\simeq
150\degs,$
$b\simeq -20\degs$ and $cz\leq 1000\ \kms,$ for example, one
sees negative peculiar velocity residuals of amplitude
$\simgt 200\ \kms,$ which is significant at such small distances.
In the same redshift range, at $l=270$--$360\degs,$
$b< 0\degs$ there are positive velocity residuals
of amplitude $\simgt 150\ \kms.$ These regions
exhibit much smaller residuals in the $\beta_I=0.5$ model.

In the higher redshift shells, the poor fit of the
$\beta_I=1.0$ model is evidenced chiefly in the
direction of the Great Attractor ($l\simeq 300\degs,$ $b\simeq 20\degs$).
For $1000 < cz \leq 2000\ \kms,$
this model predicts much too large positive
peculiar velocities, so that the data exhibit inflow
relative to the model. In the highest redshift bin,
the $\beta_I=1.0$ model exhibits both positive and
negative velocity residuals of high amplitude in the
GA direction; residuals of both signs are seen in this
region for $\beta_I=0.5$ as well, but they are of
much smaller amplitude (lower panel of
Figure~12).
The $\beta_I=0.1$ model, on the
other hand, predicts too-small positive peculiar velocities
in the GA direction at the highest redshifts. Indeed, note
that in the $2000 < cz_{LG} \leq 3000\ \kms$ shell, nearly all data
points exhibit outflow relative to the $\beta_I=0.1$
\iras\ predictions, whereas at lower velocities the
residuals typically indicate inflow. This global mismatch
is more general than the insufficient quadrupole mentioned
in the previous paragraph, showing that low $\beta_I$
could not yield a good fit even if we were to give \velmod\
full freedom in fitting the quadrupole at all $\beta_I.$

\begin{figure}[t!]
\centerline{\epsfxsize=5.5 in \epsfbox{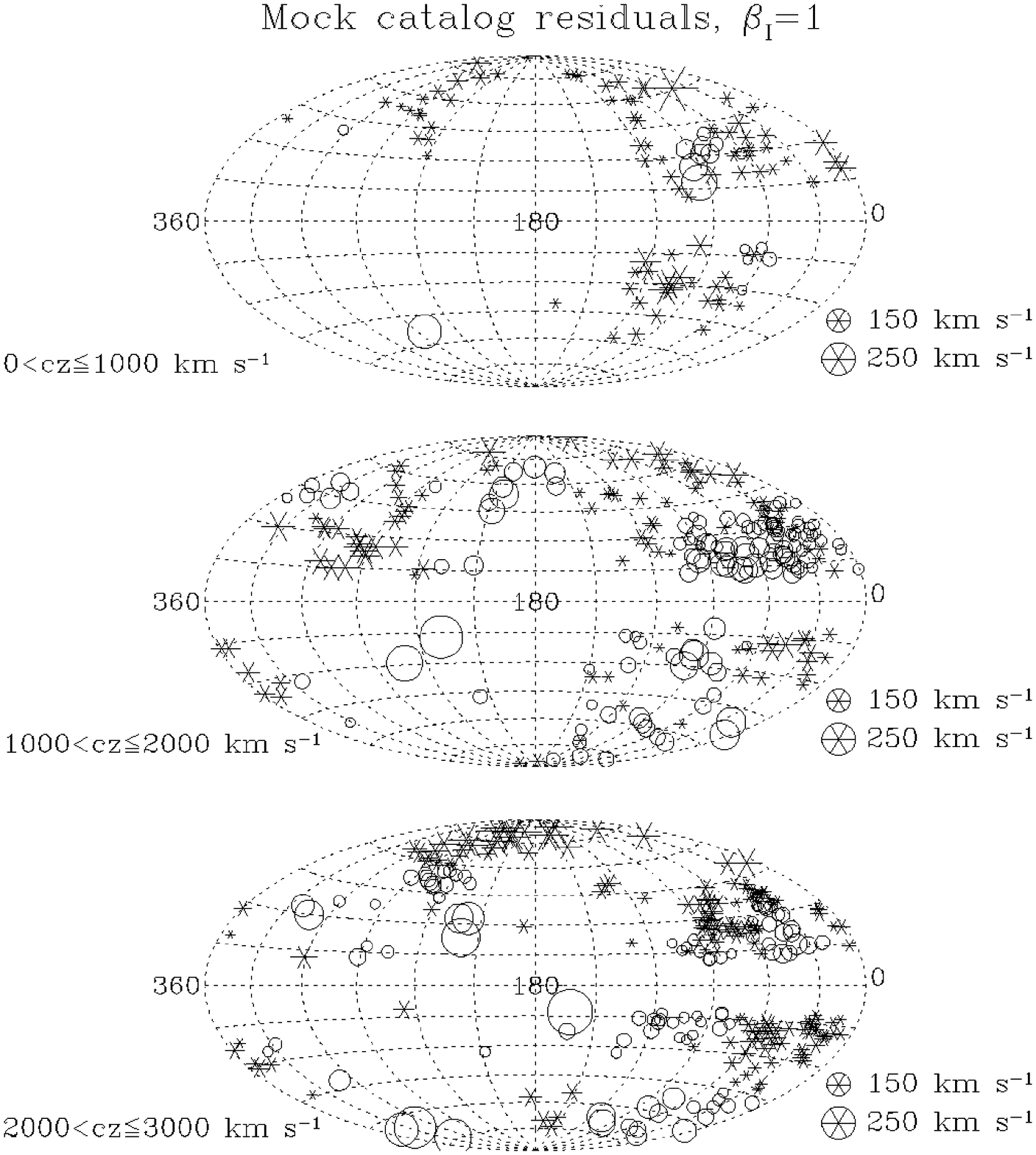}}
\caption{{\small \velmod\ velocity residuals for a
single mock catalog run using $\beta_I=1.0,$ the
true value for the mock catalog. (The particular
simulation used had a maximum likelihood value
of $\beta_I=0.963.$) The three panels show
residuals for the three redshift ranges used
in analyzing the real data.
}}
\label{fig:mockresid}
\end{figure}
Although sky plots of residuals
argue in favor of the $\beta_I=0.5$ plus quadrupole
model, the residuals from that model are not
manifestly negligible.
We will address this issue
quantitatively below. For now, however, we can demonstrate qualitatively
that the residuals seen in the $\beta_I=0.5$ plus quadrupole
model are not unexpected by comparing with the
mock catalogs, for which the \iras\ velocity predictions
are known to be a good fit. Figure~16 plots
\velmod\ velocity residuals with respect to
$\beta_I=1$ (the correct value) for a single
mock catalog. The same three redshift ranges used
for the real data are shown.
The mock catalog residuals are comparable in
amplitude and apparent coherence to the real data.
Generally speaking, velocity residuals in well-sampled
regions are $\simlt 100\ \kms$
within 1000 \kms, and are $\simlt 200\ \kms$ at
larger distances.
One also sees
apparent coherence in the mock catalog residual map,
as was the case with the real data.
The similar amount of apparent coherence in the real and mock
data indicates that the former is not a result of a poor fit.
The apparent coherence in the residual sky maps
is an artifact of the smoothing used to generate them,
as we show in the next section.

\subsection{Residual Autocorrelation Function}
\label{sec:residcorr}

\begin{figure}[t!]
\centerline{\epsfxsize=4.5 in \epsfbox{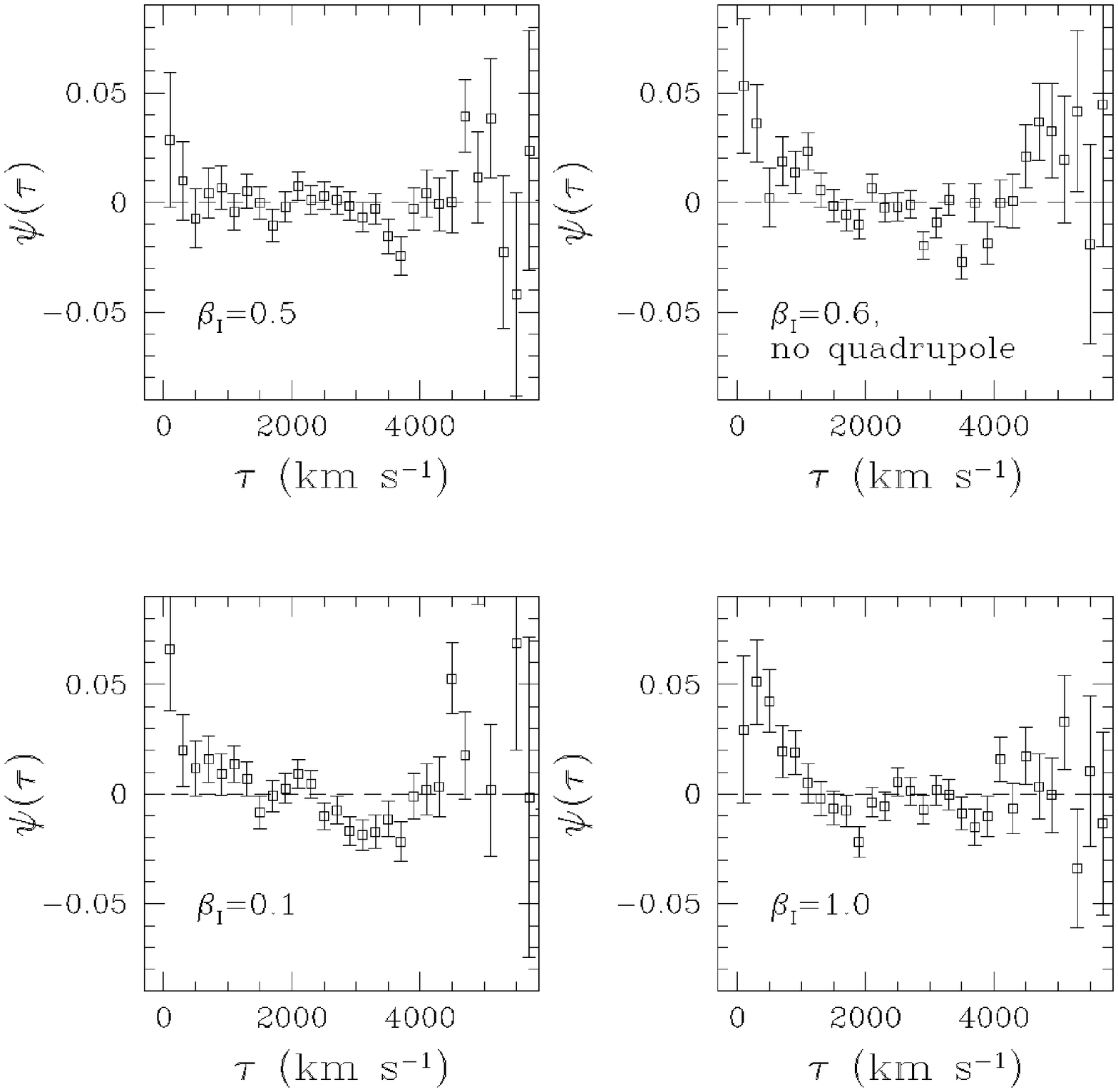}}
\caption{{\small \velmod\ residual autocorrelation
functions, $\psi(\tau),$ plotted for $\beta_I=0.5$
plus quadrupole (best fit model), $\beta_I=0.6,$
no quadrupole,
$\beta_I=0.1$ plus quadrupole, and $\beta_I=1.0$ plus quadrupole.
In the $\beta_I=0.1$ several points at large $\tau$
have residuals that are so large that they fall beyond the plot
boundaries.
%too-large positive values to seen within the chosen limits.
}}
\label{fig:autocorr1}
\end{figure}
The sky plots shown above provide visual evidence that
the $\beta_I=0.5$ plus quadrupole fit has generally
small residuals, although they are correlated to some degree. 
In this section, we quantify these correlations with the 
{\em residual autocorrelation function\/}
\begin{equation}
\psi(\tau) \equiv \frac{1}{N_p(\tau)} \sum \begin{Sb}
i < j \\
d_{ij} = \tau \pm \Delta\tau 
\end{Sb} \delta_{m,i}\delta_{m,j}\,,
\label{eq:psi-def}
\end{equation}
where $\delta_m$ was defined in Eq.~(\ref{eq:normresid}), and the sum
is over the $N_p(\tau)$ distinct pairs with \iras\ predicted separation
$d_{ij}$ within $\Delta\tau
= 100\ \kms$ of a given value $\tau$.
This definition
makes $\psi(\tau)$ insensitive
to the values of $\sigtf$ and $\sigma_v$ (because the
$\delta_{m,i}$ are themselves normalized using their
maximum likelihood values for each $\beta_I$),
but sensitive to the residual correlations that signal a poor fit.

In Figure~17, we plot $\psi(\tau)$ versus
$\tau$ for the \iras\ plus quadrupole models, with $\beta_I=0.5,$
$0.1,$ and $1.0,$ as well as the $\beta_I=0.6,$ no quadrupole
model. The error bars are described below.
The model that fits best
according to the likelihood statistic,
$\beta_I=0.5$ plus quadrupole, shows no
significant residual correlations {\em on any scale.}
The correlation function
is everywhere consistent
with zero, as we would
expect if the \iras\
velocity field plus the quadrupole is indeed a good fit to the data.
Indeed, the absence of residual correlations is the
basis for a statement made in \S~\ref{sec:velmod-detail}, 
namely, that the individual galaxy probabilities $P(m|\eta,cz)$
are independent, and thus validates the
\velmod\ likelihood statistic $\likeforw.$

The other models shown in Figure~17
all exhibit significant residual correlations. The
$\beta_I=0.6,$ no quadrupole model has noticeable
correlations on small and large scales, as does the
$\beta_I=0.1$ plus quadrupole model. Indeed, several
of the values of $\psi(\tau)$ for $\beta_I=0.1$ are
so large that they are off-scale on the plot.
The $\beta_I=1.0$ plus quadrupole model
exhibits strong correlations for $\tau\simlt 2000\ \kms$,
although it is well-behaved
on large scales.

\subsubsection{Using Residual Correlations to Identify Poor Fits
Quantitatively}
\label{sec:chistat}

In order to compare the observed residual correlations with the
results from the mock catalogs, we would like to define a single
statistic that summarizes the deviation of $\psi(\tau)$ from unity.
Let us define $\xi(\tau) \equiv N_p(\tau)\psi(\tau)$
(cf. Eq.~\ref{eq:psi-def}).
In Appendix C, we discuss the properties of this statistic in greater
detail. As we show there, $\xi(\tau)$ approximates
a Gaussian random variable of mean zero and
variance $N_p(\tau),$
if indeed the \velmod\ residuals are uncorrelated on scale $\tau.$
(This property was used to compute the error bars on $\psi(\tau)$ above.)
To the degree this approximation is a good one,
the quantity
\begin{equation}
\chi^2_\xi \equiv \sum_{k=1}^M \frac{\xi^2(\tau_k)}{N_p(\tau_k)}
\label{eq:chi2xi}
\end{equation}
will be distributed approximately as
a $\chi^2$ variable with $M$ degrees of freedom,
where $M$ is the number of separate bins in which $\xi(\tau)$ is calculated.
In contrast,
if the residuals are strongly correlated on any
scale $\tau,$ 
$\chi^2_\xi$
will significantly exceed its expected value.

However, because a single galaxy will appear in many different pairs in the
correlation statistic, both
within and between bins in $\tau$, the assumptions made above
do not hold rigorously.  In Appendix C we explore
this issue further. 
For now, we appeal to the mock catalogs to
assess how closely the quantity $\chi^2_\xi$ follows $\chi^2$ statistics. 
We computed it for
each of the 20 mock catalog
runs (\S~\ref{sec:mock}) with $\beta_I=1.$ 
We carried out the calculation 
to a maximum separation of 6400 \kms, in bins
of width $200\ \kms,$ so that $M=32,$
and found a mean value $\langle \chi^2_\xi\rangle = 
27.83\pm 1.82,$ 
which may be compared with an expected value
of $32$ for a true $\chi^2$ statistic. 
The rms scatter
in $\chi^2_\xi$ was $8.15,$ which is the same
as that expected for a true $\chi^2.$ 
The difference between the mean and expected values is
$2.3\,\sigma$, indicating that $\chi^2_\xi$ is
not exactly a $\chi^2$ statistic, for reasons
discussed in Appendix~C. 
However, because the
departure from true $\chi^2$ statstics is small, 
$\chi^2_\xi$ is a useful statistic for
measuring goodness of fit when calibrated against the mock catalogs. 

\begin{figure}[t!]
\centerline{\epsfxsize=4.5 in \epsfbox{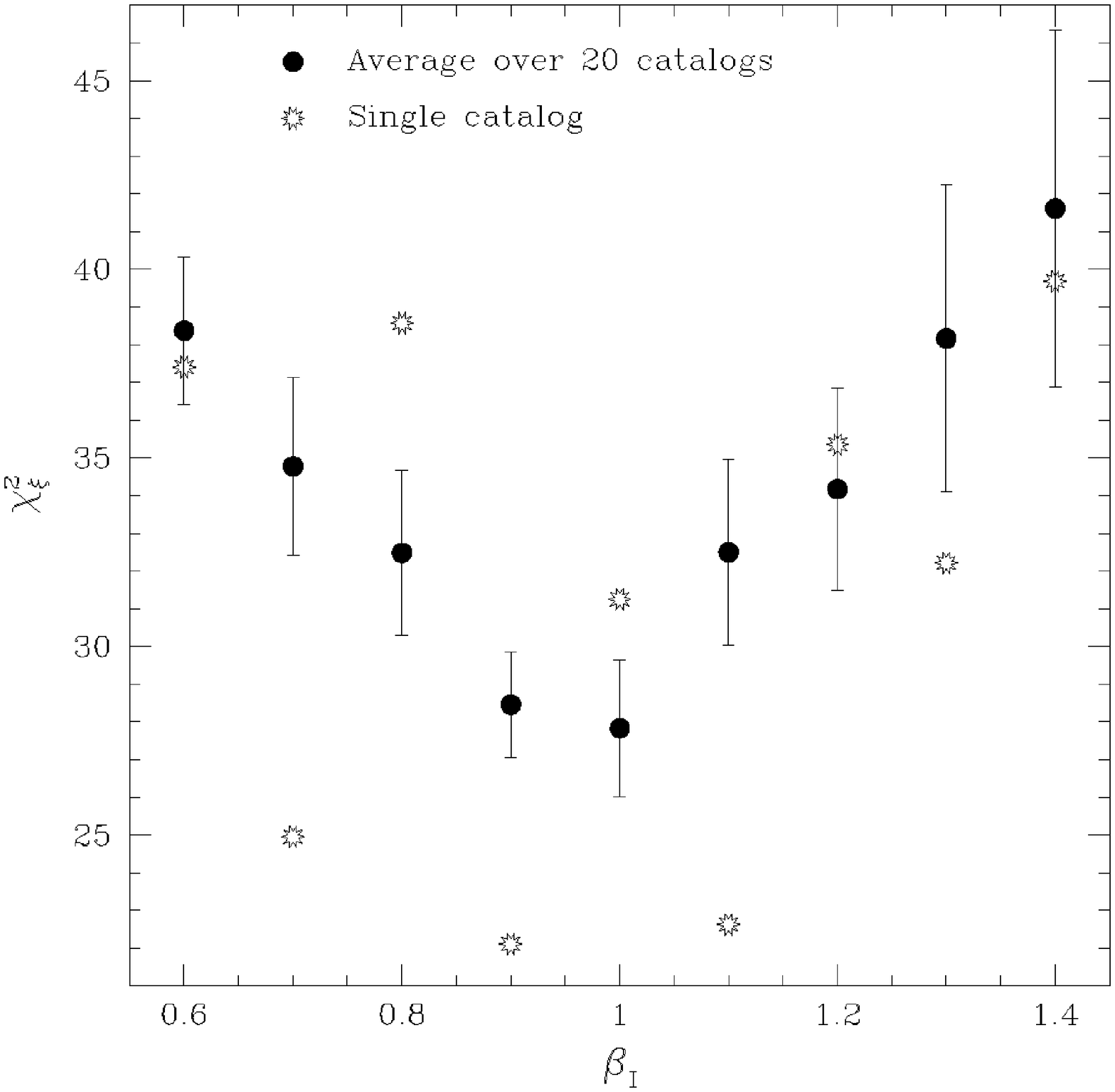}}
\caption{{\small
The residual correlation statistic
$\chi^2_\xi,$ defined by Eq.~\ref{eq:chi2xi}, plotted
as a function of $\beta_I$ for the mock catalogs.
The solid symbols show an average over 20 mock
catalogs; the open symbols show the values
obtained for a single mock catalog.
}}
\label{fig:mockcorrel}
\end{figure}
Before presenting $\chi^2_\xi$ for the
real data, we consider its variation
with $\beta_I$ for the mock catalogs.
In Figure~18
we plot the average value of $\chi^2_\xi$
over the 20 mock catalogs at each value 
of $\beta_I$ for which \velmod\ was run.
Although the minimum is at $\beta_I = 1$, it is not nearly as sharp as
is that of the likelihood as function of $\beta_I$ (e.g.,
Fig.~2); this statistic does not have the power
that the likelihood does for measuring $\beta_I$. 
Indeed, for a single realization (the open symbols),
the statistic has several local minima. 
However, it is apparent that a $\chi^2_\xi$ value
much greater than its expected true value of \sm 28
will indicate a poor fit of the model to the data.

\begin{figure}[t!]
\centerline{\epsfxsize=4.5 in \epsfbox{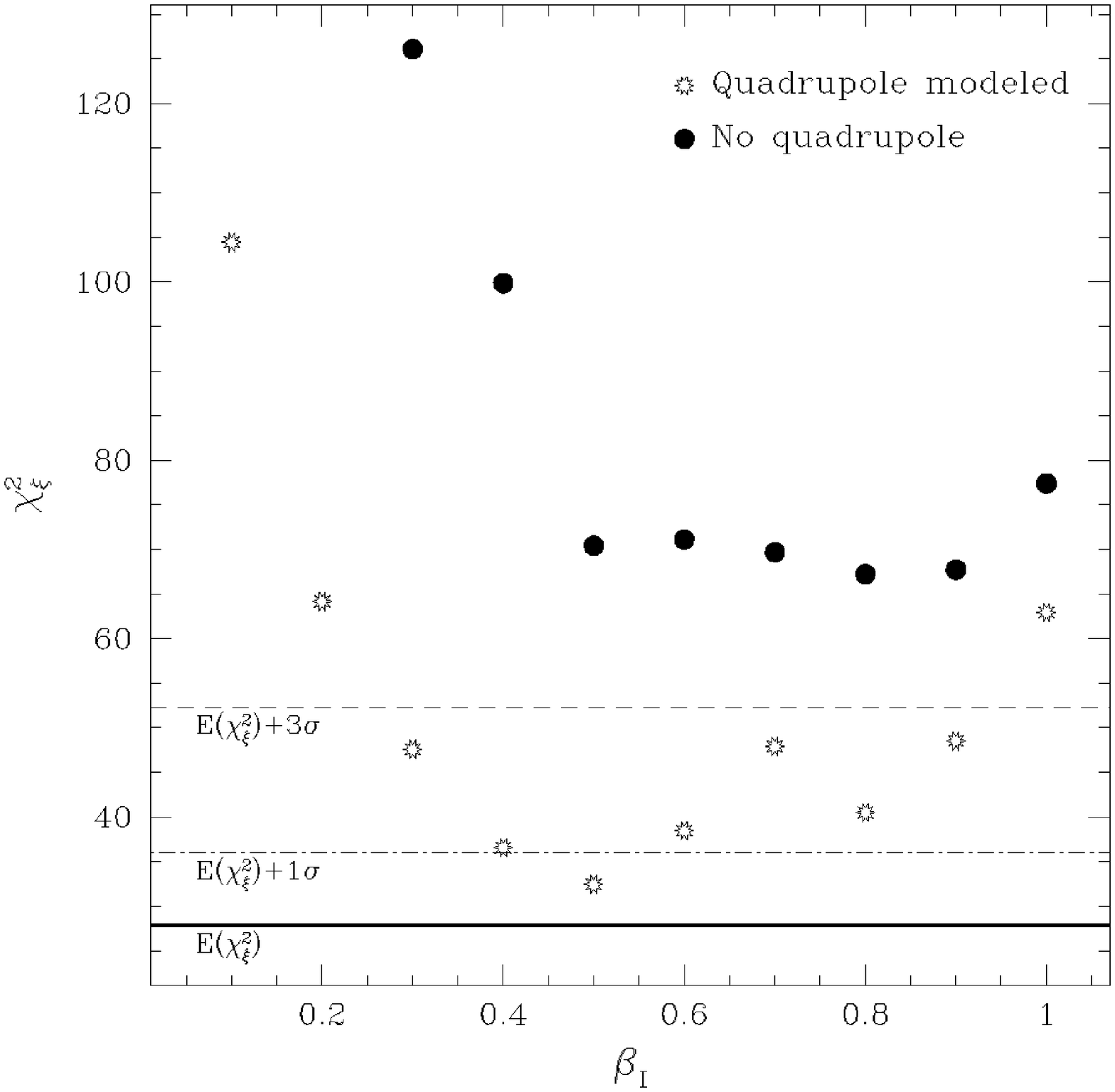}}
\caption{{\small The residual autocorrelation statistic
$\chi^2_\xi,$ defined by Eq.~\ref{eq:chi2xi}, plotted
as a function of $\beta_I$ for the real data, with
and without the
quadrupole modeled. The heavy solid line shows the
expected value of the statistic, which was determined
by averaging the derived value for 20 mock catalogs.
The two dashed lines show 1- and 3-$\sigma$
deviations from this value. Note that when
the quadrupole is not modeled, highly significant
residual correlations are detected for all values of $\beta_I.$
(The no-quadrupole points for $\beta_I=0.1$ and $0.2$ are
not shown because their $\chi^2_{\xi}$ values are
too large.)
}}
\label{fig:chi2_xi}
\end{figure}

In Figure~19, we plot the statistic $\chi^2_\xi$
as a function of $\beta_I$ for the real data, with
and without the quadrupole included. The horizontal lines indicate the
expected value of $\chi_\xi^2$, and the $1\,\sigma$ and $3\,\sigma$
deviations from it.  Note first that {\em the no-quadrupole model does
not provide 
an acceptable fit for any value of $\beta_I.$} This is
not a conclusion we could have reached on the basis of
the likelihood analysis alone. When the quadrupole is
included, the only values of $\beta_I$ that are
unambiguously ruled out are $\beta_I=0.1,$ $0.2,$ and $1.0.$
The best fit model according to \velmod,
$\beta_I=0.5$ plus quadrupole, also has the
smallest value of $\chi^2_\xi.$  Given the multiple minima seen for
one mock realization in Figure~18, this is not
necessarily deeply significant. 
The statistic $\chi^2_\xi$ is suitable
for identifying models that do {\em not\/} fit the data,
but does not have the power of the likelihood statistic for
discriminating among those models that do fit. 

In summary:
the \velmod\ likelihood maximization procedure is
the proper one for determining which value of
$\beta_I$ is better than others, but it cannot
identify poor fits to our model. The residual
correlation statistic $\chi^2_\xi$ can identify
unacceptable fits, but does not have the power to determine 
which of the acceptable fits is best.
We have found that the \iras\ velocity field with $\beta_I=0.5,$
plus the external quadrupole, is both the best fit of those
considered, and is also an acceptable fit. 
Values of $\beta_I > 0.9$ and $\beta_I<0.3$ are strongly ruled out.

\section{Discussion}
\label{sec:discussion}
\subsection{What is the value of $\beta_I$?}
\label{sec:beta-discussion}
\velmod\ recovers the correct answer, $\beta_I=1,$ to $< 10\%$
accuracy when applied to the mock catalogs. At $\beta_I=1,$ the velocity
field in the mock Virgo region is significantly triple-valued. Thus
\velmod,
despite being close in spirit to Method II,
properly treats triple-valuedness.
If the strong triple-valuedness one sees at $\beta_I =1$ were
present in the real universe,
\velmod\ would not assign it unduly small likelihood.
Nonetheless, when \velmod\ is applied to the real universe, it returns
a value of $\beta_I=0.492\pm 0.068$ (quadrupole modeled).
This value is quite insensitive to two other
quantities treated as free parameters in the velocity field model, the
Local Group random velocity $\bfwlg$ and the small-scale velocity
dispersion $\sigma_v$ (\S~\ref{sec:results}). 
Tests
with the mock catalogs demonstrated that we obtain
an unbiased $\beta_I$ using a 300 \kms-smoothed \iras\
reconstruction (\S~\ref{sec:beta-accuracy}).
However, we found that changing
to a 500 \kms-smoothed reconstruction makes relatively
little difference in $\beta_I$ (\S~\ref{sec:smoo500}).
Finally, neglecting the quadrupole causes $\beta_I$ to change
by only \sm $1\,\sigma$. 
Our conclusion that $\beta_I\simeq 0.5\pm 0.07$ is
thus robust against systematic effects internal to our
method.

The \velmod\ result is consistent with the relatively low
estimates of $\beta_I$ obtained
from the Method II analyses of
Hudson (1994),
Roth (1994),
Shaya \etal\ (1995)\footnote{The Hudson and
Shaya \etal\ papers actually derive $\beta_{{\rm opt}},$
which must be multiplied by \sm 1.3 to
obtain an equivalent $\beta_I;$ cf.\ footnote 2.},
DNW,
and Schlegel (1996), as well as those derived from comparisons of
the \iras\ density field with the motion of the Local Group (Strauss
\etal\ 1992b) and from  
some analyses of the redshift-space anisotropy
of the \iras\ density field (e.g., Hamilton 1993, 1996; Fisher \etal\
1994; Cole, Fisher, \& Weinberg 1995; Fisher \& Nusser 1996).
However, it is apparently inconsistent 
with estimates of $\beta_I$ near unity,
as have been found by the \potiras\ analysis
(Sigad \etal\ 1997),
measurements of the \potent\
fluctuation amplitude (Kolatt \& Dekel 1997, Zaroubi \etal\ 1997), and
redshift-space distortions of spherical harmonic expansions of the
density field (Fisher, Scharf, \& Lahav 1994c; Fisher 1994; Heavens \&
Taylor 1995). 

\subsubsection{Why do \velmod\ and \potiras\ yield different values
of $\beta_I$?}
\label{sec:velmod-potiras}

We do not yet have a satisfactory
explanation of why \velmod\ and
standard Method II analyses characteristically yield smaller values
of $\beta_I$ 
than the Method I 
\potiras\ approach. 
One possibility is that the differences stem from
the Method I/Method II distinction.
However, \velmod\
corrects the principal drawback of
Method II,
the inability to deal with multivalued
or flat zones in the redshift-distance relation. 
Thus, if the Method I/Method II
distinction is at the root of the discrepancy, the reason must be more
subtle than the drawbacks of standard Method II. 
Sigad
\etal\ (1997) test for biases in \potiras\ using the {\em same\/}
mock catalogs as this paper; they too find
their determination of $\beta_I$ to be essentially unbiased.
The problem
could lie with the Malmquist bias corrections that are so crucial to
Method I (cf.\ the discussion in Willick \etal\ 1997).
If these corrections are underestimated for any reason---e.g., the TF scatter
is larger than estimated, 
or the
density fluctuations are larger
than modeled---a Method I approach will produce too-strong
velocity gradients and thus
overestimate $\beta_I.$ However, the TF scatters used by
Sigad \etal\ (1997) are consistent with those obtained in this
paper, and the large \potent\ smoothing limits the
effect of Malmquist bias in any case. It is thus unlikely
that improper Malmquist bias corrections strongly affect
the value of $\beta_I$ obtained from \potiras.

An important difference between \velmod\ and \potent\ 
is the Gaussian smoothing scales
employed, 
300 and 1200 \kms\ respectively.
These very different smoothings could result in different values of $\beta_I$
if the effective bias parameters on these scales are different. 
In order to reconcile \velmod\ and \potiras, we would need the effective bias
parameter to decrease by a factor of $1.7$ 
between scales of 300 and 1200 \kms.
Such a scale-dependent biasing has been suggested by the galaxy formation 
models of Kauffman \etal\ (1996), but Weinberg (1995) and Jenkins
\etal\ (1996) do not find these trends.
A recent
analysis by Nusser \& Dekel (1997) using a spherical harmonic
expansion of the velocity field finds $\beta_I = 1.0$ for 1200 \kms\
smoothing, but only 0.6 for 600 \kms\ smoothing, approaching the
value we have found in this paper. 
Such a change of $\beta_I$ with smoothing scale
could signal scale-dependent biasing.

Still another difference is the volume considered.
We have restricted
this analysis to $cz\leq 3000\ \kms$
(\S~\ref{sec:real}), whereas the analysis of Sigad \etal\
(1997) extends to 6000 \kms; only $\sim 1/3$ of the points
used fall within 3000 \kms.  If,
for whatever reason, $b_I$ differed locally
from its global value, the \velmod\ result could be biased low.  
In a future paper we will
extend the \velmod\ analysis to larger distances; 
however, our preliminary results 
do not show an increase
in $\beta_I$ when we do so. In addition to probing
a larger volume, 
the Sigad \etal\ analysis uses the full Mark III
sample, ellipticals included; the possibility of systematic
differences between the TF subset we have used in this paper, and the
full sample, is difficult to rule out. 
Finally, it is conceivable that the requirement of pre-calibrating TF
relations (\potent), as opposed to calibrating them simultaneously
with fitting the velocity field (\velmod\ and Method II generally)
accounts for part of the discrepancy. However, 
fixing the \velmod\ TF parameters at their
Mark III values has essentially no effect on the
derived value of $\beta_I$ (\S~\ref{sec:consistency}).
This strongly argues against the notion that
a major difference between \velmod\ and \potiras\ is the TF relations
themselves.

\subsubsection{The effect of cosmic scatter}
The sphere out to 3000 \kms\ is small; the rms value of density
fluctuations within spheres of this radius is 20\% for COBE-normalized
CDM.  However, this does not propagate to a cosmic scatter error on
our derived $\beta_I$, for two reasons. First, the \iras\ velocity
field is determined within a sphere of radius 12,800 \kms, within
which the rms fluctuations are only a few percent.  Thus the peculiar
velocity field is subject to very little cosmic scatter.  Second,
this scatter primarily manifests itself as a monopole term (cf.\ the
discussion in \S~\ref{sec:quadrupole}), and therefore is 
fully absorbed into
the zero-points of the TF relations (\S~\ref{sec:tfmock}), 
having {\em no\/} effect on the
derived value of $\beta_I$.

\subsection{Do the \iras\ and TF Velocity Fields Agree?}

An important conclusion of this paper is that the agreement between the
predicted and observed peculiar velocity fields is satisfactory
(\S~\ref{sec:resid}),
as it must be if
the resulting estimate of $\beta_I$ is to be believed.
This agreement 
is consistent with the hypothesis 
that gravitational instability theory
correctly describes the relationship between the peculiar velocity
and mass density fields.
It also suggests that the linear biasing model,
Eq.~(\ref{eq:linear-bias}), is a reasonable description
of the relative distribution of \iras\ galaxies and 
all gravitating matter
Gaussian smoothed at $300\ \kms.$

\subsubsection{Comparison with Davis, Nusser, \& Willick (1996)}
DNW
reached a different conclusion.
Comparing the \iras\ and TF velocity fields with a Method II approach,
(\itf; cf.\ \S~\ref{sec:method_intro})
DNW found that the fields do not agree at a statistically
acceptable level. In particular, a $\chi^2$ statistic resulting from a
mode-by-mode comparison of the \iras\ and \itf\ velocity fields
was found to be 100 for 55 degrees of freedom.\footnote{Note that
unlike this paper, DNW {\em assumed\/} a TF scatter \apriori, which
allows them to define a goodness of fit directly from their $\chi^2$;
cf.\ the discussion in \S~\ref{sec:velmod-discussion}.} 
DNW argued that the excessive value of their $\chi^2$ statistic
resulted primarily from a dipole in the TF velocity
field that grows with scale, a feature not seen in the \iras\ predictions.
They cautioned that, as a result, their maximum likelihood
value of $\beta_I \sim 0.5$ was not
necessarily meaningful.

Why do we find agreement between the TF and \iras\ data,
while the \itf\ analysis of DNW did not? We cannot answer this
question with assurance, but
we can suggest two 
likely causes of the discrepancy.
First, 
the \itf\ analysis requires that the raw magnitude and
velocity width data of the different samples 
be placed on a single,
uniform system. This was achieved by applying linear transformations
to the magnitudes and widths of each sample 
(Willick \etal\ 1997).
Such a procedure in effect links together
the TF zero points of samples that probe different volumes.
Any systematic error in matching the data sets will
manifest itself in spurious large-scale motions;
in particular, the
scale-dependent, dipolar flow found by DNW
(see, for example, their Figures
12 and 13) is fully degenerate with a zero-point error
in the relative TF calibrations of Southern and Northern sky
samples.  
Second, DNW extended their \itf\ analysis to 6000\ \kms,
whereas we have restricted our
analysis to $\czlg\leq 3000\ \kms.$ In so doing, they 
(like \potiras) incorporated
several Mark III TF samples (W91CL, HMCL, W91PP, CF)
not included in the \velmod\ analysis.
It is possible that \iras\ and Mark III
agree locally, but progressively disagree at larger distances.
Alternatively, the possible zero point errors mentioned
above could affect mainly those Mark III samples used by DNW but
not included here, given the agreement we found between the MAT and
A82 distances with the \velmod\ calibrations
(\S~\ref{sec:consistency}). 

Since we believe that the DNW discrepancy between
the \iras\ and TF velocity fields
may well be a result of systematic errors incurred
in matching data sets, an effect to which 
\velmod\ is insensitive, we are inclined
to give more weight to our present conclusion that the \iras--TF
agreement is satisfactory.
However, if in fact the matching of
data sets by DNW is validated by ongoing
observations aimed at providing reliable North-South
homogenization (cf.\ Strauss 1996b), it will be difficult to
escape their conclusion that the predicted and observed
velocity fields do not agree on large scales. 
In that case, it will be
necessary to reexamine the conclusions of this paper
with regard to the value of $\beta_I.$

\subsubsection{The Role of the Quadrupole}

Our conclusion that the predicted and observed velocity fields
agree also depends on the validity of our adopted external quadrupole.
Figure 19 shows that only with the quadrupole does our goodness
of fit statistic $\chi^2_\xi$ take on acceptable values.  We argue in
Appendix~\ref{sec:quad-theory} that the 3.3\% residual quadrupole we
see is mostly due to the systematic difference between the true and
Wiener-filtered \iras\ density field on large scales.  The residual
quadrupole in the mock catalogs is appreciably smaller, ${}<1\%$, but
this can be understood in terms of the different amount of power on
intermediate scales ($2\,\pi/k \approx 100\ h^{-1}$ Mpc) in the mock
catalog and the real universe.  Thus, the presence of the quadrupole
residual is {\em not\/} evidence for a breakdown of our assumptions
of gravitational instability theory and linear biasing.

\subsection{What is the value of $\Omega?$}

Measuring $\beta=\Omega^{0.6}/b$ is an important objective of velocity
analysis. Of course,
the more important objective is determination of $\Omega$ itself. There
are, broadly speaking, two ways to proceed.

\subsubsection{Nonlinear Analysis}
\label{sec:nonlinear}
One may attempt to break the degeneracy between
$\Omega$ and biasing by extending gravitational instability
theory to the nonlinear dynamical
regime. In  an earlier phase of the \velmod\ project, we attempted to
do this; very preliminary results of this effort were described in SW,
\S~8.1.2. In brief, the \iras\ reconstruction was done as described in
Appendix A, but a nonlinear generalization of Eq.~(\ref{eq:vpdelta}),
\begin{equation}
\bfv(\bfr) = \frac{f(\Omega)}{4 \,\pi} \int d^3\bfr'\,
\frac{(1+a^2\nu)\delta[\delta_g(\bfr');b] + a\nu}
{1+a\delta[\delta_g(\bfr');b]}
\frac{(\bfr' - \bfr)}{\left|\bfr' - \bfr\right|^3}\,,
\label{eq:vpnonlin}
\end{equation}
was used to derive
peculiar velocities from the redshift survey density field $\delta_g.$
In Eq.~(\ref{eq:vpnonlin}), $a=0.28$ and 
$\nu \equiv \vev{\delta^2}$ is
the mean square value of $\delta$
(Ganon \etal\ 1995; cf., Nusser \etal\ 1991).
Note that 
the mass fluctuation $\delta$ is written as a generic
function of $\delta_g$ and $b,$ rather than simply as $\delta_g/b.$
This is because once we generalize to nonlinear dynamics,
we must allow for the possibility of {\em nonlinear biasing\/} as
well. There are many ways one might imagine doing this 
(SW, \S~2.5; Fry \& Gazta\~naga 1993).
Generically, however, all these complications can be expanded to
second order to yield a correction to Eq.~(\ref{eq:vpdeltag}):
\begin{equation}
\bfv(\bfr) = \frac{\beta}{4 \,\pi} \int
d^3\bfr'\,\frac{[\delta_g(\bfr') + \gamma (\delta_g^2(\bfr) - \nu)]
(\bfr' - \bfr)}{\left|\bfr' - \bfr\right|^3}\,.
\label{eq:vpdeltag-nonlin}
\end{equation}
where $\gamma$ parameterizes the combined effects of
nonlinear dynamics and
nonlinear biasing.

We carried out a suite of \velmod\ runs using predicted
peculiar velocities based on Eq.~(\ref{eq:vpdeltag-nonlin}) for a
range of values of $\gamma$, both positive and negative.
Our hope was that the \velmod\ likelihood statistic would
be significantly lower for some value of $\gamma$ than for the pure
linear case.
However, to our surprise, we found that the linear dynamics/linear
biasing reconstruction ($\gamma = 0$)
gives the best likelihood of all. We are not certain as to why this is.
Nonlinear dynamics must enter to some degree, because we
know for a fact that $\delta_g$ is not everywhere $\ll 1,$ and indeed
can be quite large with our small smoothing.
(We of course do not know
whether nonlinear biasing is important.)
Nevertheless, the small scatter between the true and \iras\ predicted
peculiar velocity fields for the mock catalogs (\S~\ref{sec:mocksigv})
confirms that the
linear \iras\ velocity field, smoothed on a 300 \kms\ scale, is a
good match to actual peculiar velocities that arise from gravitational
instability, at least in an $N$-body simulation.  

A possible explanation of this seeming contradiction is as
follows. Our method for
predicting peculiar velocities (Appendix A) entails
assigning a smooth, continuous density field from discrete redshift
survey data---a procedure which takes into account the
probability distribution of distance given redshift
(Eq.~\ref{eq:joint-probability}), smooths the data with
a 300 \kms\ Gaussian, and applies a Wiener filter---and thus
reduces small-scale density enhancements.
In doing so, this procedure mimics qualitatively
the effects of nonlinear corrections to the velocity-density
relation. The good match between the \iras\ predictions and
the actual peculiar velocities suggests that this mimicry is
in fact fortuitously good, to the degree that formal nonlinear
corrections are unnecessary. 

\subsubsection{Constraining $\Omega$  from Independent Estimates of $b_I$}
\label{sec:omega}

The second way to estimate $\Omega,$ given our measurement of 
$\beta_I,$ is to 
constrain $b_I$ using
independent information. 
If biasing is independent of scale (cf.\ the discussion in
\S~\ref{sec:velmod-potiras}),  
then $b_I$ is
the ratio of the rms fluctuations of \iras\ galaxies
on an 8\h1\ Mpc scale, $\sigma_8(\iras),$ to the
corresponding mass density fluctuations, $\sigma_8.$
Fisher \etal\ (1994a) found that $\sigma_8(\iras)=0.69\pm 0.04$ in
real space.  It follows
that $\beta_I$ can be viewed as
a prediction of $\sigma_8$ for a given value of
$\Omega:$
\begin{equation}
\sigma_8 = \frac{(0.69 \pm 0.04)\,\beta_I}{\Omega^{0.6}}\,.
\label{eq:sig8-beta}
\end{equation}
An entirely independent (though highly model-{\em dependent}) 
way to predict $\sigma_8$
as a function of $\Omega$ is to use COBE-normalized
power spectra for a range of cosmological parameters.
Liddle \etal\ (1995, 1996) have presented fitting
functions that provide the normalization of CDM power
spectra, in open and flat cosmologies, as a function
of $\Omega,$ $\Omega_\Lambda,$
the Hubble parameter $h\equiv H_0/(100\ \kms\ {{\rm Mpc}}^{-1}),$
and the primordial power spectrum index $n,$ based on the
four-year COBE observations (Bennett \etal\ 1996; G\'orski \etal\
1996). Eke, Cole, \& Frenk (1996)
used these fitting functions to obtain $\sigma_8$
by direct integration of the Liddle \etal\ power spectra,
and have kindly provided us with their code for
doing this calculation. We may thus constrain $\sigma_8$
by comparing the 
\velmod\ and COBE/CDM predictions of its value, and requiring
that they agree to within the errors.
This will be the case only for a limited range of $\Omega$ 
(the ``concordance range'').
We emphasize, however, that the discusion to follow
depends on two uncertain assumptions: first, 
that the CMB fluctuations measured by COBE
can be reliably extrapolated down to 8\h1\ Mpc scales; and
second, that the bias parameter
is scale-independent from 3\h1\ to 8\h1\ Mpc.

\begin{figure}[t!]
\plottwo{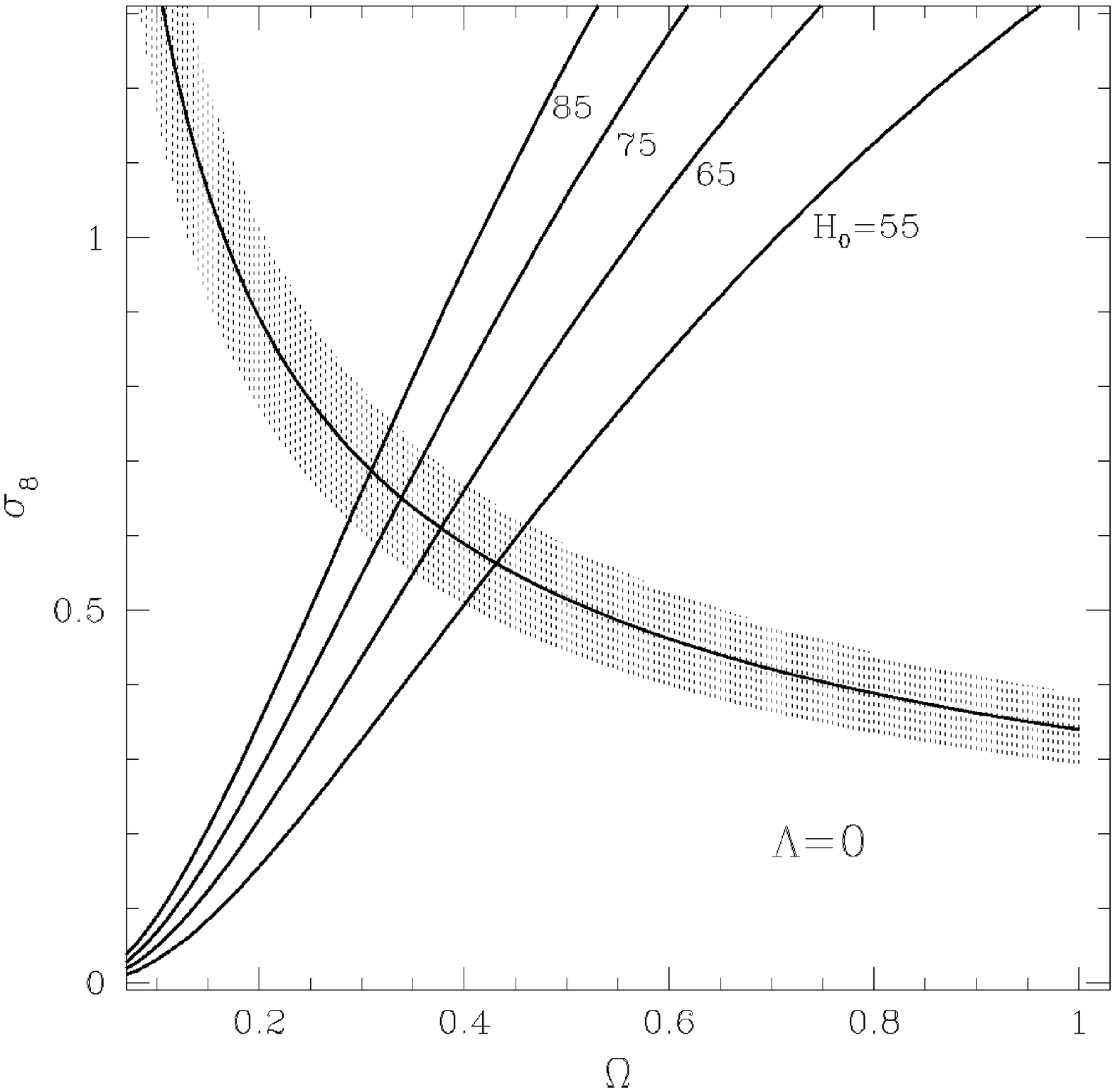}{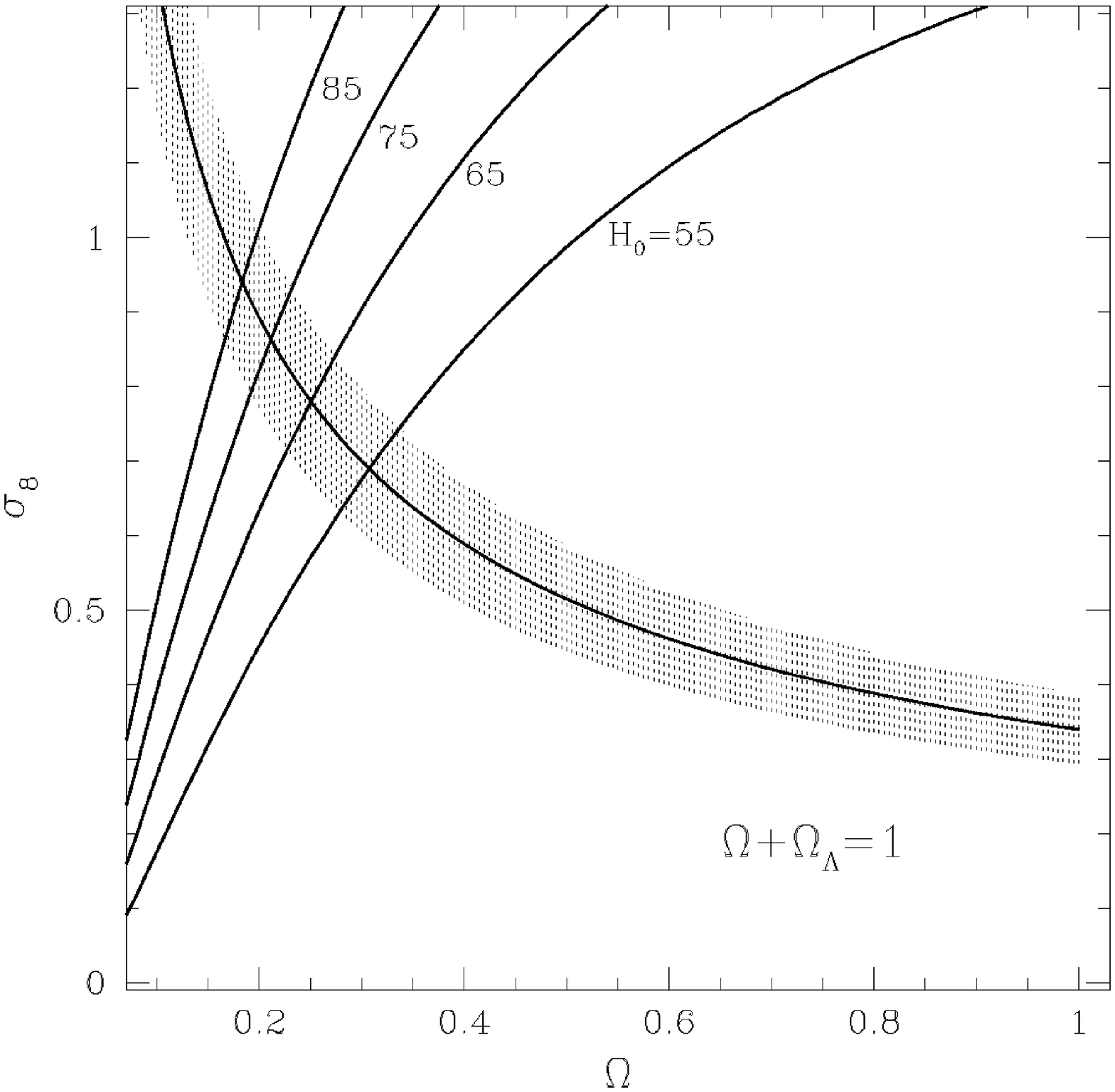}
\caption{{\small Predictions of $\sigma_8$ as a function
of $\Omega$ for open (left panel) and flat (right panel)
universes. The lines sloping  up and to the right
show the COBE-normalized CDM values of $\sigma_8$
as a function of $\Omega,$ for four values of the Hubble constant,
55, 65, 75, and 85 \kms\ Mpc$^{-1},$ under the assumption of
a scale-invariant primordial power spectrum.
The line sloping down and to the right shows
the result from this paper,
$\sigma_8=0.69\beta_I/\Omega^{0.6}.$ The shaded region represents
the 1-$\sigma$ uncertainty in our value of $\beta_I.$}}
\label{fig:cobe-scinv}
\end{figure}

In Figure~20,
we compare the two constraints on $\sigma_8$
for a scale-invariant ($n=1$) power spectrum.
The left hand panel shows results for an
open (i.e., $\Lambda=0$) universe, and the right hand panel
for a spatially flat ($\Omega+\Omega_\Lambda =1$) universe.
The COBE/CDM predictions (solid lines labeled with the values of the Hubble
constant) and the constraint from Eq.~(\ref{eq:sig8-beta}) (shaded
region) scale very differently with $\Omega$, so that
the two together give strong constraints on $\sigma_8$ and thus $\Omega.$
The shaded region represents the combined \velmod\ error on $\beta_I$
and the error in $\sigma_8(\iras)$ from Fisher \etal\ (1994a). We do
not show corresponding error regions for the COBE/CDM predictions
which result from uncertainty in the COBE normalization, because the
error in the predicted $\sigma_8$ is in fact dominated by the allowed
range of $H_0,$
which we take to be
$55 \leq H_0 \leq 85\ \kmsmpc$ based on a number of 
recent measurements (Sandage \etal\ 1996;
Freedman 1996; Riess, Press, \& Kirshner 1996; Mould \etal\
1996; Tonry \etal\ 1997; Kundi\'c \etal\ 1996).

Figure~20 gives the following constraints for $n = 1$. For 
an open model, the concordance range is $\Omega=0.28$--$0.46$
with the low (high) value corresponding to the highest (lowest)
value of $H_0$ considered.
For the flat model, it is $\Omega=0.16$--$0.34.$
Expressed in terms of the \iras\ bias parameter, these ranges
correspond to $b_I=0.92$--$1.38$ (open) and $b_I=0.68$--$1.11$ (flat).
We also considered $n \ne 1$ flat models. 
For example, with $n=0.9$ the concordance ranges 
are $\Omega=0.19$--$0.40,$ and $0.21$--$0.45,$
depending respectively 
on whether tensor fluctuations
are not, or are, included in
the COBE normalization (Liddle \etal\ 1996b).
The corresponding bias parameters are $b_I=0.74$--$1.21$ and
$b_I=0.80$--$1.29.$ 

Two salient points follow from this comparison. 
First, if $H_0\geq 60\ \kmsmpc,$ 
the concordance range for the flat, $n=1$ models requires
$\Omega\simlt 0.30,$ implying $\Omega_\Lambda\simgt 0.70.$
However, studies of gravitational lensing have placed
an upper limit of $\Omega_\Lambda\leq 0.65$ at 95\% confidence
(Maoz \& Rix 1993; Kochanek 1996), while a recent analysis of  
intermediate-redshift Type Ia Supernovae (Perlmutter \etal\ 1996) indicates
$\Omega_\Lambda\leq 0.50$ at 95\% confidence (both of
these constraints apply when a flat universe is assumed).
This contradiction 
consitutes evidence
against a flat universe with a scale-invariant
primordial power spectrum index and $H_0\geq 60\ \kmsmpc.$ 
If $n< 1.0,$ one can more easily accommodate flat universes with
$\Omega_\Lambda<0.65,$ provided the Hubble constant is $\simlt 70\
\kmsmpc.$ 
The second point is that the combined \velmod\ and
COBE/CDM predictions of $\sigma_8$
are extremely difficult to reconcile with an
Einstein-de Sitter universe for most
reasonable values of the remaining cosmological
parameters. If one assumes $n\geq 0.9,$
a Hubble constant $\simlt 30\ \kmsmpc,$
far below current observational limits, would
be required for the concordance range to include
$\Omega=1.$ 
Alternatively, if $H_0=50\ \kmsmpc,$
one would require a primordial power spectrum
index $n=0.7$ and tensor fluctuation
contributions to the CMB anisotropies.
Such a power-spectrum
index is at the lowest end of the range currently
considered plausible in inflationary universe
scenarios (e.g., Steinhardt 1996).

\subsection{Summary}
\label{sec:summary}

We have described a new maximum likelihood method, \velmod, for
comparing Tully-Fisher data with predicted peculiar
velocity fields from redshift surveys. We implemented
the method for a $cz_{{\rm LG}} \leq 3000\ \kms$ TF subsample
from the Mark III catalog (Willick \etal\ 1997), and
velocity fields predicted from the 1.2 Jy \iras\ redshift
survey (Fisher \etal\ 1995). The velocity field prediction
is dependent on the value of $\beta_I\equiv \Omega^{0.6}/b_I,$
where $b_I$ is the bias parameter for \iras\ galaxies
at 300 \kms\ Gaussian smoothing.
We maximized likelihood with respect to $\beta_I,$
the parameters of the TF relation, and several other
velocity parameters.

We applied our method to 20 mock Mark III
and \iras\ catalogs constructed to mimic
the properties of the real data. The mock catalogs were
drawn from an $\Omega=1$ $N$-body simulation 
and were constructed to ensure $b_I=1.$ Thus, the
mock catalogs satisfy $\beta_I=1.$ Our \velmod\ runs with
the twenty mock catalogs returned a mean value of
$\beta_I=0.984\pm 0.018,$ consistent with the statement
that \velmod\ yields an unbiased value of $\beta_I.$
In addition, our mock catalog tests enabled
us to assign reliable 1-$\sigma$ errors to our
estimates of $\beta_I,$  and showed that our other
derived parameters, including those of the TF relation and the
small-scale velocity noise, are also unbiased. 
Because the mock catalogs
came from an $\Omega=1$ universe, triple valued
zones in the mock Virgo region were strong, but
were properly handled by the \velmod\ analysis.

When \velmod\ was applied to the real Mark III data,
a considerably smaller value of $\beta_I$ was derived.
If we assume that the \iras-predicted velocity field fully
describes the actual one, we obtain $\beta_I=0.563\pm 0.074.$
However, the residuals from this
fit were large and coherent; fitting them 
by a quadrupolar flow 
gave a maximum likelihood value of $\beta_I=0.492\pm 0.068$.
The quadrupole points
toward the Ursa Major cluster, and has an
rms amplitude of 3.3\% of the Hubble flow.
In Appendix B, we show analytically 
that a quadrupole of this amplitude is
expected given the way that we smooth the density field; its presence
is {\em not\/} a sign that the \iras\ galaxies do
not trace the mass responsible for the local flow field. 
An analysis of the fit residuals demonstrated that
the \iras-predicted peculiar velocity field, 
with the external quadrupole,
provides a statistically acceptable fit to 
the TF data within 3000 \kms.
The data are thus
consistent with the hypothesis that the peculiar velocities are
due to the gravitational effects of a mass distribution that is
proportional to the \iras\ galaxy distribution. 
We also find that the data are consistent with a
very quiet flow field; the one-dimensional rms noise in the velocity
field relative to the \iras\ model is $125\pm 20\ \kms.$

The value of $\beta_I$ obtained here may also be
thought of as a measurement of the rms mass density
fluctuations $\sigma_8$ as a function
of $\Omega.$ Similarly, COBE-normalized
CDM power spectra predict a value of $\sigma_8$ as
a function of $\Omega$ and other cosmological parameters.
If we require that the \velmod\ and COBE-normalized
calculations agree, 
we can constrain the value of $\Omega.$ 
For scale invariant,
$\Lambda=0$ universes, we derive the constraints
$0.28 \simlt \Omega \simlt 0.46$ for for $85 \simgt H_0
\simgt 55\ \kmsmpc.$ For scale-invariant, flat universes
we find $0.16 \simlt \Omega \simlt 0.34$ for the
same range of $H_0.$ 
The constraints on $\Omega$ shift to higher values (\S~\ref{sec:omega})
if the primordial power spectra
are ``tilted,'' $n<1,$ and if
tensor fluctuations are present. 
However, both extreme tilt ($n\leq 0.7$)
and a Hubble constant at the lowest end of the
observationally allowed range ($H_0\leq 50\ \kmsmpc$) would
be required to reconcile these results with
an Einstein-de Sitter universe.

The conclusions of the previous paragraph all rest, of course,
on the validity of our measurement of $\beta_I.$
Tests with mock catalogs show that, subject to our basic assumptions,
this measurement is reliable
to within the quoted errors. We have identified
two ways these assumptions can break
down. First, the effective bias factor $b_I$
could depend on scale. In that case, our measurement of $\beta_I,$
which reflects a 300 \kms\ Gaussian smoothing scale, 
might not be the same as a measurement
obtained at larger smoothing; it would not then be
valid to equate the estimate of $\sigma_8$ obtained
from Eq.~\ref{eq:sig8-beta} with the COBE/CDM prediction.
Second, although we have found agreement between the
predicted and observed peculiar velocities within 3000 \kms,
DNW found disagreement on larger scales. If the DNW result
is validated by future observations (Strauss 1996b) aimed
at improved TF calibration across the sky, 
our present claim of TF-\iras\
agreement will be undermined. 

There are several areas for further work.  One, alluded to in
several places in this paper, is to extend our analysis to larger
redshift, using both the forward and inverse forms of the TF
relation.  This can be done both the Mark III data, and with the
extensive new TF (Mathewson \etal\ 1994; Giovanelli \etal\ 1997) and
$D_n$-$\sigma$ (Saglia \etal\ 1996) samples that are being compiled.
We should also consider extending this work to other distance
indicators; 
surface brightness fluctuation galaxies (Tonry \etal\
1997), with their accurate sampling of the nearby velocity field, are
natural candidates for the \velmod\ analysis.  
On the modeling side, this
work has left us with several conundrums, the most puzzling of which
is why the linear \iras\ model does so well with a smoothing scale of
300 \kms.  More work is needed with $N$-body simulations to understand
this.  
Finally,
we will not have a coherent picture of the relationship between the
velocity and density fields until we can 
understand the different values of $\beta_I$ obtained by \velmod\ and \potiras.

\acknowledgements
We thank Marc Davis, Carlos Frenk, and Amos Yahil for extensive discussions of
various aspects of this project, as well as the support of the entire
Mark III team: David Burstein, St\'ephane Courteau, and Sandra
Faber. JAW and MAS are grateful for the hospitality of the Hebrew
University in Jerusalem, Lick Observatory at the University of
California, Santa Cruz, and the Astronomy Department of the University
of T\-oky\-o for visits while we worked on this paper.
MAS gratefully acknowledges the support of an Alfred
P. Sloan Foundation Fellowship.
This work was supported in part by
the US National Science Foundation grant PHY-91-06678,
the US-Israel Binational Science Foundation grants 92-00355 and 95-00330, 
and the Israel Science Foundation grant 950/95. 

\appendix

\section{The \iras\ Velocity-Density Reconstruction}
\label{sec:iras}

  The redshifts of galaxies in the \iras\ sample are affected by the
same peculiar velocities that one is attempting to measure in the Mark
III dataset.  If we measure redshifts $cz$ in the rest
frame of the Local Group, then:
\begin{equation} 
cz = r + \hat \bfr\cdot\left[\bfv(\bfr) - \bfv({\bf 0})\right]\quad,
\label{eq:cz-r} 
\end{equation}
where \bfv({\bf 0}) is the peculiar velocity of the Local Group,
and $\bfv(\bfr)$ is the peculiar velocity at position $\bfr.$
Indeed, because the galaxy density field shows coherence, the galaxy
density field measured in redshift space $\delta_g(\bfs)$ differs systematically from
that in real space, $\delta_g(\bfr)$, as was first described in detail by Kaiser (1987;
cf., SW; Strauss 1996a for reviews).  
Linear perturbation theory assuming gravitational instability enable us to correct for
the effects of these velocities.  We use here the iteration technique
described by Yahil \etal\ (1991) and Strauss \etal\ (1992c), as updated by Sigad
\etal\ (1997).  The density and velocity field are calculated within a
sphere of radius 12,800 \kms; the density fluctuation
field is assumed to be zero
beyond this radius.  Here we very briefly reiterate the improvements
described in the Sigad \etal\ paper, and emphasize certain differences
from the approach there. 

  In regions in which the \iras\ velocity field model predicts a
non-monotonic relation between redshift and distance along a given
line of sight, it becomes ambiguous how to assign a distance to a
galaxy given its redshift (Figure~1).  Our approach is
similar to that used 
throughout this paper: we use our assumed density and velocity field
to calculate a probability distribution of a galaxy along a given line
of sight.

Along a given line of sight, 
we ask for the joint probability distribution of observing a galaxy
along a given line of sight, with redshift $cz$, flux density $f$
and (unknown) distance $r$:
\begin{equation} 
P(cz,f,r) = P(cz|r)\times P(f|r) \times P(r)\quad;
\label{eq:joint-probability} 
\end{equation}
compare with Eq.~(\ref{eq:pmetazr}). 
The first term is given by our velocity field model along the line of
sight, and is thus given by Eq.~(\ref{eq:pzr}). 
For the iteration code, we set $\sigma_v = 150\ \kms$, independent of
position, similar to the best fit value we find when we {\em fit\/}
for $\sigma_v$ from the velocity field data. 

The second term is given by the luminosity function of
galaxies:
\begin{equation} 
P(f|r) = \Phi(L = 4\,\pi r^2 \nu f) {d L \over d f} \propto r^2 
\Phi(L)\quad,
\label{eq:P(f|r)} 
\end{equation}
where the derivative is needed because the probability density is
defined in terms of $f$, not $L.$\footnote{Eq.~(144) of Strauss \&
Willick (1995) mistakenly left off this last term.}  
Finally, the third term in
Eq.~(\ref{eq:joint-probability}) is given by the galaxy density distribution
along the line of sight, Eq.~(\ref{eq:pofr}). 

As described in Sigad \etal\ (1997), the calculations of the velocity
and density fields are done on a Cartesian grid.  Our approach
therefore is to assign each galaxy to the grid via cloud-in-cell
(weighting by the selection function, of course), where (unlike Sigad
\etal\ 1997) we distribute
each galaxy along the line of sight according to the distribution
function of expected distance, Eq.~\ref{eq:joint-probability}.  In order 
to calculate the
selection function for an object, we of course need to have a definite
position for it; for this purpose, we assign it the expectation value
of its distance, following Sigad \etal\ (1997):
\begin{equation} 
\vev{r} = {\int r P(cz,f,r)\, d r \over \int P(cz, f, r)\, d r}. 
\label{eq:r-expect} 
\end{equation}

  Sigad \etal\ (1997) discuss the use of various filtering techniques
to suppress the shot noise in the derived density and velocity
fields.  While they argue for the use of a power-preserving filter for
the comparison of the \iras\ and \potent\ 
density fields, we have found through extensive experimentation with
mock catalogs that for the \velmod\ analysis, a Wiener filter gives the
best comparison between the density field and the peculiar velocity
data.  

  Finally, we found that when the iteration technique was run to
values of $\beta \simgt 1$, the density field became unstable
in the regions around triple-valued zones, oscillating between
iterations.  We were able to suppress these by averaging the derived
density field at each iteration with that of the iteration
preceding it.  This has no strong effect on the derived density field
for $\beta < 1$. 

\section{The Residual Quadrupole }
\label{sec:quad-theory}

In this Appendix, we calculate the expected amplitude of the
velocity quadrupole generated by density fluctuations both external to the
\iras\ sample (i.e., outside of $R = 12,800\, \kms$), and internal to
it, due to the difference between the true density field and the
noisy, smoothed estimation of the density field we have from the
\iras\ redshift survey.  The \iras\ excluded zone is another potential
source of quadrupole error, but it is filled in by interpolation from
regions above and below the excluded zone (Yahil \etal\ 1991), a
procedure which agrees well with a multipole interpolation procedure
based on spherical harmonics, at least for the $10^\circ$ wide \iras\
zone of avoidance (Lahav \etal\ 1994). 

\subsection{The Quadrupole Induced by Fluctuations Beyond the \iras\ Volume}
We express peculiar velocity in terms of a potential
function $\Phi(\bfr)$, such that the radial component of the velocity
field is given by $u(r)=-\partial\Phi/
\partial r.$  We will isolate the quadrupole component of this
potential, and calculate its angle-averaged rms contribution. 

The contribution to $\Phi$ from material at distances $>R$
is given by
\begin{equation}
\Phi(\bfr) = -\frac{f(\Omega)}{4\pi} \int_{|\bfr'|>R} 
d^3 \bfr' \,\frac{\delta(\bfr')}{|\bfr-\bfr'|}\,.
\label{eq:phi1}
\end{equation}
Here, 
$\delta$ is the
{\em mass,} not the galaxy, density fluctuation.
We now expand the denominator in the integrand in terms
of spherical harmonics (e.g., Jackson 1976, Eq.~3.70) and isolate the
quadrupole term to obtain
\begin{equation}
\Phi_Q(\bfr) = -\frac{f(\Omega) r^2}{5} \sum_{m=-2}^2 Y_{2m}(\omega)
\int_R^\infty \! \frac{dr'}{r'}  \int d\omega'\,\delta(\bfr')\,Y_{2m}^*(\omega')\,.
\label{eq:phi3} 
\end{equation}
where $\omega$ is solid angle.
Taking the radial component of the quadrupole velocity $u_Q =
-\partial\Phi_Q/\partial r$, squaring, and
averaging over solid angle gives, after several steps of algebra:
\[ u_{Q,rms}^2(r) = \frac{1}{4\pi}\int\!d\omega\,u_Q^2(\bfr) \]
\begin{equation}
= \left( \frac{2f(\Omega)r}{5}\right)^2 \sum_{mm'} C_{2m} C^*_{2m'} 
\frac{1}{4\pi}\int d\omega\,Y_{2m}(\omega) Y^*_{2m'}(\omega)
= \left( \frac{2f(\Omega)r}{5}\right)^2 \frac{1}{4\pi}\sum_m |C_{2m}|^2\,,
\label{eq:uQrms}
\end{equation}
where the last step follows from the orthonormality of
the $Y_{lm}$'s, and 
for convenience we have defined the five complex coefficients
\begin{equation}
C_{2m}[R;\delta] \equiv \int_R^\infty \! \frac{dr'}{r'}  
\int d\omega'\,\delta(\bfr')\,Y_{2m}^*(\omega')\,.
\label{eq:c2m}
\end{equation}

The expectation value of $|C_{2m}|^2$ is independent of $m$,
so when we take the expectation value of Eq.~(\ref{eq:uQrms}), we can
replace the sum with 5 times $\vev{C_{20}^2}$:
\begin{equation}
\langle  u_{Q,rms}^2(r) \rangle = \left( \frac{2f(\Omega)\,r}{5}\right)^2\!
\times \frac{5}{4\pi}\,\langle C_{20}^2 \rangle\,.
\label{eq:uqrms-stat1}
\end{equation}
Using the definition of $Y_{20}$ in terms
of the second Legendre polynomial $P_2$ in Eq.~(\ref{eq:c2m}) and
Eq.~(\ref{eq:uqrms-stat1})  gives
\begin{equation}
\vev{u_{Q,rms}^2(r)}= (f(\Omega)\,r)^2\int_R^\infty \frac{dr_1}{r_1} \int_R^\infty \frac{dr_2}{r_2}
\int_{-1}^1\int_{-1}^1 d\mu_1 d\mu_2\,P_2(\mu_1) P_2(\mu_2)
\,\xi(|\bfr_2 - \bfr_1|)\,.
\label{eq:c20}
\end{equation}
Expressing the correlation function $\xi$ as the Fourier Transform of
the power spectrum $P(k)$ (e.g., SW, Eq.~46) allows the integrals over
$\bfr_1$ and $\bfr_2$ to separate. This yields
\begin{equation}
\langle u^2_{Q,rms}(r) \rangle = \frac{(f(\Omega)\,r)^2}{(2\pi)^3} \int d^3 \bfk \,P(k)\,
\widetilde W^2(kR)\,,
\label{eq:u-with-P(k)}
\end{equation}
where the kernel is given by
\begin{equation}
\widetilde W(kR) = \int_R^\infty{d r \over r} \int_{-1}^1 \!d\mu\, e^{ikr\mu}
P_2(\mu)
 = -2 \int_R^\infty\!dr\,\frac{j_2(kr)}{r} = {2\,j_1(k R)\over k R}\,,
\label{eq:kernel}
\end{equation}
and $j_n$ is the $n$-th order spherical Bessel function. Comparison
of Eqs.~(\ref{eq:u-with-P(k)}) and~(\ref{eq:kernel})
with Eqs.~(37) and (38) of SW allows us to recast our result as
\begin{equation}
r^{-1} \vev{u_{Q,rms}^2(r)}^{1/2} = \frac{2 f(\Omega)}{3}\,\sigma_R
\label{eq:uqrms-final}
\end{equation}
for the expected rms quadrupole velocity on a sphere due
to mass density fluctuations at distances $>R,$ expressed
as a fraction of Hubble flow.  Here $\sigma^2_R$ is the variance in
the mass overdensity  within spheres of radius $R.$  As mentioned in
the text, this gives a fractional quadrupole of the order of 1-2$\,\%$
for a variety of COBE-normalized power spectra.

\subsection{The Effects of Wiener Filtering and Shot Noise}

The Wiener filter operates on the Fourier Transform of the \iras\
density field.  The final density field differs from the true density
fiels for two reasons: the discreteness of the galaxy distribution
gives rise to shot noise, and the Wiener filter, while suppressing
shot noise, also suppresses the density field itself.  We calculate
the contribution to the quadrupole from both effects. 

Let $\tilde \delta_T(\bfk)$ represent the true Fourier
component of the underlying (noiseless) density field at wavevector
$\bfk$; the quantity with which we calculate the velocity field is
the Wiener-filtered noisy image, whose Fourier modes are given by:
\begin{equation} 
\tilde \delta(\bfk) = h(k,r) [\tilde \delta_T(\bfk) + \epsilon(k)]\,,
\label{eq:delta-filter} 
\end{equation}
where the Wiener filter itself is (e.g., Zaroubi \etal\ 1995):
\begin{equation} 
h(k,r) = {P(k) \over P(k) + (n_1 \phi(r))^{-1}}\,,
\label{eq:wiener-filter} 
\end{equation}
and $P(k)$ is set {\em a priori\/}; we used a functional fit to the
\iras\ power spectrum found by Fisher \etal\ (1993).  The noise term
in the denominator of the Wiener filter is independent of $k$ (cf.,
Fisher \etal\ 1993; SW, \S~5.3); 
however, it is dependent on the density of galaxies, which is a
decreasing function of distance in the flux-limited \iras\ sample.
As explained in Sigad \etal\ (1997), we therefore calculate a {\it
series\/} of Wiener-filtered density fields for different noise
levels, and interpolate between them to find the appropriate density
field at any given distance. 

We wish to calculate the quadrupole due to the {\em error} in the
derived density field, i.e., that due to the difference between
Eq.~(\ref{eq:delta-filter}) and $\tilde \delta_T(\bfk)$.  If we expand
the density field in Eq.~(\ref{eq:c2m}) into its Fourier components,
substitute this difference for each component, and square the result,
we find the rms contribution to $u_Q$ due to the Wiener filter:
$$
\vev{u_{Q,Wiener}^2} = {(r\,f(\Omega))^2 \over (2\,\pi)^8} \int
d^3\bfk_1 d^3 \bfk_2 \int {d^3 
\bfr_1 \over r_1^3} {d^3 \bfr_2 \over r_2^3} P_2(\mu_1) P_2(\mu_2)
\exp[i(\bfk_1\cdot \bfr_1 - \bfk_2\cdot \bfr_2)] \times
$$
\begin{equation} 
\vev{
\left[(h(k_1,r_1) - 1) \tilde \delta_T(\bfk_1) + h(k_1,r_1)\epsilon(k_1)\right]
\,
\left[(h(k_2,r_2) - 1) \tilde \delta_T(\bfk_2) +
h(k_2,r_2)\epsilon(k_2)\right]}\,.
\label{eq:c20-expanded} 
\end{equation}
This rather horrific expression can be simplified by multiplying out
the term in brackets,
realizing that the cross-terms
vanish and that $\vev{\tilde\delta_T(\bfk_1)\tilde\delta_T(\bfk_2)} =
(2\,\pi)^3 P_T(k)\delta_D(\bfk_1 - \bfk_2)$, where $P_T(k)$ is the {\it
true} underlying power spectrum, not necessarily the same as that
assumed in Eq.~(\ref{eq:wiener-filter}).  We then get two terms, one
depending on the power spectrum, and the other due to shot noise.  
For the first term, the integrals over $\bfr_1$ and $\bfr_2$ separate to give:
\begin{equation} 
\langle u^2_{Q,Wiener}(r) \rangle = \frac{(f(\Omega)\,r)^2}{(2\pi)^3}
\int d^3 \bfk \,P_T(k)\, \widetilde W^2_\Delta(kR) + \vev{u^2_{Q,shot}}\,,
\label{eq:C20inner}
\end{equation}
where the new window function is given by 
\begin{equation}
\widetilde W_\Delta (k,R_1,R) = 
-2 \int_{R_1}^R\!dr\,\frac{j_2(kr)}{r} [h(k,r) - 1)]\,;
\label{eq:newkernel}
\end{equation}
compare with Eq.~(\ref{eq:kernel}). We integrate from the outer volume
of our peculiar velocity sample, $R_1=3000 \kms,$ 
to $R = 12,800$ \kms; at
smaller radii, the contribution to the quadrupole goes like $r^{-2}$,
not $r$, and this is not included in our modelling of the quadrupole
(Eq.~\ref{eq:quad}). 
The contribution to the quadrupole from this term is between 1.5 and
3\%, depending on which model we take for the true power spectrum. 
This is pleasingly close to the value we find for the real universe.
The mock catalogs have a power spectrum set by the observed \iras\
power spectrum (of course, with a cutoff at $k < 2\,\pi/L$), and thus
give a somewhat smaller contribution to this integral, about 1\%. 

  Let us now calculate the shot noise contribution to the quadrupole. 
It is given by:
\begin{equation} 
\vev{u^2_{Q,shot}} = 
{(r\,\beta)^2 \over (2\,\pi)^8} \int d^3\bfk_1 d^3 \bfk_2 \int {d^3
\bfr_1 \over r_1^3} {d^3 \bfr_2 \over r_2^3} P_2(\mu_1) P_2(\mu_2)
\exp[i(\bfk_1\cdot \bfr_1 - \bfk_2\cdot \bfr_2)] \,\vev{
h(k_1,r_1)\epsilon(k_1)\,h(k_2,r_2)\epsilon(k_2)}\,.
\label{eq:shot-mess} 
\end{equation}
Notice now the dependence on $\beta$, not $\Omega$; here we will make
no reference to a COBE-normalized power spectrum. 
The Fourier modes are calculated in a box of side $L = 25,600 \ \kms$,
and therefore are uncorrelated for $\Delta k > 2\,\pi/L$. 
Thus we can write the product of the two shot noise terms as a Dirac
delta function:
\begin{equation} 
\vev{\epsilon(\bfk_1)\epsilon(\bfk_2)} =
\vev{\epsilon^2(\bfk_1)} \left({2\,\pi \over L}\right)^3
\delta_D(\bfk_1 - \bfk_2) = 
\left({2\,\pi \over L}\right)^3 \delta_D(\bfk_1 - \bfk_2) \int d^3\bfr
{1 \over n_1 \phi(r)}\,;
\label{eq:delta-covar} 
\end{equation}
the expression for $\vev{\epsilon^2(\bfk)}$ comes from Fisher \etal\
(1993).  When we insert Eq.~(\ref{eq:delta-covar}) into
Eq.~(\ref{eq:shot-mess}), the latter simplifies dramatically.  The
integrals over $\bfr_1$ and $\bfr_2$ now separate, giving:
\begin{equation} 
 \langle u_{Q,shot}^2(r) \rangle = {(r\beta)^2 \over
 (2\,\pi)^3} \int {d^3\bfk\over L^3} \left(\int d^3\bfr' {1 \over {n_1
\phi(r')}}\right) \,W^2_{shot}(k)\,, 
\label{eq:u-q-shot-final} 
\end{equation}
where the shot noise window function looks very similar to what we have
seen before:
\begin{equation} 
W_{shot}(k,R_1,R) = 
-2 \int_{R_1}^R\!dr\,\frac{j_2(kr)}{r} h(k,r)\,.
\end{equation}
Notice that unlike the previous calculation, this result is
independent of the true power spectrum. If we calculate this using the
observed \iras\ selection function, integrating from 3000 \kms\ to
12,800 \kms, we find an rms quadrupole of 
$r^{-1} \langle u_{Q,shot}^2(r) \rangle^{1/2} = 1.7\, \beta\%$. 

We conclude that the 3.3\% quadrupole found for the real data can be
understood as a combination of the three effects discussed here: power
on scales larger than the \iras\ sample, the Wiener suppression
factor, and shot noise; the Wiener suppression factor is the dominant
one of the three.  For the mock catalogs, we still do not completely
understand why the measured residual quadrupole ($<1\%$) is smaller
than we have calculated ($\sim 2$\%). 

\section{Properties of the Statistic $\chi^2_\xi$}
\label{sec:chi2xi}

In \S\ref{sec:resid}, we introduced the statistic $\chi^2_\xi$ (Eq.~\ref{eq:chi2xi}) as
a measure of the coherence of the residual field between the \iras\ and TF data. 
Here we demonstrate that
it has approximately the properties of a true $\chi^2$ statistic, and indicate how
and why it departs from true $\chi^2$ behavior.

The measure of residual coherence at separation $\tau$ is
\begin{equation}
\xi(\tau)=\sum \begin{Sb}
i<j \\
d_{ij}=\tau\pm\Delta\tau 
\end{Sb} \delta_{m,i}\delta_{m,j}
\label{eq:defxi}
\end{equation}
where $d_{ij}$ is the separation in \iras-distance space between objects
$i$ and $j,$ and $\delta_m$ is the normalized magnitude residual, Eq.~(\ref{eq:normresid}). 
The sum runs over the $N_p(\tau)$ distinct pairs of objects with
separation $\tau\pm\Delta\tau;$ note that a given object may appear
in more than one of these pairs. The hypothesis we wish to test is that
the \iras-TF residuals are incoherent, which signifies a good fit on
all scales. A formal statement of this condition is that the individual
$\delta_{m,i}$ are independent random variables. Furthermore, the $\delta_m$
have been constructed to have mean zero and unit variance. Thus, our
hypothesis of uncorrelated residuals implies that the expectation value of
the product $\delta_{m,i}\delta_{m,j}$ vanishes for $i\neq j,$ and that
the expectation value of its square is unity. 

It follows that
\begin{equation}
E[\xi(\tau)] = \sum \begin{Sb}
i<j \\
d_{ij}=\tau\pm\Delta\tau 
\end{Sb} E(\delta_{m,i}\delta_{m,j}) = 0\,.
\label{eq:exi}
\end{equation}
The variance of $\xi(\tau)$ is
\begin{equation}
E[\xi^2(\tau)] = \sum \begin{Sb}
i<j\\
d_{ij}=\tau\pm\Delta\tau
\end{Sb} \sum \begin{Sb}
k<l\\
d_{kl}=\tau\pm\Delta\tau
\end{Sb} E(\delta_{m,i}\delta_{m,j}\delta_{m,k}\delta_{m,l})\,.
\label{eq:exi2}
\end{equation}
Now, the expectation value within the sum will vanish under our assumption
of uncorrelated residuals unless $i=k$ and $j=l.$ (Notice that we cannot
have $i=l$ and $j=k$ because of the ordered nature of the summation.) Thus,
the only nonzero terms in Eq.~(\ref{eq:exi2}) are identical pairs, and
it follows that $E[\xi^2(\tau)]=N_p(\tau).$

Because $\xi(\tau)$ is the sum of $N_p(\tau)$ random variables
each of zero mean and unit variance, we are tempted to suppose that, by
the central limit theorem, its distribution is Gaussian with mean zero
and variance $N_p(\tau)$ when $N_p(\tau)$ is large. Indeed, for the
200 \kms\ bins used in its construction (cf.\ \S\ref{sec:residcorr}), $N_p$
is typically $\simgt 10^4.$ And, as shown in the previous paragraph,
$\xi(\tau)$ does indeed have mean zero and variance $N_p(\tau).$ One may also
ask about the correlation among the $\xi(\tau)$ for different $\tau.$
Specifically, one may compute
\begin{equation}
E[\xi(\tau_1)\xi(\tau_2)] = \sum \begin{Sb}
i<j\\
d_{ij}=\tau_1\pm\Delta\tau
\end{Sb} \sum \begin{Sb}
k<l\\
d_{kl}=\tau_2\pm\Delta\tau
\end{Sb} E(\delta_{m,i}\delta_{m,j}\delta_{m,k}\delta_{m,l})\,.
\label{eq:exi12}
\end{equation}
Now, it is possible to have $i=k$ within this sum. However, because 
$\tau_1\neq \tau_2,$ if $i=k$ then $j\neq l.$ Similarly,
one may have $j=l,$ but in that case $i\neq k.$ Thus, all of the individual
expectation values in the sum vanish, and we find $E[\xi(\tau_1)\xi(\tau_2)]=0.$
To the extent the above considerations hold, the $\xi(\tau_i)$
are independent Gaussian random variables of variance $N_p(\tau_i).$
It then follows that the statistic $\chi^2_\xi$
is distributed like a $\chi^2$ variable with $M$ degrees of freedom. This
is the statistic proposed in the main text as a measure of goodness of fit.

However, the
central limit theorem applies only to sums of {\em independent\/}
random variables. The individual products $\delta_{m,i}\delta_{m,j}$ which
enter into $\xi(\tau)$ are
{\em uncorrelated\/} in the specific sense 
$E(\delta_{m,i}\delta_{m,j})E(\delta_{m,k}\delta_{m,l})
=\delta^K_{i,k}\delta^K_{j,l}$ (where $\delta^K$ is the Kronecker-delta symbol).
However, they
are not strictly {\em independent} from one another.  This
is because the same object can occur in more than one pair
at a given $\tau.$ We thus expect
the central limit to apply only approximately, and the $\xi(\tau)$
as a result are not strictly Gaussian. As a result, $\chi^2_\xi$
cannot be a true $\chi^2$ statistic. 

Furthermore, just as a single object appears in many pairs at
a given $\tau,$ it can appear in pairs at different $\tau$ as
well. Suppose object $i$ contributes to both $\xi(\tau_1)$ and
$\xi(\tau_2).$ Then the latter are not strictly independent,
even though the expectation value of their product vanishes,
as shown above. This factor, too, will result in a departure
from $\chi^2$ behavior. 

\vfill\eject

\vfill\eject

\begin{deluxetable}{rrrr}
\footnotesize
\tablecaption{Comparison of True Parameters with Means
from \velmod\
Analyses of Mock Catalogs\label{table:mock}}
\tablewidth{0pt}
\tablehead{Quantity&Input Value&Mock Results\tablenotemark{a}&Typical Error\tablenotemark{b}}
\startdata
$\beta_I$&1.0&$0.984\pm0.017$&0.08\nl
$\sigma_v$&147&$149 \pm 5$&$20\ \kms$\nl
${\bf w}_{{\rm LG,x}}$\tablenotemark{c}&$89 \pm 8$&$77\pm12$&$54\ \kms$\nl
${\bf w}_{{\rm LG,y}}$\tablenotemark{c}&$-51 \pm 10$&$-50\pm14$&$63\ \kms$\nl
${\bf w}_{{\rm LG,z}}$\tablenotemark{c}&$-57 \pm 9$&$-55\pm10$&$45\ \kms$\nl
$\rm b_{A82}$&10.0&$10.12 \pm 0.08$&0.36\nl
$\rm A_{A82}$&$-13.40$\tablenotemark{d}&$-13.44 \pm 0.02$&0.09\nl
$\rm \sigma_{TF,A82}$&0.45&$0.460 \pm 0.006$&0.026\nl
$\rm b_{MAT}$&6.71&$6.68 \pm 0.05$&0.22\nl
$\rm A_{MAT}$&$-5.86$\tablenotemark{d}&$-5.92 \pm 0.02$&0.09\nl
$\rm \sigma_{TF,MAT}$&0.42&$0.419 \pm 0.003$&0.013\nl
\enddata
\tablenotetext{a} {Errors given are in the mean.}
\tablenotetext{b} {Errors in a single realization.}
\tablenotetext{c} {Cartesian coordinates defined by Galactic
coordinates.}
\tablenotetext{d} {These true zero points differ
from those reported by Kolatt \etal\ (1996), Table 1,
because they measured distances in Mpc, whereas
we use \kms.}
\end{deluxetable}

\begin{deluxetable}{rrl}
\footnotesize
\tablecaption{Numerical Results from \velmod\ analysis of Real Data
\label{table:real}}
\tablewidth{0pt}
\tablehead{Quantity&\multicolumn{1}{c}{Value}&Comments}
\startdata
${\cal V}_Q(1,1)$&$37\  \kms$& at 2000 \kms; cf.\ Eq.~\ref{eq:quad}\nl
${\cal V}_Q(2,2)$&$36\  \kms$&$"$\nl
${\cal V}_Q(1,2)$&$15\  \kms$&$"$\nl
${\cal V}_Q(1,3)$&$113\ \kms$&$"$\nl
${\cal V}_Q(2,3)$&$-24\ \kms$&$"$\nl
$\sigma_v$&125 \kms\nl
${\bf w}_{{\rm LG,x}}$&$-30\ \kms$\nl
${\bf w}_{{\rm LG,y}}$&$-10\ \kms$\nl
${\bf w}_{{\rm LG,z}}$&$30\ \kms$\nl
$\rm b_{A82}$&$10.36\pm 0.36$&$\;10.29\pm 0.22$ (Mark III value)\nl
$\rm A_{A82}$&$-5.96\pm 0.09$&$-5.95\pm 0.04$ (Mark III value)\nl
$\rm \sigma_{TF,A82}$&$0.464\pm 0.026$&$\;\;\;0.47\pm 0.03$ (Mark III value)\nl
$\rm b_{MAT}$&$7.12\pm 0.22$&$\;\;\;6.80\pm 0.08$ (Mark III value)\nl
$\rm A_{MAT}$&$-5.75\pm 0.09$&$-5.79\pm 0.03$ (Mark III value)\nl
$\rm \sigma_{TF,MAT}$&$0.453\pm 0.013$&$\;\;\;0.43\pm 0.02$ (Mark III value)\nl
$\beta_I$&$0.492\pm0.068$&With Quadrupole\nl
$\beta_I$&$0.563\pm0.074$&Without Quadrupole\nl
$\beta_I$&$0.489\pm0.084$&A82 data only\nl
$\beta_I$&$0.498\pm0.107$&MAT data only\nl
$\beta_I$&$0.453\pm0.093$&$0 < \czlg \le 1350\ \kms$\nl
$\beta_I$&$0.495\pm0.133$&$1350 < \czlg \le 2150\ \kms$\nl
$\beta_I$&$0.573\pm0.142$&$2150 < \czlg \le 3000\ \kms$\nl
$\beta_I$&$0.521\pm0.050$&$\bfwlg = 0;\ \sigma_v$ fixed to 250 \kms\nl
$\beta_I$&$0.491\pm0.045$&$\bfwlg = 0;\ \sigma_v$ fixed to 150 \kms\nl
$\beta_I$&$0.544\pm0.071$&With Quadrupole; 500 \kms\ smoothing\nl
$\beta_I$&$0.635\pm0.083$&Without Quadrupole; 500 \kms\ smoothing\nl
$\beta_I$&$0.510\pm0.038$&TF parameters fixed at Mark III values; with quadrupole\nl
$\beta_I$&$0.517\pm0.039$&TF parameters fixed at Mark III values; without quadrupole\nl

\enddata
\tablenotetext{}{We did not do a likelihood search in parameter space to find
formal error bars on quantities other than $\beta_I$.  Error estimates
for the TF parameters come from averaging over the
mock catalog \velmod\ runs;
see Table~\ref{table:mock}.}
\end{deluxetable}

\end{document}